\documentclass[12pt]{amsart}
\usepackage{amssymb}
\usepackage{amsmath}
\usepackage{amsfonts}
\usepackage{mathrsfs}
\usepackage{graphicx}
\usepackage{color}
\usepackage[onehalfspacing]{setspace}
\usepackage{caption}
\usepackage{natbib}
\usepackage{enumerate}
\usepackage[utf8]{inputenc}
\usepackage{charter}
\usepackage[colorlinks=true,citecolor=blue,urlcolor=blue,pdfpagemode=UseNone,pdfstartview=FitH]{hyperref}
\usepackage{apptools}
\usepackage{placeins}
\usepackage{multibib}
\newcites{main,supp}{References,References}
\makeatletter
\def\section{\@startsection{section}{1}
	\z@{1.0\linespacing\@plus\linespacing}{.8\linespacing}{\Large}}

\def\subsection{\@startsection{subsection}{2}
	\z@{.8\linespacing\@plus.7\linespacing}{.7\linespacing}{\large}}

\def\subsubsection{\@startsection{subsubsection}{3}
	\z@{.5\linespacing\@plus.7\linespacing}{-.5em}{\normalfont\bfseries}}
\makeatother

\numberwithin{equation}{section}

\newtheorem{theorem}{Theorem}[section]
\newtheorem{lemma}{Lemma}[section]

\theoremstyle{definition}
\newtheorem{definition}{Definition}[section]

\theoremstyle{definition}
\newtheorem{assumption}{Assumption}[section]

\theoremstyle{definition}

\title{}
\begin{document}
	\vspace*{3ex minus 1ex}
	\begin{center}
		\Large \textsc{Estimating Local Interactions Among Many Agents Who Observe Their Neighbors}
		\bigskip
	\end{center}
	
	\date{%
		%TCIMACRO{\TeXButton{Today}{\today}}%
		%BeginExpansion
		\today%
		%EndExpansion
	}
	
	\vspace*{3ex minus 1ex}
	\begin{center}
		Nathan Canen, Jacob Schwartz, and Kyungchul Song\\
		\textit{University of Houston, University of Haifa, and University of British Columbia}\\
		\medskip
		
	\end{center}
	
	\thanks{This research has benefited from conversations with Qingmin Liu and Jo\~{a}o Ramos at its initial stage. We also thank Don Andrews, Khai Chiong, Tim Conley, Sokbae Lee, Michael Leung, Xiaodong Liu, and seminar participants at University of Colorado at Boulder, UBC, University of Southern California, and participants at Cemmap Conference in Kyoto and Canadian Econometrics Study Group Meeting for valuable comments and questions. We would like to thank Mike Peters and Li Hao for valuable conversations and suggestions. We also thank a co-editor and three anonymous referees for valuable comments, criticisms, and suggestions. All errors are ours. Song acknowledges that this research was supported by Social Sciences and Humanities Research Council of Canada. Corresponding address: Kyungchul Song, Vancouver School of Economics, University of British Columbia, 6000 Iona Drive, Vancouver, BC, Canada, V6T 1L4. Email address: kysong@mail.ubc.ca.}
	
	%		\fontsize{13}{14} \selectfont
	
	\begin{abstract}
		In various economic environments, people observe other people with whom they strategically interact. We can model such information-sharing relations as an information network, and the strategic interactions as a game on the network. When any two agents in the network are connected either directly or indirectly in a large network, empirical modeling using an equilibrium approach can be  cumbersome, since the testable implications from an equilibrium generally involve all the players of the game, whereas a researcher's data set may contain only a fraction of these players in practice. This paper develops a tractable empirical model of linear interactions where each agent, after observing part of his neighbors' types, not knowing the full information network, uses best responses that are linear in his and other players' types that he observes, based on simple beliefs about the other players' strategies. We provide conditions on information networks and beliefs such that the best responses take an explicit form with multiple intuitive features. Furthermore, the best responses reveal how local payoff interdependence among agents is translated into local stochastic dependence of their actions, allowing the econometrician to perform asymptotic inference without having to observe all the players in the game or having to know the precise sampling process.
		\medskip
		
		{\noindent \textsc{Key words.} Strategic Interactions; Behavioral Modeling; Information Sharing; Games on Networks; Cross-Sectional Dependence}
		\medskip
		
		{\noindent \textsc{JEL Classification: C12, C21, C31}}
	\end{abstract}
	
	\maketitle
	
	\section{Introduction}
	Interactions between agents - for example, through personal or business relations - generally lead to their actions being correlated. In fact, such correlated behaviors form the basis for identifying and estimating peer effects, neighborhood effects, or more generally, social interactions in the literature. (See \citemain{Blume/Brock/Durlauf/Ioannides:11:HandbookSE} and \citemain{Durlauf/Ioannides:10:ARE} for a review of this literature.)
	
	Empirical modeling becomes nontrivial when one takes seriously the fact that people are often connected directly or indirectly on a large complex network, and observe some of their neighbors' types. Such strategic environments may be highly heterogeneous across agents, with each agent occupying a nearly ``unique" position in the network. Information sharing potentially creates a complex form of cross-sectional dependence among the observed actions of agents, yet the econometrician typically observes only a fraction of the agents on the network, and rarely observes the entire network which governs the cross-sectional dependence structure.
	
	The main contribution of this paper is to develop a tractable empirical model of linear interactions among agents with the following three major features. First, assuming a large game on a complex, exogenous network, our empirical model does not require the agents to observe the full network. Instead, we assume that each agent observes only a local network around herself and only part of the type information of those who are local to her.\footnote{For example, a recent paper by \citemain{Breza/Chandrasekhar/Tahbaz-Salehi:18:WP} documents that people in a social network may lack substantial knowledge of the network and that such informational assumptions may have significant implications for the predictions of network models. Models assuming that agents possess only local knowledge have drawn interest in the literature on Bayesian learning on networks. For example, see a recent contribution by \citemain{Li/Tan:19:TE} and references therein.}
	
	Second, our model explains strategic interdependence among agents through correlated observed behaviors. In this model, the cross-sectional local dependence structure among the observed actions reflects the network of strategic interdependence among the agents. Most importantly, unlike most incomplete information game models in the literature, our set-up allows for \textit{information sharing on unobservables}, i.e., each agent is allowed to observe his neighbors' payoff-relevant signals that are not observed by the econometrician.
	
	Third, the econometrician does not need to observe the whole set of players in the game for inference. It suffices that he observe many (potentially) non-random samples of local interactions. The inference procedure that this paper proposes is asymptotically valid \textit{independently} of the actual sampling process, as long as the sampling process satisfies certain weak conditions. Accommodating a wide range of sampling processes is useful because random sampling is rarely used for the collection of network data, and a precise formulation of the actual sampling process is often difficult in practice.
	
	A standard approach for studying social interactions is to model them as a game, and use the game's equilibrium strategies to derive predictions and testable implications. Such an approach is cumbersome in our set-up. Since a particular realization of any agent's type affects all the other agents' equilibrium actions through a chain of information sharing, each agent needs to form a ``correct" belief about the entire information graph. Apart from such an assumption being highly unrealistic, it also implies that predictions from an equilibrium that generate testable implications usually involve all the players in the game, when it is often the case that only a fraction of the players are observed in practice. Thus, an empirical analysis which regards the players in the researcher's sample as coincident with the actual set of players in the game may suffer from a lack of external validity when the target population is a large game involving many more players than those present in the actual sample.
	
	Instead, this paper adopts an approach of behavioral modeling, where it is assumed that each agent, not knowing fully the information sharing relations, optimizes according to simple beliefs about the other players' strategies. The crucial part of our behavioral assumption is a primitive form of \textit{belief projection} which says that each agent, not knowing the full set of information-sharing relations, projects his own beliefs about other players onto his payoff neighbors. More specifically, if agent $i$ gives more weight to agent $j$ than to agent $k$, agent $i$ believes that each of his payoff neighbors $s$ does the same in comparing agents $j$ and $k$. Here the ``weights" represent the strategic importance of other players, and belief projection can be viewed as a rule-of-thumb for an agent who needs to form expectations of the actions of the players, not knowing who they observe. When the strategic importance of one player to another is based primarily on ``vertical" characteristics such as skills or assets, the assumption of belief projection does not seem unrealistic.\footnote{Belief projection in our paper can be viewed as connected, though loosely, to inter-personal projection studied in behavioral economics. A related behavioral concept is projection bias of \citemain{Loewenstein/ODonohue/Rabin:03:QJE} which refers to the tendency of a person projecting his own current taste to his future taste. See also \citemain{vanBoven/Loewenstein/Dunning:03:JEBO} who reported experimental results on the interpersonal projection of tastes onto other agents. Since an agent's belief formation is often tied to their information, belief projection is closely related to information projection in \citemain{Madarasz:12:ReStud}, who focuses on the tendency of a person to project his information to other agents'. The main difference here is that our focus is to formulate the assumption in a way that is useful for inference using observational data on the actions of agents who interact on a network.}
	
	Our belief projection approach yields an explicit form of the best response which has intuitive features. For example, the best response is such that each agent $i$ gives more weight to those agents with a higher \textit{local centrality} to him, where the local centrality of agent $j$ to agent $i$ is said to be high if and only if a high fraction of agents whose actions affect agent $i$'s payoff have their payoffs affected by agent $j$'s action. Also, the best response is such that each agent responds to a change in his own type more sensitively when there are stronger strategic interactions, due to what we call the \textit{reflection effect}. The reflection effect of player $i$ captures the way that player $i$'s type affects his own action through his payoff neighbors whose payoffs are affected by player $i$'s types and actions.
	
	The best responses reveal an explicit form of local dependence among the observed actions from which we can derive minimal conditions for feasible asymptotic inference. It turns out that the econometrician does not need to observe all the players in the game, nor does he need to know the precise sampling process. Furthermore, the best response from the belief assumption provides a testable implication for information sharing on unobservables in data. In fact, the cross-sectional correlation of residuals indicates information sharing on unobservables. (See the Supplemental Note for details on the testing procedure based on the cross-sectional correlation of residuals.)
	
	It is instructive to compare the predictions from our behavioral model to those from an equilibrium model. When the payoff graph is comprised of multiple disjoint subgraphs that are complete, the behavioral strategies and equilibrium strategies coincide. Moreover, for a game on a general payoff graph, we show that as the rationality of agents deepens and their information expands, the behavioral strategies converge to the equilibrium strategies of an incomplete information game where each agent observes all the \textit{sharable} types of every other agent. 
	
	We provide conditions under which the parameters are locally identified, but propose asymptotic inference in a general setting that does not require such conditions. We also investigate the finite sample properties of our asymptotic inference through Monte Carlo simulations using various payoff graphs. The results show reasonable performance of the inference procedures. In particular, the size and the power of the test for the strategic interaction parameter are good in finite samples. We apply our methods to an empirical application which studies the decision of state presence by municipalities, revisiting \citemain{Acemoglu/GarciaJimeno/Robinson:15:AER}. We consider an incomplete information game model which permits information sharing on unobservables. The fact that our best responses explicitly reveal the local dependence structure means that it is unnecessary to separately correct for spatial correlation following, for example, the  procedure of \citemain{Conley:99:JOE}.
	
	The literature on social interactions often looks for evidence of interactions through correlated behaviors. For example, linear interactions models investigate correlation between the outcome of an agent $i$ and the average outcome over agent $i$'s neighbors. See for example \citemain{Manski:93:Restud}, \citemain{DeGiorgi/Pellizzari/Redaelli:10:AEJ}, \citemain{Bramoulle/Djebbari/Fortin:09:JOE} and \citemain{Blume/Brock/Durlauf/Jayaraman:15:JPE} for identification analysis in linear interactions models, and see \citemain{Calvo-Armengol/Pattacchini/Zenou:09:ReStud} for an application to the study of peer effects. \citemain{Goldsmith-Pinkham/Imbens:13:JBES} considers nonlinear interactions on a social network and discusses endogenous network formation. Such models often assume that the researcher observes many independent samples of such interactions, where each independent sample constitutes a game containing the entire set of the players in the game.
	
	In the context of a complete information game, a linear interaction model on a large social network can generally be estimated without assuming independent samples. The outcome equations in such a setting frequently take the form of spatial autoregressive models, which have been actively studied in the spatial econometrics literature (\citemain{Anselin:88:SpatialEconometrics}). A recent study by \citemain{Johnsson/Moon:16:WP} considers a model of linear interactions on a large social network which allows for endogenous network formation. Developing inference on a large game model with nonlinear interactions is more challenging. See \citemain{Menzel:16:ReStud},  \citemain{Xu:15:WP}, \citemain{Song:14:WP}, \citemain{Xu/Lee:15:WP}, and \citemain{Yang/Lee:16:JOE} for a large game model of nonlinear interactions. This large game approach is suitable when the data set does not have many independent samples of interactions. One of the major issues in the large game approach is that the econometrician often observes only a subset of the agents from the original game of interest.\footnote{\citemain{Song:14:WP}, \citemain{Xu:15:WP}, \citemain{Johnsson/Moon:16:WP}, \citemain{Xu/Lee:15:WP} and \citemain{Yang/Lee:16:JOE} assume that all the players in the large game are observed by the researcher. In contrast, \citemain{Menzel:16:ReStud} allows for observing i.i.d. samples from the many players, but assumes that each agent's payoff involves all the other agents' actions exchangeably.}
	
	Our empirical approach is based on a large game model which is close to models of linear interactions in the sense that it attempts to explain strategic interactions through the correlated behavior of neighbors. In our set-up, the cross-sectional dependence of the observed actions is not merely a nuisance that complicates asymptotic inference; it provides the very information that reveals the nature of strategic interdependence among agents. Such correlated behavior also arises in equilibrium in models of complete information games or games with types that are either privately or commonly observable. (See \citemain{Bramoulle/Djebbari/Fortin:09:JOE} and \citemain{Blume/Brock/Durlauf/Jayaraman:15:JPE}.) However, as emphasized before, such an approach can be cumbersome in our context of a large game primarily because the testable implications from the model typically involve the entire set of players, when in many applications the econometrician observes only a small subset of the game's players.  After finishing the first draft of paper, we learned of a recent paper by \citemain{Eraslan/Tang:17:WP} who model the interactions as a Bayesian game on a large network with private link information. They do not require the agents to observe the full network, and show identification of the model primitives adopting a Bayesian Nash equilibrium as a solution concept. One of the major differences of our paper from theirs is that our paper permits information sharing on unobservables, so that the actions of neighboring agents are potentially correlated even after controlling for observables.  
	
	A departure from the equilibrium approach in econometrics is not new in the literature. \citemain{Aradillas-Lopez/Tamer:08:JBES} studied implications of various rationality assumptions for identification of the parameters in a game. Unlike their approach, our focus is on a large game where many agents interact with each other on a single complex network, and, instead of considering all the  beliefs which rationalize observed choices, we consider a particular set of beliefs that satisfy a simple rule and yield an explicit form of best responses. (See also \citemain{Goldfarb/Xiao:11:AER} and \citemain{Hwang:17:WP} for empirical research adopting behavioral modeling for interacting agents.)
	
	This paper is organized as follows. In Section 2, we introduce an incomplete information game of interactions with information sharing. This section derives the crucial result of best responses under simple belief rules. We also show the convergence of behavioral strategies to equilibrium strategies as the rationality of agents becomes higher and their information sets expand. Section 3 focuses on econometric inference. It explains the data set-up and a method for constructing confidence intervals. Section 4 investigates the finite sample properties of our inference procedure through a Monte Carlo study. Section 5 presents an empirical application on state capacity among municipalities. Section 6 concludes. Due to the space constraints, the technical proofs of the results are found in the Supplemental Note of this paper. The Supplemental Note also contains other materials including extensions to a model of information sharing among many agents over time, testing for information sharing on unobservables, and a model selection procedure for choosing among different behavioral models.
	
	\section{Strategic Interactions with Information Sharing}
	\label{sec: model}
	
	\subsection{A Model of Interactions with Information Sharing}
	
	Strategic interactions among a large number of information-sharing agents can be modeled as an incomplete information game. Let $N$ be the set of a finite yet large number of players. Each player $i \in N$ is endowed with his type vector $(\tau_i,\eta_i)$, where $\eta_i$ is a private type and $\tau_i$ a sharable type.\footnote{Later in a section devoted to econometric inference, we specify the sharable type $\tau_i$ to be a linear index of $(X_i',\varepsilon_i)'$, where $X_i$ is a covariate vector observed by the econometrician and $\varepsilon_i$ (together with the private type $\eta_i$) is not observed. Thus our framework permits \textit{information sharing on unobservables} in the sense that ``neighbors" of an agent $i$ observe $\varepsilon_i$.}  As we will elaborate later, information $\eta_i$ is kept private to player $i$ whereas $\tau_i$ is observed by his neighbors in a network which we define below.
	
	To capture strategic interactions among players, let us introduce an \textit{undirected} graph $G_P=(N,E_P)$, where $E_P$ denotes the set of edges $ij$, $i,j \in N$ with $i \ne j$, and each edge $ij \in E_P$ represents that the action of player $i$ affects player $j$'s payoff.\footnote{A graph $G=(N,E)$ is undirected if $ij \in E$ whenever $ji \in E$ for all $i,j \in N$.} We denote $N_P(j)$ to be the $G_P$-neighborhood of player $j$, i.e., the collection of players whose actions affect the payoff of player $j$:
	\begin{eqnarray*}
		N_P(j) = \{i \in N:ij \in E_P\},
	\end{eqnarray*}
	and let $n_P(j)=|N_P(j)|.$ We define $\overline N_P(j) = N_P(j) \cup \{j\}$ and let $\overline n_P(j) = |\overline N_P(j)|$.
	
	Player $i$ choosing action $y_i \in \mathcal{Y}$ with the other players choosing $y_{-i} = (y_j)_{j \ne i}$ obtains payoff:
	\begin{eqnarray}
	\label{payoff}
	u_i(y_i,y_{-i},\tau,\eta_i) &=& y_i \left(\tau_i + \beta_0 \overline{y}_i +\eta_i \right) -\frac{1}{2}y_i^2,
	\end{eqnarray}
	where $\tau = (\tau_i)_{i \in N}$, and
	\begin{eqnarray*}
		\overline{y}_i = \frac{1}{n_P(i)} \sum_{k \in N_P(i)} y_k,
	\end{eqnarray*}
	if $N_P(i) \ne \varnothing$, and $\overline{y}_i = 0$ otherwise. Thus the payoff depends on other players' actions and types only through those of his $G_P$-neighbors. We call $G_P$ the \textit{payoff graph}.
	
	The parameter $\beta_0$ measures the payoff externality among agents. As for $\beta_0$, we make the following assumption:
	\begin{assumption}
		\label{assump: beta}
		$-1 < \beta_0 < 1.$
	\end{assumption} 
	This assumption is commonly used to characterize a pure strategy equilibrium in the social interactions literature. (See e.g. \citemain{Bramoulle/Djebbari/Fortin:09:JOE} and \citemain{Blume/Brock/Durlauf/Jayaraman:15:JPE} for examples of its use.) When $\beta_0 >0$, the game is called a game of strategic complements and, when $\beta_0 <0$, a game of strategic substitutes.
	
	Let us introduce information sharing relations in the form of a \textit{directed} graph (or a network) $G_I=(N,E_I)$ on $N$ so that each $ij$ in $E_I$ represents the edge \textit{from} player $i$ \textit{to} player $j$, where the presence of edge $ij$ joining players $i$ and $j$ indicates that $\tau_i$ is observed by player $j$. Hence the  presence of an edge $ij$ between agents $i$ and $j$ represents information flow from $i$ to $j$. This paper calls graph $G_I$ the \textit{information graph}. For each $j \in N$, define
	\begin{eqnarray*}
		N_I(j) = \{i \in N: ij \in E_I\},
	\end{eqnarray*}
	that is, the set of $G_I$-neighbors observed by player $j$.\footnote{More precisely, the neighbors in $N_I(j)$ are called \textit{in-neighbors} and $n_I(j) = |N_I(j)|$ \textit{in-degree}. Throughout this paper, we simply use the term neighbors and degrees, unless specified otherwise.} Also let $\overline N_I(i) = N_I(i) \cup \{i\}$, and $\overline n_I(i) = |\overline N_I(i)|$.
	
	In this paper, we do not assume that each agent knows the whole information graph $G_I$ and the payoff graph $G_P$. To be precise about each agent's information set, let us introduce some notation. For each $i \in N$, we set $\overline N_{P,1}(i) = \overline N_P(i)$ and $\overline N_{I,1}(i) = \overline N_I(i)$, and for $m \ge 2$, define recursively
	\begin{eqnarray*}
		\overline N_{P,m}(i) = \bigcup_{j \in \overline N_P(i)}\overline N_{P,m-1}(j), \textnormal{ and }
		\overline N_{I,m}(i) = \bigcup_{j \in \overline N_I(i)}\overline N_{I,m-1}(j).
	\end{eqnarray*}
	Thus $\overline N_{P,m}(i)$ denotes the set of players which consist of player $i$ and those players who are connected to player $i$ through at most $m$ edges in $G_P$, and similarly with $\overline N_{I,m}(i)$. Also, define $N_{P,m}(i) = \overline N_{P,m}(i)\setminus\{i\}$ and $N_{I,m}(i) = \overline N_{I,m}(i)\setminus\{i\}$.
	
	For each player $i \in N$, let us introduce a local payoff graph $G_{P,m}(i) = (\overline N_{P,m}(i), E_{P,m}(i))$, where for $k_1, k_2 \in \overline N_{P,m}(i)$, $k_1 k_2 \in E_{P,m}(i)$ if and only if $k_1 k_2 \in E_P$. Define for $m \ge 1$,\footnote{The graph $G_{P,m}(i)$ is an \textit{induced subgraph} of $G_P$ induced by the vertex set $\overline N_{P,m}(i)$. Note also that while $\overline N_{P,1}(i) \subset \overline N_{P,2}(i)$, this does not imply that a player who knows the set $\overline N_{P,2}(i)$ knows what the set $\overline N_{P,1}(i)$ is. Our information set assumption requires them to know the local graph $G_{P,m}(i)$ rather than just $\overline N_{P,m}(i)$.}
	\begin{eqnarray}
	\label{inf set}
	\mathcal{I}_{i,m-1} = (G_{P,m+1}(i),\overline N_{I,m}(i),\tau_{\overline N_{I,m}(i)},\eta_i),
	\end{eqnarray}
	where $\tau_{\overline N_{I,m}(i)} = (\tau_j)_{j \in \overline N_{I,m}(i)}$. We use $\mathcal{I}_{i,m}$ to represent the information set of agent $i$. For example, when agent $i$ has $\mathcal{I}_{i,0}$ as his information set, it means that agent $i$ knows the payoff subgraph $G_{P,2}$ among the agents $\overline N_{P,2}(i)$, the set of agents whose types he observes (i.e., $N_I(i)$), and his own private signal $\eta_i$. As for the payoff graph and information graph, we make the following assumption.
	
	\begin{assumption}
		\label{assump: limited spillover}
		For each $i \in N$ and $m \ge 1$,
		\begin{eqnarray*}
			\overline N_{P,m+1}(i) \subset \overline N_{I,m}(i).
		\end{eqnarray*}
	\end{assumption}
	This assumption requires for example that an agent with information $\mathcal{I}_{i,0}$ observes their $G_P$ neighbors and their payoff relevant neighbors. The assumption on $G_I$ only requires what each set $\overline N_{I,m}(i)$ should at least include but not what it should exclude. Hence all the results of this paper carry through even if we have $\overline N_{I,m}(i) = N$ for all $i \in N$, as in a complete information game. In other words, the incomplete information feature of our game is \textit{permitted but not required} for our framework.
	
	\subsection{Predictions from Rationality}
	Each player chooses a strategy that maximizes his expected payoff according to his beliefs. Given player $i$'s strategy, information set $\mathcal{I}_i$, and his beliefs on the strategy of other players $s_{-i}^i = (s_k^i)_{k \in N \setminus \{i\}}$, the (interim) expected payoff of player $i$ is defined as
	\begin{eqnarray*}
		U_i(s_i,s_{-i}^i;\mathcal{I}_i) = \mathbf{E}[u_i(s_i(\mathcal{I}_i),s_{-i}^i(\mathcal{I}_{-i}),\tau,\eta_i)|\mathcal{I}_i],
	\end{eqnarray*}
	where $s_{-i}^i(\mathcal{I}_{-i}) = (s_{k}^i(\mathcal{I}_k))_{k \in N \setminus \{i\}}$, $\mathcal{I}_{-i} = (\mathcal{I}_k)_{k \ne i}$ and $\tau=(\tau_i)_{i \in N}$. A \textit{best response} $s_i^{\textsf{BR}}$ of player $i$ corresponding to the strategies $s_{-i}^i$ of the other players as expected by player $i$ is such that for any strategy $s_i$,
	\begin{eqnarray*}
		U_i(s_i^\textsf{BR},s_{-i}^i;\mathcal{I}_i) \ge U_i(s_i,s_{-i}^i;\mathcal{I}_i), \textnormal{ a.e.}
	\end{eqnarray*} 
	The quadratic payoff function and the information structure of the game implies that if player $i$ has information set $\mathcal{I}_i$ and believes that each of her $G_P$-neighbors, say, $k$, plays a strategy $s_k^i(\mathcal{I}_k)$, her best response is given by
	\begin{eqnarray}
	\label{BR linear}
		s_i^\textsf{BR}(\mathcal{I}_i) = \tau_i +  \frac{\beta_0}{n_P(i)}\sum_{k \in N_P(i)} \mathbf{E}[s_k^i(\mathcal{I}_k)|\mathcal{I}_i] + \eta_i.
	\end{eqnarray}
    This implies that the best responses will be linear in types $\tau_j$ as long as the conditional expectation is.
    
    In order to generate predictions, one needs to deal with the beliefs (i.e., $s_k^i(\mathcal{I}_k)$) in the conditional expectation. There are three approaches. The first approach is an equilibrium approach, where we take the predicted strategies as a set of best response strategies $s_i^{\textsf{BNE}}$ such that for any strategy $s_i$,
    \begin{eqnarray}
    	\label{BNE}
    	U_i(s_i^\textsf{BNE},s_{-i}^{\textsf{BNE}};\mathcal{I}_i) \ge U_i(s_i,s_{-i}^{\textsf{BNE}};\mathcal{I}_i), \textnormal{ a.e.}
    \end{eqnarray} 
    Hence in equilibrium strategies, each player believes that the other players' strategies coincide with the best response strategies by the agents in equilibrium. The second approach, rationalizability, considers all strategies that are rationalizable given some belief. The third approach is a behavioral approach where one considers a set of simple behavioral assumptions on the beliefs and focuses on the best responses to these beliefs.
	
	There are pros and cons with each of the three approaches. The equilibrium approach requires that the beliefs of all the players be ``correct" in equilibrium. However, since each player $i$ generally does not know who each of his $G_P$-neighbors observes, a Bayesian player in an incomplete information game with rational expectations would need to know the distribution of the entire information graph $G_I$ (or at least have a common prior on the information graph commonly agreed upon by all the players) to form a ``correct" belief given his information. Given that the players are only partially observed and $G_I$ is rarely observed with precision, producing a testable implication from such an equilibrium model appears far from a trivial task.
	
	The rationalizability approach can be used to relax this rational expectations assumption by eliminating the requirement that the beliefs be correct. Such an approach considers all the predictions that are rationalizable given \textit{some} beliefs. However, the set of predictions from rationalizability can potentially be large and may fail to produce sharp predictions useful in practice.
	
	This paper takes the approach of behavioral modeling. We adopt a set of simple behavioral assumptions on players' beliefs which can be incorrect from the viewpoint of a person with full knowledge of the distribution of the information graph, yet useful as a rule-of-thumb for an agent in a complex decision-making environment such as the one in our model. As we shall see later, this approach can give a sharp prediction that is intuitive and analytically tractable. Furthermore, the best responses from this approach coincide with equilibrium strategies for a special class of payoff graphs, and converge to equilibrium strategies for any payoff graph as the rationality of agents becomes deeper and their information set expands.
	
	\subsection{Belief Projection and Best Responses}
	
	\subsubsection{A Game with Complete Payoff Subgraphs}
	\label{subsubsec: complete payoff subgraphs}
	Let us consider first a special case where the payoff graph $G_P = (N,E_P)$ is such that $N$ is partitioned into subsets $N_1,...,N_Q$, and for all $i,j \in N$ with $i \ne j$, $ij \in E_P$ if and only if $i,j \in N_q$ for some $q = 1,...,Q$. In this game, there are multiple strategically disjoint subgames and agents in each subgame observe all other players' sharable types $\tau_j$'s in the subgame, but do not observe their private types $\eta_j$.\footnote{Each subgame is a special case of the Bayesian game in \citemain{Blume/Brock/Durlauf/Jayaraman:15:JPE}.} In this case, there exists a unique Bayesian Nash equilibrium where the equilibrium strategies take the explicit form of a linear strategy: for each player $i$,
	\begin{eqnarray*}
		\label{lin strategy0}
		s_i(\mathcal{I}_{i,0}) = \sum_{j \in \overline N_P(i)} \tau_j w_{ij} + \eta_i,
	\end{eqnarray*}
	and $w_{ij}$ are weights such that for all $i,j,k,\ell$ in the same cluster,
	\begin{eqnarray}
	\label{symmetry}
		w_{kk} = w_{ii}, \text{ and }
		w_{k\ell} = w_{ij}, \text{ whenever } i \ne j \text{ and } k \ne \ell.
	\end{eqnarray}
	Thus the equilibrium is within-cluster symmetric in the sense that the equilibrium strategy is the same across all the agents in the same cluster. However, this symmetric equilibrium does not extend to a general payoff graph $G_P$. 
	
	\subsubsection{Belief Projection}
	
	 Our approach uses a weaker version of the symmetry restrictions (\ref{symmetry}) to specify initial beliefs, so that a best response function exists uniquely for any payoff graph configuration.\footnote{A best response function satisfying the symmetry restrictions in (\ref{symmetry}) may not exist for a general payoff graph $G_P$.} More specifically, we introduce the following symmetry restrictions on the belief formation.
	
	\begin{definition}
		\label{def: BP}
		We say that a player $i$ with information set $\mathcal{I}_{i,0}$ \textit{does BP (Belief Projection)}, if she believes that each of her $G_P$-neighbors, say, $k$, plays a linear strategy as:
		\begin{eqnarray}
		\label{lin strategy}
		s_k^i(\mathcal{I}_{k,0}) = \sum_{j \in \overline N_P(k)} \tau_j w_{kj}^i + \eta_k,
		\end{eqnarray}
		for some nonnegative weights $w_{kj}^i$ in (\ref{lin strategy}) that satisfy the following conditions:
		
		(BP-a) $w_{kk}^i = w_{ii}$ and $w_{kj}^i = w_{ij}$ for $j \ne i,k$;
		
		(BP-b) $w_{ki}^i = w_{ik}$.
	\end{definition}
	
	Condition (\ref{lin strategy}) assumes that player $i$ believes that player $k$ responds only to the types of those players $j \in N_P(k)$. This is a rule-of-thumb for player $i$ to form expectations about player $k$ while not observing $N_I(k)$.
	
	In forming beliefs about other players' strategies, not knowing who they observe, BP assumes that each player \textit{projects his own beliefs about himself and other players onto }his $G_P$ neighbors, as epitomized by Conditions (BP-a)-(BP-b). The ranking of weights $w_{ij}$ over $j$ represents the relative strategic importance of player $j$ to player $i$.
	
	More specifically, Condition (BP-a) says that each player $i$ believes that the weight his $G_P$-neighbor $k$ attaches to himself or player $j \in N_P(k)$ is the same as the weight player $i$ attaches to himself or the same player $j$. Thus player $i$'s belief on his $G_P$ neighbor $k$'s weight to player $j$ is formed in reference to his own weight to player $j$. In other words, without any information on how his $G_P$-neighbors rank other players, each player simply takes himself as a benchmark to form beliefs about his $G_P$ neighbors' ranking of other players. This assumption does not seem unreasonable if the weights $w_{ij}$ are based on the vertical characteristics such as skills or assets of agent $j$. Condition (BP-b) imposes a symmetry restriction that player $i$ believes that player $k$ gives the same weight to player $i$ as the weight player $i$ gives to player $k$.\footnote{It is important to note that we do not impose BP directly on the strategies of the players as predicted outcomes of the game. Instead, BP is used as an initial input for each player to form best response strategies with limited information on networks. It is these best response strategies that constitute the predicted outcomes from the game.}
	
	\subsubsection{Belief Projection as a Prior Specification for a Bayesian Decision Maker}
	\label{subsubsec:Bayesian DM}
	The best response of a player with quadratic utility and beliefs as in Definition \ref{def: BP} can be viewed as arising from a decision maker with Bayesian rationality with a prior that satisfies certain symmetry restrictions. More specifically, let $W = (W_{k,j})_{k,j \in N}$ be a weight matrix such that $W_{k,j}$ is the weight player $k$ attaches to player $j$. Each player $i$ believes that other players $k$ play a linear strategy, say, $s_k(\mathcal{I}_{k,0};w_k)$, $w_k = (w_{kj})_{j \in N}$, of the form in (\ref{lin strategy}) of Definition \ref{def: BP}. (We are simply making explicit the dependence of the strategy on $w_k$ in our notation.)  Suppose that each agent $i$ is given information set $\mathcal{I}_{i,0}$ and a prior $Q_i$ over $(\mathcal{I}_{j,0})_{j \in N}$ and $W$ in the game. We assume that $W$ and $(\mathcal{I}_{j,0})_{j \in N}$ are independent under $Q_i$.\footnote{This independence simply reflects that the strategies as a function of signals are distinct from the signals themselves.} Then, the decision maker with Bayesian rationality proceeds as follows.
	\medskip
	
	(Step 1) A mediator (or Nature) suggests a weight vector $w_i$ to each player $i$.
	
	(Step 2) Player $i$ forms a posterior mean of his payoff given $w_i$ and $\mathcal{I}_{i,0}$.
	
	(Step 3) Unless there is an action that gives a strictly better posterior mean of his payoff than the action $s_i(\mathcal{I}_{i,0};w_i)$, the player accepts $w_i$ and chooses the latter action.
	\medskip
	
	\noindent Thus, the best response of this Bayesian decision maker is given by $s_i(\mathcal{I}_{i,0};w_i)$ such that
	\begin{eqnarray*}
		&& \mathbf{E}_{Q_i}\left[ u_i(y_i,s_{-i}(\mathcal{I}_{-i,0};W_{-i});\mathcal{I}_{i,0}) |W_i = w_i, \mathcal{I}_{i,0}\right] \\
		&\le& \mathbf{E}_{Q_i}\left[ u_i(s_i(\mathcal{I}_{i,0};W_i),s_{-i}(\mathcal{I}_{-i,0};W_{-i});\mathcal{I}_{i,0}) |W_i = w_i, \mathcal{I}_{i,0}\right], \text{ for all } y_i \in \mathcal{Y},
	\end{eqnarray*} 
	where $\mathbf{E}_{Q_i}$ denotes the conditional expectation under $Q_i$. 
	
	Now, the belief projection of a player corresponds to a set of restrictions on the prior $Q_i$, where we specify $Q_i$ as a Gaussian prior (centered at zero) over the weight matrix $W$ such that the following restrictions hold for all $i=1,...,n$ and $k \in N_P(i)$:\medskip
	
	(a) $\mathbf{E}_{Q_i}[W_{kk}|W_i] = W_{ii}$ and $\mathbf{E}_{Q_i}[W_{kj}|W_i] = W_{ij}$ for all $j \in N_P(k)$.
	
	(b) $\mathbf{E}_{Q_i}[W_{ki}|W_i] = W_{ik}$.
	
	(c) $\mathbf{E}_{Q_i}[W_{kj}|W_i] = 0$ for all $j \notin \overline N_P(k)$.
	\medskip
	
	\noindent Then, it is not hard to see that the best response of this Bayesian player is the same as that of a simple type agent who behaves according to the belief projection assumption. (See the Supplemental Note for details.) The restriction (a) says that player $i$ \textit{projects} his weight $W_{ii}$ and $W_{ij}$ to $W_{kk}$ and $W_{kj}$. To see the meaning of this restriction, suppose that
	\begin{eqnarray*}
		\text{Var}_{Q_i}(W_{kk}) = \text{Var}_{Q_i}(W_{ii}),
	\end{eqnarray*}
    and that $\mathbf{E}_{Q_i}[W_{kk}|W_i] = \mathbf{E}_{Q_i}[W_{kk}|W_{ii}]$. Then the first statement of (a) requires that $W_{kk}$ and $W_{ii}$ are perfectly positively correlated under $Q_i$. The perfect correlation seems the only reasonable specification in this setting because if $W_{kk}$ and $W_{ii}$ are not perfectly correlated, it yields the odd implication that player $i$ believes $W_{kk}$ to be less than $W_{ii}$ for all $k \in N_P(i)$ in absolute value. A similar observation can be made for the second statement of (a). The restriction (b) imposes that the weight that agent $k$ is expected to attach to $i$ is the same as the weight that agent $i$ gives to the agent $k$. The restriction (c) means that player $i$ does not consider the weight of player $k$ attaches to player $j$ when $j$ is outside of $\overline N_P(k)$, because player $i$ may not even know who $j$ is, and from player $i$'s perspective, player $j$ can be any player in the large population outside  of $\overline N_P(k)$.
	
	\subsubsection{Best Responses} Let us first define the type of players and games for which we derive best responses.
	\begin{definition}
		\label{def: simple and sophisticated players}
		\noindent (i) Player $i \in N$ is said to be \textit{of simple type} if she has information set $\mathcal{I}_{i,0}$, and does BP. Let $\Gamma_0$ denote the game populated by $n$ players who are of simple type and have payoff functions in (\ref{payoff}) and payoff graph $G_P$.
		
		(ii) For each $m \ge 1$, player $i \in N$ is said to be \textit{of the $m$-th order sophisticated type}, if she has information set $\mathcal{I}_{i,m}$ and believes that the other players play the best response strategies from $\Gamma_{m-1}$. Let $\Gamma_m$ denote the game populated by $n$ players who are of the $m$-th order sophisticated type and have payoff functions in (\ref{payoff}) and payoff graph $G_P$.
	\end{definition}
	
	The higher-order sophisticated type agents are analogous to agents in level-$k$ models in behavioral economics. (See Chapter 5 of \citemain{Camerer:03:BehavioralGameTheory} for a review.\footnote{Note that level-$k$ models in behavioral economics are different from $k$-rationalizability models of \citemain{Bernheim:84:Eca} and \citemain{Pearce:84:Eca} which are studied by \citemain{Aradillas-Lopez/Tamer:08:JBES} in the context of econometrics.}) In experiments, a level-$0$ player chooses an action without considering strategic interactions, making them much simpler than our simple types. Our simple-type player already considers strategic interdependence and forms a best response. On the other hand, the level-$k$ models allows the agents to be of different orders of rationality within the same game. In our set-up centered on observational data, identification of the unknown proportion of each rationality type appears far from trivial. Hence in this paper, we consider a game where all the agents have the same order of sophistication.\footnote{See \citemain{Gillen:10:WP} and \citemain{An:17:JOE} for an application of level-$k$ models to observational data from first-price auctions. One of the major distinctions between our model and their level-$k$ models is that we focus on a set-up of a single large game populated by many players occupying strategically heterogenous positions, whereas their research centers on a set-up where the econometrician observes the same game played by a fixed number of agents many times.} 
	
	Unlike in level-$k$ models, the difference between the simple type and the first-order sophisticated type lies not only on their degree of rationality but also on their information set.  In particular, the information requirements are stronger for more sophisticated agents. For example, the first-order sophisticated type knows who belongs to $\overline N_{P,3}(i)$, whereas a simple-type does not need to.
	
	Below we give a unique explicit form of best responses from game $\Gamma_0$. First, define for all $i,j \in N$ with $ i \ne j$,
	\begin{eqnarray*}
		\lambda_{ij} \equiv \frac{1}{1 - \beta_0 c_{ij}}, \text{ and } \overline \lambda_i \equiv \frac{1}{n_P(i)}\sum_{j \in N_P(i)} \lambda_{ij},
	\end{eqnarray*}
    where
	\begin{eqnarray}
	\label{c_ij}
	\quad c_{ij} \equiv \frac{|N_P(i) \cap N_P(j)|}{n_P(i)}.
	\end{eqnarray}
	The theorem below gives the explicit form of the best response in this game. 
	\begin{theorem}
		\label{thm: best response}
		Suppose that Assumptions \ref{assump: beta} - \ref{assump: limited spillover} hold and for each $i \in N$, and any $k \ne i$, $\mathbf{E}[\eta_k|\mathcal{I}_{i,0}] = 0$. Then each player $i$'s best response $s_i^{\textsf{BR}}$ from game $\Gamma_0$ takes the form $s_i^{\textsf{BR}} = s_i^{[0]}$ with
		\begin{eqnarray*}
			s_{i}^{[0]}(\mathcal{I}_{i,0})=w_{ii}^{[0]}\tau_{i}+\sum_{j\in N_{P}(i)}w_{ij}^{[0]}\tau_{j}+\eta_{i},
		\end{eqnarray*}
		where, if $n_P(i) \ge 1$, 
		\begin{eqnarray*}
			w_{ii}^{[0]} & \equiv & 1+\frac{\beta_{0}^{2}\overline{\lambda}_{i} }{n_{P}(i)-\beta_{0}^{2}\overline{\lambda}_{i}},\text{ and }\\
			w_{ij}^{[0]} & \equiv & \frac{\beta_{0}\lambda_{ij}}{n_{P}(i)}w_{ii}^{[0]},\text{ for }i\ne j,
		\end{eqnarray*}
	and if $n_P(i) = 0$, $w_{ii}^{[0]} = 1$ and $w_{ij}^{[0]} = 0$.
	\end{theorem} 
    It is worth noting that the unique best response in Theorem \ref{thm: best response} always exists regardless of the configurations of the payoff graph $G_P$. (Indeed, we always have $n_P(i) - \beta_0^2 \overline \lambda_i >0$ as long as $\beta_0 \in (-1,1)$. See Lemma \ref{lemm: exist BR} in the Supplemental Note.) Furthermore, the best responses have several intuitive features. First, note that the behavioral strategies $s_i^{[0]}(\mathcal{I}_{i,0})$ maintain strategic interactions to be local around each player's $G_P$-neighbors, regardless of the magnitude of $\beta_0$, so that a player can have a strong interaction with his $G_P$-neighbors without being influenced by a change in the type of a far-away player.\footnote{This prediction is in contrast with that from the equilibrium strategies of a complete information version of the game. According to the equilibrium strategies, the influence of one player can reach a far-away player when $\beta_0$ is high. See Section \ref{subsec: comparing eq str and behav str}.}
	
	Second, the best response captures the network externality in an intuitive way. The quantity $c_{ij}$ measures the proportion of player $i$'s $G_P$-neighbors whose payoffs are influenced by the type and action of player $j$. Hence if $c_{ij}< c_{ik}$, player $k$ is ``strategically more important" to player $i$ than player $j$. Note that $\lambda_{ij}$ is an increasing function of $c_{ij}$ with its slope increasing in $\beta_0$. We take $\lambda_{ij}$ to represent the strategic  \textit{local centrality} of player $j$ to player $i$. Then we have 
	\begin{eqnarray}
	\label{externality}
	\frac{\partial s_i^{[0]}(\mathcal{I}_{i,0})}{\partial \tau_i} &=& 1 + \frac{\beta_0^2 \overline \lambda_i}{n_P(i) - \beta_0^2 \overline \lambda_i} \text{ and }\\ \notag
	\frac{\partial s_i^{[0]}(\mathcal{I}_{i,0})}{\partial \tau_j} &=& \frac{\beta_0  \lambda_{ij} }{n_P(i)} \left( 1 + \frac{\beta_0^2 \overline \lambda_i}{n_P(i) - \beta_0^2 \overline \lambda_i} \right), \text{ for } j \in N_P(i),
	\end{eqnarray}
	both of which measure the response of actions of agent $i$ to a change in the observed type change of his own and his $G_P$-neighbors. The second quantity captures the \textit{network externality} in the strategic interactions.
	
	The network externality for agent $i$ from a particular agent $j$ decreases in $n_P(i)$ and increases in $\beta_0$. More importantly, the network externality from one player to another is heterogeneous, depending on each player's ``importance" to others in the payoff graph. This is seen from the network externality (\ref{externality}) being an increasing function of agent $j$'s local centrality to agent $i$, i.e., $\lambda_{ij}$, when the game is that of strategic complements (i.e., $\beta_0 >0$). In other words, the larger the fraction of agent $i$'s $G_P$-neighbors whose payoff is affected by agent $j$'s action, the higher the network externality of agent $i$ from agent $j$'s type change becomes.
	
	It is interesting to note that the network externality for agent $i$ with respect to his own type $\tau_i$ is greater than $1$. We call the additive term in (\ref{externality}),
	\begin{eqnarray*}
		\frac{\beta_0^2 \overline \lambda_i}{n_P(i) - \beta_0^2 \lambda_i},
	\end{eqnarray*}
	the \textit{reflection effect} which captures the way player $i$'s type affects his own action through his $G_P$ neighbors whose payoffs are affected by player $i$'s types and actions. The reflection effect arises because each agent, in decision making, considers the fact that his type affects other $G_P$-neighbors' decision making. When there is no payoff externality (i.e., $\beta_0 = 0$), the reflection effect is zero. However, when there is a strong strategic interactions or when a majority of player $i$'s $G_P$-neighbors have a small $G_P$-neighborhood (i.e., for a majority of $j \in N_P(i)$, $N_{P}(j)$'s in the definition of $c_{ij}$ in (\ref{c_ij}) have few elements), the reflection effect is large.
	
	When $G_P$ consists of disconnected complete subgraphs as in (\ref{subsubsec: complete payoff subgraphs}), the best responses $s_i^{[0]}$ coincide with Bayesian Nash equilibrium strategies. More specifically, let $s_i^{\textsf{BNE}}(\mathcal{I}_{i,0})$ be the Bayesian Nash equilibrium strategies which satisfy (\ref{BNE}). Then, we can show that
	\begin{eqnarray}
	\label{BNE2}
		s_i^{\textsf{BNE}}(\mathcal{I}_{i,m}) = w_{ii}^{\textsf{BNE}} \tau_i + \frac{\beta_0}{n_P(i)}\sum_{j \in N_P(i)} w_{ij}^{\textsf{BNE}} \tau_j + \eta_i,
	\end{eqnarray}
	where 
	\begin{eqnarray*}
		w_{ii}^{\textsf{BNE}} &\equiv& 1 + \frac{\beta_0^2}{(n_P(i) + \beta_0)(1 - \beta_0)}, \text{ and } \\
		w_{ij}^{\textsf{BNE}} &\equiv& \frac{\beta_0}{(n_P(i) + \beta_0)(1 - \beta_0)} , \text{ for } i \ne j.
	\end{eqnarray*}
    In this case with disconnected complete subgraphs, $n_P(i)$'s are all equal for $i$'s in the same cluster, and $c_{ij} = (n_P(i) - 1)/n_P(i)$, yielding 
    \begin{eqnarray*}
    	\lambda_{ij} = \frac{n_P(i)}{n_P(i) - \beta_0(n_P(i)-1)}, \text{ for all } j \in N_P(i).
    \end{eqnarray*}
Using this, it is not hard to check that
\begin{eqnarray*}
	w_{ii}^{[0]} = w_{ii}^{\textsf{BNE}}, \text{ and } w_{ij}^{[0]} = w_{ij}^{\textsf{BNE}}.
\end{eqnarray*}
(See the Supplemental Note for the derivations.)

	The following theorem shows that each $\Gamma_m$ yields a unique, explicit form of best responses.
	
	\begin{theorem}
		\label{thm: best response2}
		Suppose that Assumptions \ref{assump: beta} - \ref{assump: limited spillover} hold and for each $i\in N$ and $k\ne i$, $\mathbf{E}[\eta_{k}|\mathcal{I}_{i,m}]=0$ for $m\ge1$.
		Then each player $i$'s best response $s_{i}^{\textsf{BR}}$ from
		game $\Gamma_{m}$ takes the form $s_{i}^{\textsf{BR}}=s_{i}^{[m]}$
		with 
		\begin{eqnarray*}
			s_{i}^{[m]}(\mathcal{I}_{i,m})=w_{ii}^{[m]}\tau_{i}+\sum_{j\in N_{P,m+1}(i)}w_{ij}^{[m]}\tau_{j}+\eta_{i},
		\end{eqnarray*}
		where 
		\begin{eqnarray*}
			w_{ii}^{[m]}&=& 1+\frac{\beta_{0}}{n_{P}(i)}\sum_{k\in N_{P}(i)}w_{ki}^{[m-1]}\text{ and }\\
			w_{ij}^{[m]}&=& \frac{\beta_{0}}{n_{P}(i)}\sum_{k\in N_{P}(i)}w_{kj}^{[m-1]}1\{j\in\overline{N}_{P,m}(k)\}.
		\end{eqnarray*}
	\end{theorem} 

	As compared to game $\Gamma_0$, game $\Gamma_1$ predicts outcomes with broader network externality. Indeed, when $m =1$,
	\begin{eqnarray}
	\label{FOS BR}
		s_{i}^{[1]}(\mathcal{I}_{i,1})&=&\left(1+\frac{\beta_{0}}{n_{P}(i)}\sum_{k\in N_{P}(i)}w_{ki}^{[0]}\right)\tau_{i}\\ \notag
		&& +\sum_{j\in N_{P,2}(i)}\left(\frac{\beta_{0}}{n_{P}(i)}\sum_{k\in N_{P}(i)}w_{kj}^{[0]}1\{j\in\overline{N}_{P}(k)\}\right)\tau_{j}+\eta_{i}.
	\end{eqnarray}
	For example, the types of neighbors whose actions do not affect player $i$'s payoff can affect his best response. More specifically, note that for $j \in N_{P,2}(i) \setminus N_P(i)$, 
	\begin{eqnarray*}
		\frac{\partial s_{i}^{[1]}(\mathcal{I}_{i,1})}{\partial \tau_j} = \frac{\beta_0}{n_P(i)} \sum_{k \in N_P(i)} 1\{j \in \overline N_P(k)\} w_{kj}^{[0]}.
	\end{eqnarray*}
	The externality from player $j$ to player $i$ is strong when player $j$ has a high local centrality $\lambda_{kj}$ to a large fraction of player $i$'s $G_P$-neighbors $k$.
	
	\subsection{Comparing Equilibrium Strategies and Behavioral Strategies}
	\label{subsec: comparing eq str and behav str}
	
	\subsubsection{Convergence of Behavioral Strategies to Equilibrium Strategies}
	\label{subsubsec: convergence of behav str}
	
	We show that as the information set expands and the order of sophistication becomes higher, the behavioral strategies converge to the equilibrium strategies from a game where all players observe all other players' sharable types. Let $\Gamma_\infty$ be the game where players have the same payoff function and the same payoff graph as in $\Gamma_0$ except that the information set for each player $i$ is given by $\mathcal{I}_{i,\infty} = (G_P,\tau,\eta_i)$. Thus each player $i$ knows the whole payoff graph $G_P$, all sharable types, $\tau =(\tau_i)_{i \in N}$, and private information $\eta_i$. (This information structure is similar to \citemain{Blume/Brock/Durlauf/Jayaraman:15:JPE}.) Let $s^{\textsf{BNE}} =(s_i^{\textsf{BNE}})_{i \in N}$ be the Bayesian Nash equilibrium strategy profile from the game $\Gamma_\infty$.
	
	Below, we give a theorem which shows that the sequence of behavioral strategies $s_i^{[m]}$ converges to the equilibrium strategies $s_i^{\textsf{BNE}}$ as $m \rightarrow \infty$.
	
	\begin{theorem}
		\label{thm: conv}
		Suppose that the conditions of Theorem \ref{thm: best response} hold and that
		\begin{eqnarray}
		\label{moment bound}
		\max_{i \in N} \mathbf{E}[\|\tau_i\|^2] < \infty.
		\end{eqnarray}
		Then, as $m\rightarrow\infty$, 
		\begin{eqnarray*}
			\mathbf{E}\left[\max_{i \in N}\left(s_i^{[m]}(\mathcal{I}_{i,m}) - s_i^{\textsf{BNE}}(\mathcal{I}_{i,\infty}))\right)^2 \right] \rightarrow 0.
		\end{eqnarray*}
	\end{theorem}
	
	Theorem \ref{thm: conv} shows that as the order of sophistication deepens, the best response strategies from the behavioral model become closer to the equilibrium strategies. It is not hard to check that $s_i^{[m]}(\mathcal{I}_{i,m}) = s_i^{[m]}(\mathcal{I}_{i,\infty})$, i.e., the best response remains the same if we expand the information set $\mathcal{I}_{i,m}$ to $\mathcal{I}_{i,\infty}$. Therefore, the convergence in Theorem \ref{thm: conv} can be viewed as the convergence of the best responses $s_i^{[m]}(\mathcal{I}_{i,\infty})$ to equilibrium strategies $s_i^{\textsf{BNE}}(\mathcal{I}_{i,\infty})$ while the information set is fixed to be $\mathcal{I}_{i,\infty}$.
	
	\subsubsection{Comparison in Terms of Network Externality}
	We compare the behavioral strategies and equilibrium strategies in terms of network externality which measures how sensitively an agent's action responds to a change in her neighbor's types. We also compare how this network externality changes as the network grows. Let $Y_i$ be the observed outcome of player $i$ as predicted from either of the two game models. For simplicity, we remove $\eta_i$'s from the models so that the game $\Gamma_\infty$ now becomes a complete information game.
	
	The complete information game gives the following prediction for action $Y_i$ of agent $i$:
	\begin{eqnarray*}
		Y_i = \frac{\beta_0}{n_P(i)}\sum_{j \in N_P(i)} Y_j + \tau_i,
	\end{eqnarray*}
	where $Y_i$ denotes the action of player $i$ in equilibrium. Then the reduced form for $Y_i$'s can be written as
	\begin{eqnarray}
	\label{comp}
		y = (I - \beta_0 A)^{-1} \tau,
	\end{eqnarray}
	where $y=(Y_1,...,Y_n)',\tau = (\tau_1,...,\tau_n)'$, and $A$ is a row-normalized adjacency matrix of the payoff graph $G_P$, i.e., the $(i,j)$-th entry of $A$ is $1/n_P(i)$ if $j \in N_P(i)$ and zero otherwise. Thus when $\beta_0$ is close to one (i.e., the local interaction becomes strong), the equilibrium outcome can exhibit extensive cross-sectional dependence.
	
	On the other hand, our behavioral model predicts the following:
	\begin{eqnarray*}
		Y_i = \left( 1+\frac{\beta_{0}^{2}\overline{\lambda}_{i} }{n_{P}(i)-\beta_{0}^{2}\overline{\lambda}_{i}} \right)\left(\tau_{i}+\sum_{j\in N_{P}(i)}\frac{\beta_{0}\lambda_{ij}}{n_{P}(i)} \tau_j \right),
	\end{eqnarray*} 
	which comes from Theorem \ref{thm: best response} without $\eta_i$'s. When we compare this with (\ref{comp}), it is clear that the cross-sectional dependence structure of our behavioral model is different from that from the complete information equilibrium model. In the case of the complete information equilibrium model, it is possible that two actions $Y_i$ and $Y_j$ between two agents $i$ and $j$ can be correlated even if $i$ and $j$ are very far from each other in graph $G_P$. However, the cross-sectional dependence structure of the actions from the behavioral model closely follows the graph $G_P$: $Y_i$ and $Y_j$ can be correlated only if their $G_P$ neighbors overlap.
	
	For comparison purposes, for a given strategy $s_i(\mathcal{I}_i)$ for an agent $i$ with information set $\mathcal{I}_i$, we introduce the average network externality (ANE):
	\begin{eqnarray}
	\label{ane}
	\frac{1}{n} \sum_{j \in N} \sum_{i \in N: i \ne j} \frac{\partial s_i(\mathcal{I}_i)}{\partial \tau_j}. 	
	\end{eqnarray}
	The ANE measures the average impact of a change in the neighbors' type on the actions of the player. The ANE from equilibrium strategies of the complete information game is  $\frac{1}{n} \sum_{j \in N} \sum_{i \in N: i \ne j} [(I - \beta_0 A)^{-1}]_{ij}$, where $[(I - \beta_0 A)^{-1}]_{ij}$ denotes the $(i,j)$-th entry of the matrix $(I - \beta_0 A)^{-1}$.

	\begin{table}[t]
		\caption{\small The Characteristics of the Payoff Graphs}
		
		\begin{centering}
			\small
			\begin{tabular}{c|ccc|ccc}
				\hline 
				\hline 
			    & \multicolumn{3}{c|}{{\small{}Erdös-Rényi}} & \multicolumn{3}{c}{{\small{}Barabási-Albert}}\tabularnewline
				& {\small{}Network A} & {\small{}Network B} & {\small{}Network C} & {\small{}Network A} & {\small{}Network B} & {\small{}Network C}\tabularnewline
				\hline 
				{\small{}$n$} & {\small{}164.9 } & {\small{}783.4 } & {\small{}3116.8} & {\small{}236.1 } & {\small{}1521.0 } & {\small{}4773.8}\tabularnewline
				{\small{}$d_{mx}$} & {\small{}11.14} & {\small{}12.74} & {\small{}14.12} & {\small{}70.00} & {\small{}124.4 } & {\small{}135.4}\tabularnewline
				{\small{}$d_{av}$} & {\small{}2.046} & {\small{}2.307 } & {\small{}3.198} & {\small{}1.563 } & {\small{}2.057 } & {\small{}2.568}\tabularnewline
				\hline
			\end{tabular}
			\par\end{centering}
		\bigskip
		\parbox{6.2in}{\footnotesize
			Notes: This table gives average characteristics of the payoff graphs,
			$G_{P}$, used in the simulation study, where the average was over 50 simulations. $d_{av}$ and $d_{mx}$ denote
			the average and maximum degrees of the payoff graphs.}
	\end{table}
	
	We consider the average of the ANE's over simulated payoff graphs. For the payoff graph $G_P$, we considered two different models for random graph generation. The first kind of random graphs are Erd\"{o}s-R\'{e}nyi (ER) random graph with the probability equal to $5/n$ and the second kind of random graphs are Barab\'{a}si-Albert (BA) random graph such that beginning with an Erd\"{o}s-R\'{e}nyi random graph of size 20 with each link forming with equal probability 1/19 and grows by including each new node with two links formed with the existing nodes with probability proportional to the degree of the nodes.
	
	For each random graph, we first generate a random graph of size 10,000, and then construct three subgraphs $A,B,C$ such that network $A$ is a subgraph of network $B$ and the network $B$ is a subgraph of network $C$. We generate these subgraphs as follows. First, we take a subgraph $A$ to be one that consists of agents within distance $k$ from agent $i=1$. Then network $B$ is constructed to be one that consists of the neighbors of the agents in network $A$ and network $C$ is constructed to be one that consists of the neighbors of the agents in network $B$. For an ER random graph, we took $k=3$ and for a BA random graph, we took $k=2$. We repeated the process 50 times to construct an average behavior of network externality as we increase the network. Table 1 shows the average network sizes and degree characteristics as we move from Networks A, B to C.
	
	\begin{figure}[t]
		\caption{\small The Average Network Externality Comparison Between Equilibrium and Behavioral Models: Erd\"{o}s-R\'{e}nyi Graphs}
		\includegraphics[scale=0.55]{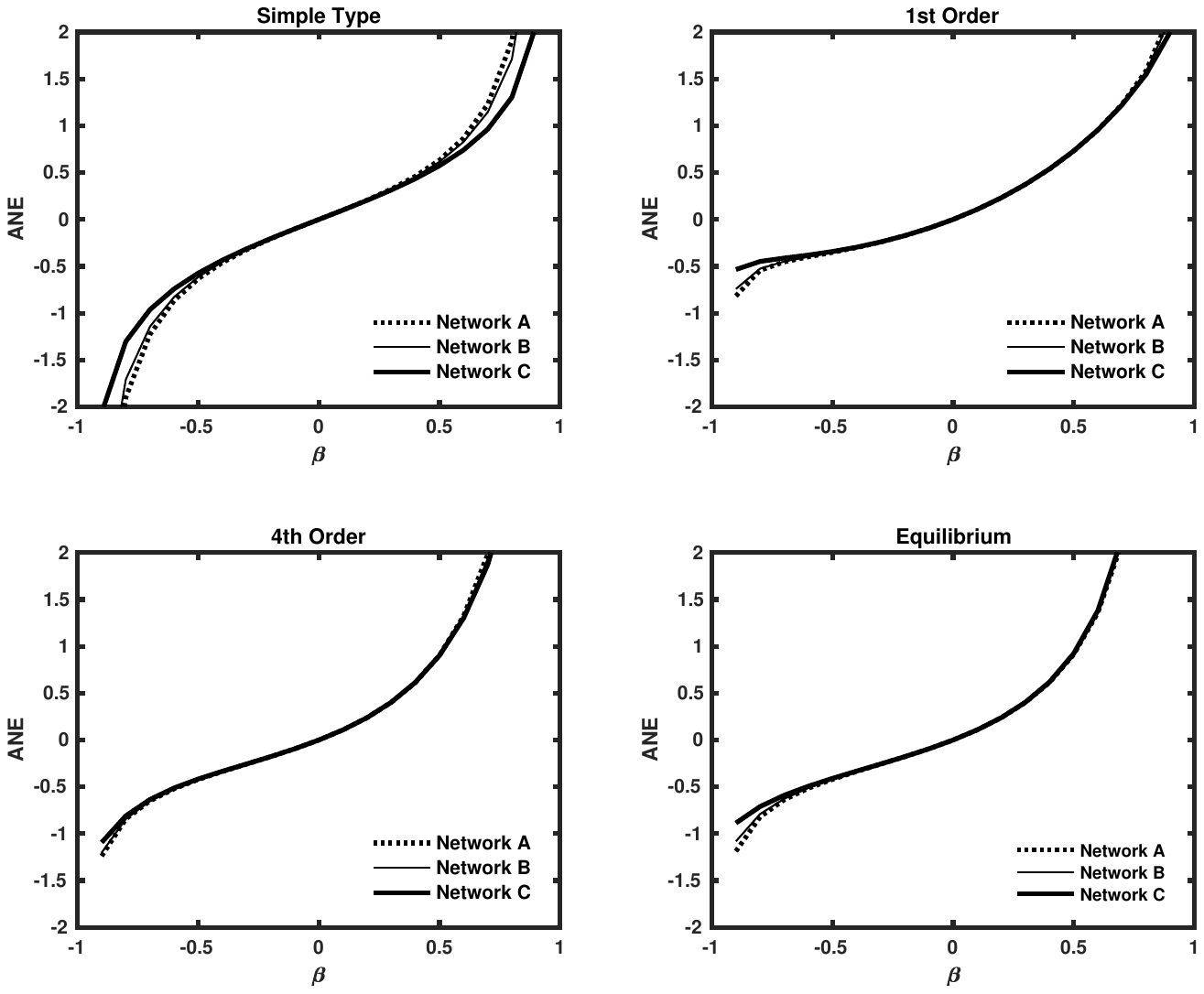}
		\parbox{6.2in}{\footnotesize Notes: Each line represents the average network externality (ANE) as a function of $\beta_0$. Each panel shows multiple lines representing ANE as we expand the graph from a subgraph of agents within distance $k$ from the agent 1. (Networks A, B, and C correspond to networks with $k=3,4,5$ from a small graph to a large one.) The figures show that the ANE is stable across different networks, and that the ANE from the behavioral model converges to that from the equilibrium model as the order of sophistication becomes higher.}
	\end{figure}
	
	\begin{figure}[t]
		\caption{\small The Average Network Externality Comparison Between Equilibrium and Behavioral Models: Barab\'{a}si-Albert Graphs}
		\includegraphics[scale=0.55]{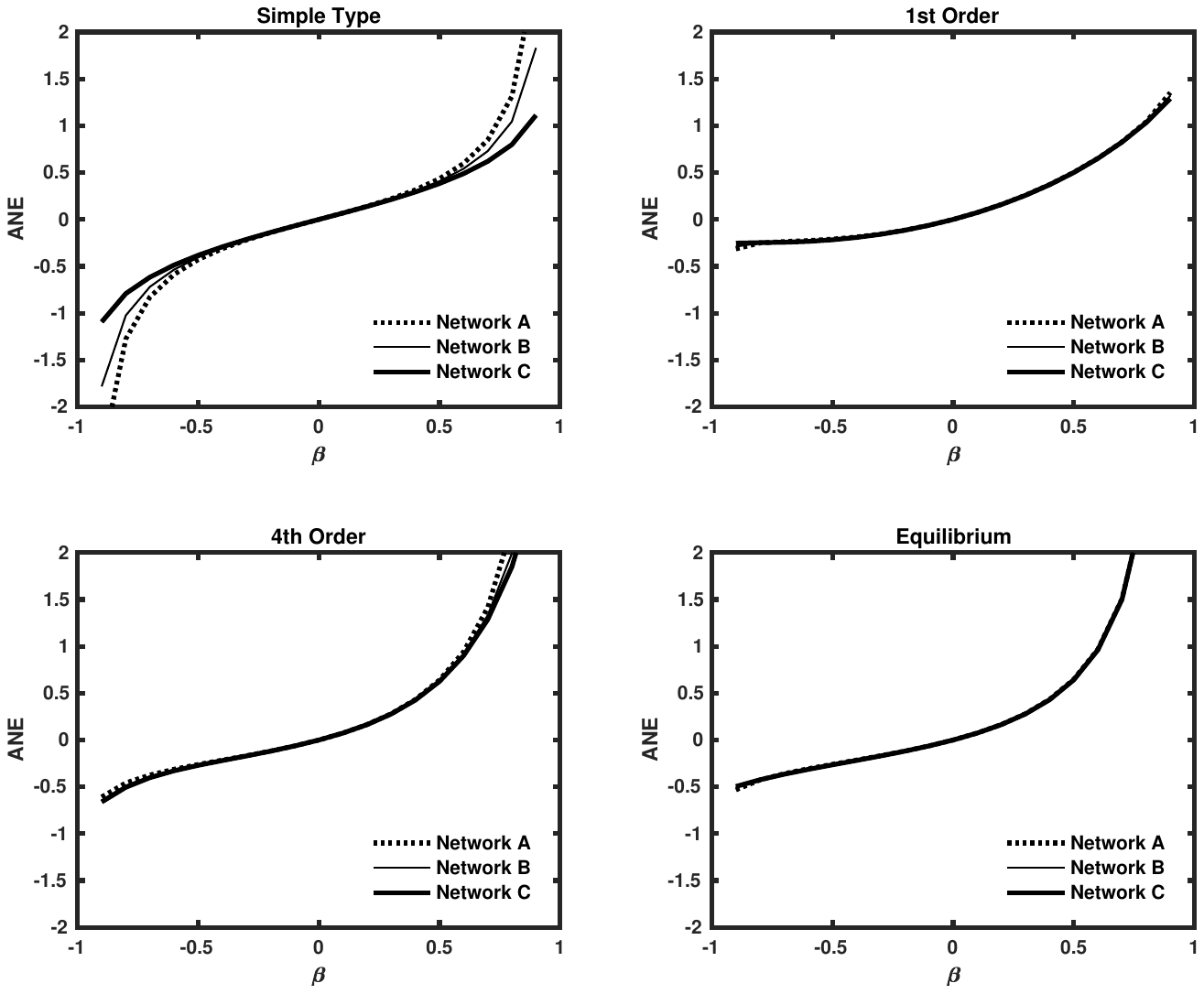}
		\parbox{6.2in}{\footnotesize Notes: Each line represents the average network externality (ANE) as a function of $\beta_0$. Each panel shows multiple lines representing ANE as we expand the graph from a subgraph of agents within distance $k$ from the agent 1. (Networks A, B, and C correspond to networks with $k=2,3,4$ from a small graph to a large one.)}
	\end{figure}
	The ANEs from the equilibrium strategies from game $\Gamma_\infty$, and the behavioral strategies from games $\Gamma_m$ as $m$ becomes higher are shown in Figures 1 and 2.\footnote{We provide conditions for the local identification of $\beta$ in Section \ref{subsubsec: local identification} below.} First, as $m$ becomes larger, the ANEs from $\Gamma_\infty$ and those from $\Gamma_m$ get closer, as predicted by Theorem \ref{thm: conv}. Furthermore, the ANEs from the behavioral model are similar to that from the equilibrium model especially when $\beta_0$ is between $-0.5$ and $0.5$. Finally, the network externalities from the games with simple type players are somewhat sensitive to the size of the networks when $\beta_0$ is very high or very low. This sensitivity is reduced substantially when we consider the game with first order sophisticated agents. Finally, network externality tends to be much higher for equilibrium models than the behavioral models when $\beta_0$ is high. Hence using our behavioral approach as a proxy for an equilibrium approach makes sense only when strategic interdependence is not too high. 
	
	\section{Econometric Inference}
	\label{sec: moment restrictions}
	\subsection{General Overview}
	\subsubsection{Partial Observation of Interactions}
	A large network data set is often obtained through a non-random sampling process. (See e.g. \citemain{Kolaczyk:09:SAND}.) The actual sampling process of network data is often unknown to the researcher. Our approach of empirical modeling can be useful in a situation where only a fraction of the players are observed through a certain non-random sampling scheme that is not precisely known to the researcher. In this section, we make explicit the data requirements for the econometrician and propose inference procedures. We mainly focus on the game where all the players in the game are of simple type. We develop inference for games with agents of first-order sophisticated type in the Supplemental Note.
	
	Suppose that the original game of interactions consists of a large number of agents whose set we denote by $N$. Let the set of players be on a payoff graph $G_P$ and an information graph $G_I$, facing the strategic environment as described in the preceding section. Denote the best response as an observed dependent variable $Y_i$: for $i \in N$,
	\begin{eqnarray*}
		Y_i = s_i^{[0]}(\mathcal{I}_{i,0}),
	\end{eqnarray*}
where the sharable type $\tau_i$ is specified as:
\begin{eqnarray}
\label{tau spec}
	\tau_i = X_i' \rho_0 + \varepsilon_i,
\end{eqnarray} 
and $X_i$ is a $d$-dimensional vector of covariates pertaining to agent $i$ observed by the econometrician, $\rho_0 \in \mathbf{R}^d$ is a coefficient vector, and $\varepsilon_i$ is unobserved heterogeneity. The covariate $X_i$ can contain $G_P$-neighborhood averages of individual covariates. Let us make the following additional assumption on this original large game. Let us first define
	\begin{eqnarray*}
		\mathcal{F} = \sigma(X,G_P,G_I) \vee \mathcal{C},
	\end{eqnarray*}
	i.e., the $\sigma$-field generated by $X= (X_i)_{i \in N}$, $G_P$, $G_I$ and $\mathcal{C}$ is a given common shock which is to be explained below.
	
	\begin{assumption}
		\label{assump: unobserved heterogeneity} (i) $\varepsilon_i$'s and $\eta_i$'s are conditionally i.i.d. across $i$'s given $\mathcal{F}$.
		
		\noindent (ii) $\{\varepsilon_i\}_{i=1}^n$ and $\{\eta_i\}_{i=1}^n$ are conditionally independent given $\mathcal{F}$.
		
		\noindent (iii) For each $i \in N$, $\mathbf{E}[\varepsilon_i|\mathcal{F}] = 0$ and $\mathbf{E}[\eta_i|\mathcal{F}] = 0$.
	\end{assumption}
	
	Condition (i) excludes pre-existing cross-sectional dependence of unobserved heterogeneity in the payoffs once conditioned in $\mathcal{F}$. This condition implies that conditional on $\mathcal{F}$, the cross-sectional dependence of observed actions is due solely to the information sharing among the agents. Condition (ii) requires that conditional on $\mathcal{F}$, the unobserved payoff heterogeneities observed by other players and those that are private are independent. Condition (iii) excludes endogenous formation of $G_P$ or $G_I$, because the condition requires that the unobserved type components $\varepsilon_i$ and $\eta_i$ be conditionally mean independent of these graphs, given $X = (X_i)_{i \in N}$ and $\mathcal{C}$. However, the condition does not exclude the possibility that $G_P$ and $G_I$ are exogenously formed based on $(X,\mathcal{C})$. For example, suppose that $ij \in E_P$ if and only if 
	\begin{eqnarray*}
		f_{ij}(X_i,X_j,a_i,a_j,u_{ij}) \ge 0,
	\end{eqnarray*}
	where $a_i$ represents degree heterogeneity, $u_{ij}$'s errors, and $f_{ij}$ a given nonstochastic function. In this set-up, the econometrician does not observe $a_i$'s or $u_{ij}$'s. This nests the dyadic regression model of \citemain{Graham:17:Eca} as a special case. Condition (iii) accommodates such a set-up, as long as $\{a_i\}_{i = 1}^n$ and $\{u_{ij}\}_{i,j = 1}^n$ are conditionally independent of $\{\varepsilon_i\}_{i=1}^n$ and $\{\eta_i\}_{i=1}^n$ given $X$. One simply has to take $\mathcal{C}$ to contain $a_i$'s and $u_{ij}$'s. 
	
	The econometrician observes only a subset $N^* \subset N$ of agents and part of $G_P$ through a potentially stochastic sampling process of unknown form. We assume for simplicity that $n^* \equiv |N^*|$ is nonstochastic. This assumption is satisfied, for example, if one collects the data for agents with predetermined sample size $n^*$. We assume that though being a small fraction of $N$, the set $N^*$ is still a large set justifying our asymptotic framework that sends $n^*$ to infinity. Most importantly, constituting only a small fraction of $N$, the observed sample $N^*$ of agents induces a payoff subgraph which one has no reason to view as ``approximating" or ``similar to" the original payoff graph $G_P$. Let us make precise the data requirements. 
	\smallskip
	
	\noindent \textbf{Condition A:} The stochastic elements of the sampling process are conditionally independent of $\{(\tau_i,\eta_i)'\}_{i \in N}$ given $\mathcal{F}$.
	\smallskip
	
	\noindent \textbf{Condition B:} For each $i \in N^*$, the econometrician observes $N_P(i)$ and $(Y_i,X_i)$, and for each $j \in N_P(i)$, the econometrician observes $|N_P(i)\cap N_P(j)|$, $n_P(j)$ and $X_j$.
	\smallskip
	
	\noindent \textbf{Condition C:} Either of the following two conditions is satisfied:

	(a) For $i,j \in N^*$ such that $i \ne j$, $N_P(i) \cap N_P(j) = \varnothing$.
	
	(b) For each agent $i \in N^*$, and for any agent $j \in N^*$ such that $N_P(i) \cap N_P(j) \ne \varnothing$, the econometrician observes $Y_j$, $|N_P(j)\cap N_P(k)|$, $n_P(k)$ and $X_k$ for all $k \in N_P(j)$.
	\smallskip
	
	Before we discuss the conditions, it is worth noting that these conditions are trivially satisfied when we observe the full payoff graph $G_P$ and $N^* = N$. Condition A is satisfied, for example, if the sampling process is based on observed characteristics $X$ and some characteristics of the strategic environment that is commonly observed by all the players. This condition is violated if the sampling is based on the outcomes $Y_i$'s or unobserved payoff-relevant signals such as $\varepsilon_i$ or $\eta_i$. Condition B essentially requires that in the data set, we observe $(Y_i,X_i)$ of many agents $i$, and for each $G_P$-neighbor $j$ of agent $i$, observe the number of the agents who are common $G_P$-neighbors of $i$ and $j$ and the size of $G_P$-neighborhood of $j$ along with the observed characteristics $X_j$.\footnote{Note that this condition is violated when the neighborhoods are top-coded in practice. For example, the maximum number of friends in the survey for a peer effects study can be set to be lower than the actual number of friends for many students. The impact of this top-coding upon the inference procedure is an interesting question on its own which deserves exploration in a separate paper.} As for a $G_P$-neighbor $j$ of agent $i \in N^*$, this condition does not require that the agent $j$'s action $Y_j$ or the full set of his $G_P$-neighbors are observed. Condition C(a) is typically satisfied when an initial sample of agents is randomly selected from a much larger set of agents so that no two agents have overlapping $G_P$-neighbors in the sample, and then their $G_P$ neighbors are selected for each agent in the sample to constitute $N^*$.\footnote{This random selection does not need to be a random sampling from the population of agents. Note that the random sampling is extremely hard to implement in practice in this situation, because one needs to use the equal probability for selecting each agent into the collection $N^*$, but this equal probability will be feasible only when one has at least the catalog of the entire population $N$.} In practice for use in inference, one can take the set $N^*$ to include only those agents that satisfy Conditions A-C as long as $N^*$ thereof is still large and the selection is based only on $(X,G_P)$. One can simply use only those agents whose $G_P$-neighborhoods are not overlapping, as long as there are many such agents in the data.
	
	\subsubsection{Moment Conditions}
	In order to introduce inference procedures for $\beta_0$ and other payoff parameters, let us define for $i \in N$,
	\begin{eqnarray}
	\label{Z_i}
	Z_i = \left(1 + \frac{\beta_0^2 \overline \lambda_i}{n_P(i) - \beta_0^2 \overline \lambda_i} \right)\left( X_i + \frac{\beta_0}{n_P(i)}\sum_{j \in N_P(i)} \lambda_{ij} X_j\right).
	\end{eqnarray}
	(Note that $Z_i$ relies on $\beta_0$ although it is suppressed from notation for simplicity as we do frequently below for other quantities.) By Theorem \ref{thm: best response} and (\ref{tau spec}), we can write
	\begin{eqnarray}
	\label{linear reg}
		Y_i = Z_i'\rho_0 + v_i,
	\end{eqnarray}
	where
	\begin{eqnarray*}
		v_i = \left(1 + \frac{\beta_0^2 \overline \lambda_i}{n_P(i) - \beta_0^2 \overline \lambda_i} \right)\left(\varepsilon_i +  \frac{\beta_0}{n_P(i)}\sum_{j \in N_P(i)} \lambda_{ij} \varepsilon_j\right) + \eta_i.
	\end{eqnarray*}
	Note that the observed actions $Y_i$ are cross-sectionally dependent (conditional on $\mathcal{F}$) due to information sharing on unobservables $\varepsilon_i$.
	
	Suppose that $\varphi_i$ is $M \times 1$ vector of instrumental variables (which potentially depend on $\beta_0$) with $M > d$ such that for all $i \in N$,
	\begin{eqnarray}
	\label{moment fn}
		\mathbf{E}[v_i \varphi_i] = 0.
	\end{eqnarray}
	Note that the orthogonality condition above holds for any $\varphi_i$ as long as for each $i \in N$, $\varphi_i$ is $\mathcal{F}$-measurable, i.e., once $\mathcal{F}$ is realized, there is no extra randomness in $\varphi_i$. This is the case, for example, when $\varphi_i$ is a function of $X=(X_i)_{i \in N}$ and $G_P$.
	
	While the asymptotic validity of our inference procedure admits a wide range of choices for $\varphi_i$'s, one needs to choose them with care to obtain sharp inference on the payoff parameters. Especially, it is important to consider instrumental variables which involve the characteristics of $G_P$-neighbors to obtain sharp inference on payoff externality parameter $\beta_0$. This is because the cross-sectional dependence of observations carries substantial information for strategic interdependence among agents.
	
	\subsubsection{Local Identification}
	\label{subsubsec: local identification}
	
	It is not hard to see that under regularity conditions (such as those preventing multicollinearity in $Z_i$), $\rho_0$ is identified up to $\beta_0$.\footnote{A standard identification analysis centers on a ``representative probability" from which we observe i.i.d. draws. A parameter is identified if it is uniquely determined under each representative probability. However, in our set-up, there is no such probability, as all observations exhibit heterogeneity and local dependence along a large, complex network. Here, ``identification" simply means ``consistent estimability" and ``local identification" means ``consistent estimability around a neighborhood of the true parameter".} However, the moment function in (\ref{moment fn}) is nonlinear in $\beta_0$, and hence even local identification of $\beta_0$ is not guaranteed unless we impose further assumptions. Here we provide conditions for local identification, but for inference we propose later, we pursue asymptotically valid inference allowing the parameters to be only partially identified.
	
	Let $\theta\equiv[\beta,\rho']'$ and write $v_{i}(\theta)=Y_{i}-Z_i '(\beta)\rho$, where $Z_i(\beta)$ is the same as $Z_i$ except that $\beta_0$ is replaced by $\beta$. Let  $\Theta$ be the parameter space for $\theta_0$.
	\begin{assumption}
		\label{assump: ident}
        (i) For all $i = 1,...,n$, $\varphi_i$ does not depend on $\theta \in \Theta$, and the parameter space $\Theta$ is compact, and $\beta \in [-1+\nu, 1 - \nu]$ for all $\beta$ such that $[\beta,\rho']' \in \Theta$, for some small $\nu>0$.
        
        (ii) There exists $C>0$ such that for all $n \ge 1$,
		\begin{eqnarray*}
			\frac{1}{n}\sum_{i=1}^n \mathbf{E}\left[ \left.\|X_{i,n}\|^4 + \frac{1}{n_P(i)}\sum_{j \in N_P(i)} \|X_{j,1}\|^4\right\vert G_P \right] + \frac{1}{n}\sum_{i=1}^n \mathbf{E}\left[\|\varphi_i\|^4|G_P\right] < C.
		\end{eqnarray*}
	
	   (iii) There exists $c>0$ such that the minimum eigenvalue of the matrix
	\begin{eqnarray*}
		\sum_{m=1}^M \left(\frac{1}{n}\sum_{i=1}^n H_{i,m}(\theta_0) \right) \left(\frac{1}{n}\sum_{i=1}^n H_{i,m}(\theta_0) \right)'
	\end{eqnarray*}
   is bounded from below by $c$ for all $n \ge 2$, where
   \begin{eqnarray*}
   	  H_{i,m}(\theta) \equiv \mathbf{E}\left[ \left. \left[ \binom{\partial v_i(\theta)/\partial \beta}{-Z_i(\beta)} \varphi_{i,m}\right]\right\vert G_P\right],
   \end{eqnarray*}
    and $\varphi_{i,m}$ is the $m$-th entry of $\varphi_i$. 
    
    (iv) $\max_{i \in N} n_P^2(i) /\sqrt{n} \rightarrow 0$, as $n \rightarrow \infty$.
	\end{assumption}

    Assumption \ref{assump: ident}(i) in regards to $\varphi_i$ simplifies the identification arguments and is satisfied when the ``instruments" $\varphi_i$ consist only of observed variables. Assumption \ref{assump: ident}(ii) is a moment condition for the covariates. Assumption \ref{assump: ident}(iii) is a nontrivial condition and is violated if the parameter space for $\rho_0$ includes zero, because we have $\partial v_{i}(\beta,0)/\partial \beta = 0$. Thus this assumption requires the researcher to know that the true parameter $\rho_0$ is away from zero. Assumption \ref{assump: ident}(iv) is a mild condition that requires that the payoff graph $G_P$ is not overly dense.
    
    Under this assumption, in combination with the conditions of Theorem \ref{thm: best response}, we can show that $\theta_0$ is locally identified (i.e., consistently estimable over a neighborhood of $\theta_0$.)
    	\begin{theorem}
    	\label{thm: cons estimability}
    	Suppose that Assumption \ref{assump: ident} and the conditions of Theorem \ref{thm: best response} hold. Then, there exists $\varepsilon>0$ such that if $\Theta = \overline B(\theta_0;\varepsilon)$, $\theta_0$ is consistently estimable, where $B(\theta_0;\varepsilon)$ is the $\varepsilon$-neighborhood of $\theta_0$ and $\overline B(\theta_0;\varepsilon)$ is its closure.
    \end{theorem}
    
    Since consistent estimability of $\theta_0$ requires that $\rho_0$ be away from zero, it is expected that as $\rho_0$ gets close to zero, $\beta_0$ is only ``weakly (locally) identified". As a researcher is rarely \textit{a priori} certain that $\rho_0$ is away from zero, we pursue inference that does not require this.
     
	\subsubsection{Estimation and Inference}
	\label{subsubsec: est and inf}
	We first estimate $\rho_0$ assuming knowledge of $\beta_0$. Define
	\begin{eqnarray*}
		S_{\varphi \varphi} = \varphi'\varphi/n^*, \text{ and } \tilde \varphi = \varphi S^{-1/2}_{\varphi \varphi},
	\end{eqnarray*}
	where $\varphi$ is an $n^* \times M$ matrix whose $i$-th row is given by $\varphi_i'$, $i \in N^*$. Define
	\begin{eqnarray}
	\label{lambda}
	\Lambda = \frac{1}{n^*}\sum_{i \in N^*} \sum_{j \in N^*} \mathbf{E}[v_i v_j|\mathcal{F}] \tilde \varphi_i \tilde \varphi_j',
	\end{eqnarray}
	where $\tilde \varphi_i$ represents the transpose of the $i$-th row of $\tilde \varphi$, and let $\hat \Lambda$ be a consistent estimator of $\Lambda$. (We will explain how we construct this estimator in Section \ref{subsubsec:asym cov} below.) Define
	\begin{eqnarray*}
		S_{Z \tilde \varphi} = Z'\tilde \varphi/n^*, \text{ and }
		S_{\tilde \varphi y} = \tilde \varphi'y/n^*,
	\end{eqnarray*}
	where $Z$ is an $n^* \times d$ matrix whose $i$-th row is given by $Z_i'$ and $y$ is an $n^* \times 1$ vector whose $i$-th entry is given by $Y_i$, $i \in N^*$. Then we estimate
	\begin{eqnarray}
	\label{rho hat}
	\hat \rho = \left[S_{Z \tilde \varphi}\hat \Lambda^{-1} S_{Z \tilde \varphi}'\right]^{-1} S_{Z \tilde \varphi}\hat \Lambda^{-1} S_{\tilde \varphi y}.
	\end{eqnarray}
	Using this estimator, we construct a vector of residuals $\hat v = [\hat v_i]_{i\in N^*}$, where
	\begin{eqnarray}
	\label{hat v beta}
	\hat v_i = Y_i - Z_i'\hat \rho.
	\end{eqnarray}
	Finally, we form a profiled test statistic as follows:
	\begin{eqnarray}
	\label{Pop Obj} \quad \quad \quad
	T(\beta_0) = \frac{\hat v' \tilde \varphi \hat \Lambda^{-1} \tilde \varphi' \hat v}{n^*},
	\end{eqnarray}
	making it explicit that the test statistic depends on $\beta_0$. Later we show that 
	\begin{eqnarray*}
		T(\beta_0) \rightarrow_d \chi_{M-d}^2, \text{ as } n^* \rightarrow \infty,
	\end{eqnarray*}
	where $\chi_{M-d}^2$ denotes the $\chi^2$ distribution with degree of freedom $M-d$. Let $C_{1-\alpha}^\beta$ be the $(1 - \alpha)100$\% confidence set for $\beta_0$ defined as
	\begin{eqnarray*}
		C_{1-\alpha}^\beta \equiv \{\beta \in (-1,1): T(\beta) \le c_{1-\alpha} \},
	\end{eqnarray*}
	where $T(\beta)$ is computed as $T(\beta_0)$ with $\beta_0$ replaced by $\beta$ and the critical value $c_{1-\alpha}$ is the $(1-\alpha)$-quantile of $\chi_{M-d}^2$.
	
	Let us now construct a confidence set for $\rho_0$. First, we establish that under regularity conditions,
	\begin{eqnarray*}
		\sqrt{n^*} \hat V^{-1/2}(\hat \rho - \rho_0) \rightarrow_d N(0,I_d),
	\end{eqnarray*}
	as $n^* \rightarrow \infty$, where 
	\begin{eqnarray*}
		\hat V = \left[S_{Z \tilde \varphi}\hat \Lambda^{-1} S_{Z \tilde \varphi}'\right]^{-1}.
	\end{eqnarray*}
	(See Section \ref{subsec: asymptotic theory} below for conditions and formal results.) Using this estimator $\hat \rho$, we can construct a $(1-\alpha)100\%$ confidence interval for $a'\rho_0$ for any non-zero vector $a$. For this define
	\begin{eqnarray*}
		\hat \sigma^2(a) = a' \hat V a.
	\end{eqnarray*}
	Let $z_{1-(\alpha/4)}$ be the $(1-(\alpha/4))$-percentile of $N(0,1)$. Define for a vector $a$ with the same dimension as $\rho$,
	\begin{eqnarray*}
		C_{1-(\alpha/2)}^\rho(\beta_0,a) = \left[a'\hat \rho - \frac{z_{1-(\alpha/4)} \hat \sigma(a)}{\sqrt{n}} ,a'\hat \rho + \frac{z_{1-(\alpha/4)} \hat \sigma(a)}{\sqrt{n}}\right].
	\end{eqnarray*}
	Then the confidence set for $a'\rho$ is given by\footnote{Instead of the Bonferroni approach here, one could consider a profiling approach where one uses $T(\rho) = \sup_{\beta}T(\beta,\rho)$ as the test statistic, where $T(\beta,\rho)$ is the test statistic constructed using $\rho$ in place of $\hat \rho$. The profiling approach is cumbersome to use here because one needs to simulate the limiting distribution of $T(\rho)$ for each $\rho$, which can be computationally complex when the dimension of $\rho$ is large. Instead, this paper's Bonferroni approach is simple to use because $\beta_0$ takes values from $(-1,1)$.} 
	\begin{eqnarray*}
		C_{1-\alpha}^\rho(a) = \bigcup_{\beta \in C_{1-(\alpha/2)}^\beta} C_{1-(\alpha/2)}^\rho(\beta,a).
	\end{eqnarray*}
	Notice that since $\beta$ runs in $(-1,1)$ and the estimator $\hat \rho$ has an explicit form, the confidence interval is not computationally costly to construct in general.
	
	Often the eventual parameter of interest is one that captures how strongly the agents's decisions are interdependent through the network. For this, we can use the average network externality (ANE) introduced in (\ref{ane}). Let $s_i^{[0]}(\mathcal{I}_{i,0})$ be the best response of agent $i$ having information set $\mathcal{I}_i$. Then the ANE with respect to $X_{i,r}$ (where $X_{i,r}$ represents the $r$-th entry of $X_i$) is given by $\theta_1(\beta_0,\rho_{0,r})$, where
	\begin{eqnarray*}
		\theta_1(\beta_0,\rho_{0,r}) &=& \frac{1}{n^*}\sum_{i \in N^*} \sum_{j \in N_P(i)}\frac{\partial s_i^{[0]}(\mathcal{I}_{i,0})}{\partial x_{j,r}}\\
		&=& \frac{1}{n^*}\sum_{i \in N^*} \sum_{j \in N_P(i)}\frac{\beta_0  \lambda_{ij} }{n_P(i)} \left( 1 + \frac{\beta_0^2 \overline \lambda_i}{n_P(i) - \beta_0^2 \overline \lambda_i} \right)\rho_{0,r},
	\end{eqnarray*}
	and $\rho_{0,r}$ denotes the $r$-th entry of $\rho_0$. See (\ref{externality}). Thus the confidence interval for $\theta_1(\beta_0,\rho_{0,r})$ can be constructed from the confidence interval for $\beta_0$ and $\rho_0$ as follows:
	\begin{eqnarray}
	\label{CI network extern}
		C^{\theta_1}_{1-\alpha} = \left\{\theta_1(\beta,\rho_r):  \beta \in C^\beta_{1-\alpha/2}, \text{ and }  \rho_r \in C^{\rho_r}_{1-\alpha/2} \right\},
	\end{eqnarray}
	where $C^\beta_{1-\alpha/2}$ and $C^{\rho_r}_{1-\alpha/2}$ denote the $(1-\alpha/2)100\%$ confidence intervals for $\beta_0$ and $\rho_{0,r}$ respectively.
	
	\subsubsection{Downweighting Players with High Degree Centrality}
	When there are players who are linked to many other players in $G_P$, the graph $G_P$ tends to be denser, and it becomes difficult to obtain good variance estimators that perform stably in finite samples. (In particular, obtaining an estimator of $\Lambda$ in (\ref{lambda}) which performs well in finite samples can be difficult.) To remedy this situation, this paper proposes a downweighting of those players with high degree centrality in $G_P$. More specifically, in choosing an instrument vector $\varphi_i$, we may consider the following:
	\begin{eqnarray}
	\label{downweighting}
	\varphi_i(X) = \frac{1}{\sqrt{\overline n_P(i)}}g_i(X),
	\end{eqnarray}
	where $g_i(X)$ is a function of $X$. This choice of $\varphi_i$ downweights players $i$ who have a large $G_P$-neighborhood. Thus we rely less on the variations of the characteristics of those players who have many neighbors in $G_P$.
	
	Downweighting agents too heavily may hurt the power of inference because the actions of agents with high centrality contain information about the parameter of interest through the moment restrictions. On the other hand, downweighting them too lightly may hurt the finite sample stability of inference due to strong cross-sectional dependence they cause to the observations. Since a model with agents of higher-order sophisticated type results in observations with more extensive cross-sectional dependence, the role of downweighting can be important for finite sample stability of inference in such a model.
	
	\subsubsection{Comparison with Linear-in-Means Models}
	
	Let us compare our model with a linear-in-means model used in the literature, which is specified as follows:
	\begin{eqnarray}
	\label{linear in means}
	Y_i = X_{i,1}'\gamma_0 + \overline X_{i,2}'\delta_0 + \beta_0 \mu_i^e(\overline Y_i) + v_i,
	\end{eqnarray}
	where $\mu_i^e(\overline Y_i)$ denotes the player $i$'s expectation of $\overline Y_i$, and
	\begin{eqnarray*}
		\overline{Y}_i = \frac{1}{n_P(i)}\sum_{j \in N_P(i)} Y_j \text{ and } \overline{X}_{i,2} = \frac{1}{n_P(i)}\sum_{j \in N_P(i)} X_{j,2}.
	\end{eqnarray*}
	The literature assumes rational expectations by equating $\mu_i^e(\overline Y_i)$ to $\mathbf{E}[\overline Y_i|\mathcal{I}_i]$, and then proceeds to identification analysis of parameters $\gamma$, $\delta_0$ and $\beta_0$. For actual inference, one needs to use an estimated version of $\mathbf{E}[\overline Y_i|\mathcal{I}_i]$. One standard way in the literature is to replace it by $\overline Y_i$ so that we have
	\begin{eqnarray*}
		Y_i = X_{i,1}'\gamma_0 + \overline X_{i,2}'\delta_0 + \beta_0 \overline Y_i + \tilde v_i,
	\end{eqnarray*}
	where $\tilde v_i$ is an error term defined as $\tilde v_i = \beta_0(\mathbf{E}[\overline Y_i|\mathcal{I}_i] - \overline Y_i) + v_i$. The complexity arises due to the presence of $\overline Y_i$ which is an endogneous variable that is involved in the error term $\tilde v_i$.\footnote{A similar observation applies in the case of a complete information version of the model, where one directly uses $\overline Y_i$ in place of $\mu_i^e(\overline Y_i)$ in (\ref{linear in means}). Still due to simultaneity of the equations, $\overline Y_i$ necessarily involve error terms $v_i$ not only of agent $i$'s own but other agents' as well.}
	
	As for dealing with endogeneity, there are two kinds of instrumental variables proposed in the literature. The first kind is a peers-of-peers type instrumental variable which is based on the observed characteristics of the neighbors of the neighbors. This strategy was proposed by \citemain{Kelejian/Robinson:1993:PRS}, \citet*{Bramoulle/Djebbari/Fortin:09:JOE} and \citet*{DeGiorgi/Pellizzari/Redaelli:10:AEJ}. The second kind of an instrumental variable is based on observed characteristics excluded from the group characteristics as instrumental variables. (See \citemain{Brock/Durlauf:01:ReStud} and \citemain{Durlauf/Tanaka:08:EI}.) However, finding such an instrumental variable in practice is not always a straightforward task in empirical research.
	
	Our approach of empirical modeling is different in several aspects. Our modeling uses behavioral assumptions instead of rational expectations, and produces a reduced form for observed actions $Y_i$ from using best responses. This reduced form gives a rich set of testable implications and makes explicit the source of cross-sectional dependence in relation to the payoff graph. Our inference approach permits any nontrivial functions of $\mathcal{F}$ to serve as instrumental variables. Furthermore, one does not need to observe many independent interactions for inference.
	
	\subsubsection{Estimation of Asymptotic Covariance Matrix}
	\label{subsubsec:asym cov}
	One needs to find estimators $\hat \Lambda$ and $\hat V$ to perform inference. First, let us find an expression for their population versions. After some algebra, it is not hard to see that 
	the population version (conditional on $\mathcal{F}$) of $\hat V$ is given by
	\begin{eqnarray}
	\label{Omega and V}
	V = \left[S_{Z \tilde \varphi} \Lambda^{-1} S_{Z \tilde \varphi}'\right]^{-1}.
	\end{eqnarray}
	
	For estimation, it suffices to estimate $\Lambda$ defined in (\ref{lambda}). For this, we need to incorporate the cross-sectional dependence of the residuals $v_i$ properly. From the definition of $v_i$, it turns out that $v_i$ and $v_j$ can be correlated if $i$ and $j$ are connected indirectly through two edges in $G_P$. One may construct an estimator of $\Lambda$ that is similar to the HAC (Heteroskedasticity and Autocorrelation Consistent) estimator, simply by imposing the dependence structure and replacing $v_i$ by $\hat v_i$. However, this standard method can lead to conservative inference with unstable finite sample properties, especially when each player has many players connected through two edges. Instead, this paper proposes an alternative estimator of $\Lambda$ as follows. (See the Supplemental Note for more explanations for this estimator.)
	
	Fixing a value for $\beta_0$, we first obtain a first-step estimator of $\rho$ as follows:
	\begin{eqnarray}
	\label{rho tilde}
	\tilde \rho = \left[S_{Z \tilde \varphi} S_{Z \tilde \varphi}'\right]^{-1} S_{Z \tilde \varphi} S_{\tilde \varphi y}.
	\end{eqnarray}
	(Compare this with (\ref{rho hat}).) Using this estimator, we construct a vector of residuals $\tilde v = [\tilde v_i]_{i \in N^*}$, where
	\begin{eqnarray}
	\label{tilde v beta}
	\tilde v_i = Y_i - Z_i'\tilde \rho.
	\end{eqnarray}
	Then we define
	\begin{eqnarray*}
		\hat \Lambda_1 &=& \frac{1}{n^*} \sum_{i \in N^*} \tilde v_i^2 \tilde \varphi_i \tilde \varphi_i', \text{ and }\\
		\hat \Lambda_2 &=& \frac{\hat s_\varepsilon}{n^*} \sum_{i \in N^*} \sum_{j \in N^*_{-i}: N_P(i) \cap N_P(j) \ne \varnothing} q_{\varepsilon,ij} \tilde \varphi_i \tilde \varphi_j',
	\end{eqnarray*}
	where
	\begin{eqnarray}
	\label{hat s}
	\hat s_\varepsilon = \frac{\displaystyle  \sum_{i \in N^*} \sum_{j \in N_P(i)\cap N^*} \tilde v_i \tilde v_j}{\displaystyle  \sum_{i \in N^*} \sum_{j \in N_P(i)\cap N^*} q_{\varepsilon,ij}},
	\end{eqnarray}
	and
	\begin{eqnarray*}
		q_{\varepsilon,ij} =  w_{ii}^{[0]}w_{jj}^{[0]}\left(\frac{\lambda_{ji}1\{i \in N_P(j)\}}{n_P(j)} 
		+ \frac{\lambda_{ij}1\{j \in N_P(i)\}}{n_P(i)} 
		+ \frac{\beta_0}{n_P(i)n_P(j)} \sum_{k \in N_P(i) \cap N_P(j)} \lambda_{ik} \lambda_{jk}\right).
	\end{eqnarray*}
	(Note that the quantity $q_{\varepsilon,ij}$ can be evaluated once $\beta_0$ is fixed.) 
	We construct an estimator of $\Lambda$ as follows:\footnote{Under Condition C(a) for sample $N^*$, we have $\Lambda_2 = 0$ because the second sum in the expression for $\Lambda_2$ is empty. Hence in this case, we can simply set $\hat \Lambda_2 = 0$.}
	\begin{eqnarray*}
		\hat \Lambda = \hat \Lambda_1 + \hat \Lambda_2.
	\end{eqnarray*}
	Using $\hat \Lambda$, we take the estimator for the covariance matrix $V$ to be\footnote{In finite samples, $\hat V$ is not guaranteed to be positive definite. We can modify the estimator by using spectral decomposition similarly as in \citemain{Cameron/Gelbach/Miller:11:JBES}. More specifically, we first take a spectral decomposition $\hat V = \hat B \hat A \hat B'$, where $\hat A$ is a diagonal matrix of eigenvalues $\hat a_j$ of $\hat V$. We replace each $\hat a_j$ by the maximum between $\hat a_j$ and some small number $c>0$ in $\hat A$ to construct $\hat A_*$. Then the modified version $\tilde V \equiv \hat B \hat A_* \hat B'$ is positive definite. For $c>0$, one may take $c=0.005$. In our simulation studies, this modification does not make much difference after all.}
	\begin{eqnarray}
	\label{hat Omega and hat V}
	\hat V  =  \left[S_{Z \tilde \varphi} \hat \Lambda^{-1} S_{Z \tilde \varphi}'\right]^{-1}.
	\end{eqnarray}

	\subsection{Asymptotic Theory}
	\label{subsec: asymptotic theory}
	In this section, we present the assumptions and formal results of asymptotic inference. We introduce some technical conditions.  
	\begin{assumption}
		\label{assump: nondeg} There exists $c>0$ such that for all $n^* \ge 1$, $\lambda_{\min}(S_{\varphi \varphi}) \ge c$, $\lambda_{\min} (S_{Z \tilde \varphi} S_{Z \tilde \varphi}') \ge  c$, $\lambda_{\min} (S_{Z \tilde \varphi} \Lambda^{-1} S_{Z \tilde \varphi}') \ge  c$, $\lambda_{\min} (\Lambda) \ge  c$, $\sigma_\eta^2>0$, and
		\begin{eqnarray*}
			\frac{1}{n^*}\sum_{i \in N^*} \frac{1}{n_P(i)} \sum_{j \in N_P(i) \cap N^*}\lambda_{ij} > c,
		\end{eqnarray*}
		where $\lambda_{\min}(A)$ for a symmetric matrix $A$ denotes the minimum eigenvalue of $A$.
	\end{assumption}
	
	\begin{assumption}
		\label{assump: moment bound}
		\noindent There exists a constant $C>0$ such that for all $n^* \ge 1$,
		\begin{eqnarray*}
			\max_{i \in N^\circ} ||X_i|| +  \max_{i \in N^\circ} ||\tilde \varphi_i|| \le C
		\end{eqnarray*}
		and $\mathbf{E}[\varepsilon_i^4|\mathcal{F}] + \mathbf{E}[\eta_i^4|\mathcal{F}] <C$,
		where $n^\circ = |N^\circ|$ and
		\begin{eqnarray*}
			N^\circ = \bigcup_{ i \in N^*} \overline{N}_P(i).
		\end{eqnarray*}
	\end{assumption}
	
	Assumption \ref{assump: nondeg} is used to ensure that the asymptotic distribution is nondegenerate. This regularity condition is reasonable, because an asymptotic scheme that gives a degenerate distribution would not be adequate for approximating a finite sample, nondegenerate distribution of an estimator. Assumption \ref{assump: moment bound} can be weakened at the expense of added complexity in the conditions and the proofs.
	
	We introduce an assumption which requires the payoff graph to have a bounded degree over $i$ in the observed sample $N^*$.
	\begin{assumption}
		\label{assump: degree and moment}	
		There exists $C>0$ such that for all $n^* \ge 1$,
		\begin{eqnarray*}
			\max_{i \in N^*} |N_P(i)| \le C.
		\end{eqnarray*}
	\end{assumption}
	We may relax the assumption to a weaker, yet more complex condition at the expense of longer proofs, but in our view, this relaxation does not give additional insights. When $N^*$ is large, one can remove very high-degree nodes to obtain stable inference. As such removal is solely based on the payoff graph $G_P$, the removal does not lead to any violation of the conditions in the paper. 
	
	The following theorem establishes the asymptotic validity of inference based on the best responses in  Theorem \ref{thm: best response}, without using Assumption \ref{assump: ident}, i.e., without requiring local identification of $\theta_0$. 

	\begin{theorem}
		\label{thm: theta hat exog}
		Suppose that the conditions of Theorem \ref{thm: best response} and Assumptions \ref{assump: unobserved heterogeneity} - \ref{assump: degree and moment} hold. Then,
		\begin{eqnarray*}
			T(\beta_0) \rightarrow_d \chi^2_{M-d}, \text{ and } \hat V^{-1/2} \sqrt{n^*}\left( \hat \rho  - \rho_0 \right) \rightarrow_d N(0,I_d),
		\end{eqnarray*}
		as $n^* \rightarrow \infty$.
	\end{theorem}
	
	The theorem yields that the confidence sets for $\beta_0$ and $\rho_0$ that we proposed earlier are asymptotically valid. The proof of both theorems are found in the Supplemental Note. At the center of the asymptotic derivation is noting first that $v_i$'s have a conditional depenency graph in the sense that two sets $(v_j)_{j \in A}$ and $(v_j)_{j \in B}$ with $G_P$ neighborhoods of $A$ and $B$ nonoverlapping are conditionally independent given $\mathcal{F}$) and then applying the Central Limit Theorem for a sum of random variables that has a sparse conditional dependency graph. For the  proof, we use a version of such a central limit theorem in \citemain{Penrose:03:RandomGeometricGraphs}. The sparsity of such a graph is ensured by the bounded degree assumption (\ref{assump: degree and moment}). (Note that we can relax this assumption by letting the maximum degree increase slowly with $n$, but as mentioned before, this relaxation does not add additional insights only lengthening the mathematical proofs.) The local dependence structure coming from the conditional dependency graph affects our inference through the estimated variance via the way $\hat \Lambda$ is constructed. If the payoff graph $G_P$ is not sparse enough, the asymptotic approximation in Theorem \ref{thm: theta hat exog} may perform poorly in finite samples.
		
\section{A Monte Carlo Simulation Study}

\label{sec: Monte Carlo}

\subsection{Simulation Design}
In this section, we investigate the finite sample properties of the
asymptotic inference across various configurations of the payoff graph,
$G_{P}$. (We present Monte Carlo simulation
results for the game with the first-order sophisticated types in Appendix \ref{App sec:Inf FOS} in the
Supplemental Note.) The payoff graphs are generated according to two
models of random graph formation, which we call Specifications 1 and
2. Specification 1 uses the Barabási-Albert model of preferential
attachment, with $m$ representing the number of edges each new node
forms with existing nodes. The number $m$ is chosen from $\{1,2,3\}$.
Specification 2 is the Erdös-Rényi random graph with probability $p=\lambda/n$,
where $\lambda$ is also chosen from $\{1,2,3\}$.\footnote{Note that in Specification 1, the Barabási-Albert graph is generated
	with an Erdös-Rényi seed graph, where the number of nodes in the seed
	is set to equal the smallest integer above $5\sqrt{n}$. All graphs
	in the simulation study are undirected.} In Table \ref{table graph characteristics MC}, we report degree characteristics of the payoff
graphs used in the simulation study.

For the simulations, we set the following: 
\begin{align*}
\tau_{i} & =X_{i}'\rho_{0}+\varepsilon_{i},
\end{align*}
where $\rho_{0}=(2,4,1,3,4)'$ and $X_{i}=(X_{i,1},\overline{X}_{i,2})'$,
and 
\begin{eqnarray*}
	\overline{X}_{i,2}=\frac{1}{n_{P}(i)}\sum_{j\in N_{P}(i)}X_{j,2}.
\end{eqnarray*}
We generate $Y_{i}$ from the best response function in Theorem \ref{thm: best response} (or as in (\ref{linear reg})). We set and $a$ to be a column of ones so that $a'\rho_0=14$. The
variables $\varepsilon$ and $\eta$ are drawn i.i.d. from $N(0,1)$.
The first column of $X_{i,1}$ is a column of ones, while remaining
columns of $X_{i,1}$ are drawn independently from $N(1,1)$. The
columns of $X_{i,2}$ are drawn independently from $N(3,1)$.

For instruments, we use downweighting (\ref{downweighting}) as follows: 
\begin{eqnarray*}
	\varphi_i(X) = \frac{1}{\sqrt{\overline n_P(i)}} g_i(X),
\end{eqnarray*}
where 
\[
g_i(X) =[\tilde{Z}_{i,1},X_{i,1}^{2},\overline{X}_{i,2}^{2},\overline{X}_{i,2}^{3}]',
\]
where we define 
\[
\tilde{Z}_{i,1}\equiv\frac{1}{n_{P}(i)}\sum_{j\in N_{P}(i)}\lambda_{ij}X_{j,1}.
\]
While the instruments $X_{i,1}^{2},\overline{X}_{i,2}^{2},\overline{X}_{i,2}^{3}$
capture the nonlinear impact of $X_{i}$'s, the instrument $\tilde{Z}_{i,1}$
captures the cross-sectional dependence along the payoff graph. The
use of this instrumental variable is crucial in obtaining a sharp
inference for $\beta_{0}$. Note that since we have already concentrated
out $\rho_0$ in forming the moment conditions, we cannot use linear
combinations of $X_{i,1}$ and $\overline{X}_{i,2}$ as our instrumental
variables. The nominal size in all the experiments is set at $\alpha=0.05$. The Monte Carlo simulation number is set to 5000.

\begin{table}
	\caption{\small The Average and Maximum Degrees of Graphs in the Simulations}
	
	\begin{centering}
		
		\small
		\begin{tabular}{cc|cccccc}
			\hline 
			\hline 
			&  & \multicolumn{3}{c}{Specification 1} & \multicolumn{3}{c}{Specification 2}\tabularnewline
			$n$ &  & $m=1$ & $m=2$ & $m=3$ & $\lambda=1$ & $\lambda=2$ & $\lambda=3$\tabularnewline
			\hline 
			$500$ & $d_{mx}$ & 17 & 21 & 30 & 5 & 8 & 11\tabularnewline
			& $d_{av}$ & 1.7600 & 3.2980 & 4.8340 & 0.9520 & 1.9360 & 2.9600\tabularnewline
			\hline 
			$1000$ & $d_{mx}$ & 18 & 29 & 34 & 6 & 7 & 9\tabularnewline
			& $d_{av}$ & 1.8460 & 3.5240 & 5.2050 & 0.9960 & 1.9620 & 3.0020\tabularnewline
			\hline 
			$\text{5000}$ & $d_{mx}$ & 32 & 78 & 70 & 7 & 10 & 11\tabularnewline
			& $d_{av}$ & 1.9308 & 3.7884 & 5.6466 & 0.9904 & 2.0032 & 3.0228\tabularnewline
			\hline 
			& \multicolumn{1}{c}{} &  &  &  &  &  & \tabularnewline
		\end{tabular}
	\end{centering}
		\par
	\parbox[c]{6.2in}{ \footnotesize
		Notes: This table gives characteristics of the payoff graphs, $G_{P}$,
		used in the simulation study. $d_{av}$ and $d_{mx}$ represent the
		average and maximum degrees of the networks respectively; that is,
		$d_{av}\equiv\frac{1}{n}\sum_{i\in N}n_{P}(i)$ and $d_{mx}\equiv\max_{i\in N}n_{P}(i)$.%
	}
\label{table graph characteristics MC}

\bigskip
\bigskip
\end{table}
\begin{table}
	\caption{{\small{}The Empirical Coverage Probability and Average Length of
			Confidence Intervals for $\beta_{0}$ at 95\% Nominal Level.}}
	\small
	\begin{centering}
		\begin{tabular}{cc|cccccc}
			\multicolumn{8}{c}{Coverage Probability}\tabularnewline
			\hline 
			\hline 
			& \multicolumn{1}{c}{} & \multicolumn{3}{c}{Specification 1} & \multicolumn{3}{c}{Specification 2}\tabularnewline
			\cline{3-8} \cline{4-8} \cline{5-8} \cline{6-8} \cline{7-8} \cline{8-8} 
			$\beta_{0}$ & \multicolumn{1}{c|}{} & $m=1$ & $m=2$ & $m=3$ & $\lambda=1$ & $\lambda=2$ & $\lambda=3$\tabularnewline
			\hline 
			$-0.5$ & $n=500$ & 0.9642  & 0.9580  & 0.9648 & 0.9686  & 0.9638  & 0.9622\tabularnewline
			& $n=1000$ & 0.9638  & 0.9634  & 0.9574  & 0.9650  & 0.9644  & 0.9604\tabularnewline
			& $n=5000$ & 0.9596 & 0.9560  & 0.9530  & 0.9704  & 0.9608  & 0.9596 \tabularnewline
			\hline 
			$-0.3$ & $n=500$ & 0.9540  & 0.9536  & 0.9612  & 0.9608  & 0.9546  & 0.9568 \tabularnewline
			& $n=1000$ & 0.9566 & 0.9568  & 0.9566  & 0.9564  & 0.9578  & 0.9548 \tabularnewline
			& $n=5000$ & 0.9534 & 0.9548  & 0.9542  & 0.9636 & 0.9568  & 0.9546 \tabularnewline
			\hline 
			$0$ & $n=500$ & 0.9504  & 0.9464  & 0.9554  & 0.9474 & 0.9478  & 0.9490 \tabularnewline
			& $n=1000$ & 0.9486  & 0.9508  & 0.9514  & 0.9498  & 0.9510  & 0.9526 \tabularnewline
			& $n=5000$ & 0.9440  & 0.9490  & 0.9546  & 0.9516  & 0.9482 & 0.9478 \tabularnewline
			\hline 
			$0.3$ & $n=500$ & 0.9548 & 0.9512  & 0.9584 & 0.9562  & 0.9552  & 0.9556 \tabularnewline
			& $n=1000$ & 0.9600  & 0.9558 & 0.9524  & 0.9598  & 0.9592 & 0.9590 \tabularnewline
			& $n=5000$ & 0.9524  & 0.9536  & 0.9574 & 0.9604 & 0.9544  & 0.9522 \tabularnewline
			\hline 
			$0.5$ & $n=500$ & 0.9648 & 0.9574 & 0.9618 & 0.9640 & 0.9610 & 0.9620\tabularnewline
			& $n=1000$ & 0.9630 & 0.9604 & 0.9534 & 0.9710 & 0.9648 & 0.9634\tabularnewline
			& $n=5000$ & 0.9564 & 0.9598 & 0.9612 & 0.9700 & 0.9632 & 0.9584\tabularnewline
			\hline 
			& \multicolumn{1}{c}{} &  &  &  &  &  & \tabularnewline
			\multicolumn{8}{c}{Average Length of CI}\tabularnewline
			\hline 
			\hline 
			& \multicolumn{1}{c}{} & \multicolumn{3}{c}{Specification 1} & \multicolumn{3}{c}{Specification 2}\tabularnewline
			\cline{3-8} \cline{4-8} \cline{5-8} \cline{6-8} \cline{7-8} \cline{8-8} 
			$\beta_{0}$ & \multicolumn{1}{c|}{} & $m=1$ & $m=2$ & $m=3$ & $\lambda=1$ & $\lambda=2$ & $\lambda=3$\tabularnewline
			\hline 
			$-0.5$ & $n=500$ & 0.0834  & 0.1307 & 0.1947  & 0.1089  & 0.0751  & 0.0750 \tabularnewline
			& $n=1000$ & 0.0490  & 0.0794  & 0.1038  & 0.0630  & 0.0438  & 0.0463\tabularnewline
			& $n=5000$ & 0.0053 & 0.0203 & 0.0303 & 0.0108 & 0.0026 & 0.0024\tabularnewline
			\hline 
			$-0.3$ & $n=500$ & 0.0799  & 0.1216  & 0.1639  & 0.1083  & 0.0865  & 0.0910 \tabularnewline
			& $n=1000$ & 0.0464  & 0.0758  & 0.0990 & 0.0639  & 0.0519  & 0.0577 \tabularnewline
			& $n=5000$ & 0.0034 & 0.0187 & 0.0296 & 0.0116 & 0.0060 & 0.0075\tabularnewline
			\hline 
			$0$ & $n=500$ & 0.0785  & 0.1212  & 0.1572  & 0.1070  & 0.0970 & 0.1087 \tabularnewline
			& $n=1000$ & 0.0452  & 0.0753  & 0.0996  & 0.0638 & 0.0597  & 0.0700 \tabularnewline
			& $n=5000$ & 0.0024 & 0.0182 & 0.0298 & 0.0113 & 0.0106 & 0.0155\tabularnewline
			\hline 
			$0.3$ & $n=500$ & 0.0713  & 0.1062  & 0.1384  & 0.0983 & 0.0685  & 0.0676 \tabularnewline
			& $n=1000$ & 0.0404  & 0.0640  & 0.0872  & 0.0562  & 0.0389  & 0.0412 \tabularnewline
			& $n=5000$ & 0.0017 & 0.0155 & 0.0262 & 0.0076 & 0.0015 & 0.0013\tabularnewline
			\hline 
			$0.5$ & $n=500$ & 0.0495  & 0.0738  & 0.1085  & 0.0666  & 0.0289  & 0.0240 \tabularnewline
			& $n=1000$ & 0.0252  & 0.0337  & 0.0657  & 0.0328  & 0.0089 & 0.0079 \tabularnewline
			& $n=5000$ & 0.0001 & 0.0055 & 0.0147 & 0.0004 & 0.0000 & 0.0000\tabularnewline
			\hline 
			& \multicolumn{1}{c}{} &  &  &  &  &  & \tabularnewline
		\end{tabular}
		\par\end{centering}
	\parbox[c]{6.2in}{\footnotesize
		Notes: The first half of the table reports the empirical coverage
		probability of the asymptotic confidence interval for $\beta_{0}$
		and the second half reports its average length. The simulated rejection
		probability at the true parameter is close to the nominal size of
		$\alpha=0.05$ and the average lengths decrease with $n$. The simulation
		number is $R=5000$.%
	}
\label{table: beta MC}
\end{table}

\begin{table}
	\caption{{\small{}The Empirical Coverage Probability and Average Length of
			Confidence Intervals for $a'\rho_{0}$ at 95\% Nominal Level.}}
	\small
	\begin{centering}
		\begin{tabular}{cc|cccccc}
			\multicolumn{8}{c}{Coverage Probability}\tabularnewline
			\hline 
			\hline 
			& \multicolumn{1}{c}{} & \multicolumn{3}{c}{Specification 1} & \multicolumn{3}{c}{Specification 2}\tabularnewline
			\cline{3-8} \cline{4-8} \cline{5-8} \cline{6-8} \cline{7-8} \cline{8-8} 
			$\beta_{0}$ & \multicolumn{1}{c|}{} & $m=1$ & $m=2$ & $m=3$ & $\lambda=1$ & $\lambda=2$ & $\lambda=3$\tabularnewline
			\hline 
			$-0.5$ & $n=500$ & 0.9848  & 0.9802  & 0.9860  & 0.9862  & 0.9834  & 0.9740 \tabularnewline
			& $n=1000$ & 0.9616  & 0.9610  & 0.9680  & 0.9682  & 0.9670 & 0.9596 \tabularnewline
			& $n=5000$ & 0.9596 & 0.9548 & 0.9606 & 0.9706 & 0.9668 & 0.9614\tabularnewline
			\hline 
			$-0.3$ & $n=500$ & 0.9802  & 0.9772  & 0.9858  & 0.9832  & 0.9826  & 0.9794 \tabularnewline
			& $n=1000$ & 0.9620  & 0.9756  & 0.9796  & 0.9772 & 0.9682  & 0.9692 \tabularnewline
			& $n=5000$ & 0.9544 & 0.9510 & 0.9562 & 0.9588 & 0.9568 & 0.9556\tabularnewline
			\hline 
			$0$ & $n=500$ & 0.9740  & 0.9754  & 0.9868  & 0.9828  & 0.9792  & 0.9786 \tabularnewline
			& $n=1000$ & 0.9668  & 0.9770  & 0.9798 & 0.9746  & 0.9738  & 0.9756\tabularnewline
			& $n=5000$ & 0.9430 & 0.9500 & 0.9524 & 0.9546 & 0.9496 & 0.9494\tabularnewline
			\hline 
			$0.3$ & $n=500$ & 0.9804  & 0.9804  & 0.9866  & 0.9810  & 0.9778  & 0.9788 \tabularnewline
			& $n=1000$ & 0.9698  & 0.9794  & 0.9812  & 0.9816  & 0.9718  & 0.9758 \tabularnewline
			& $n=5000$ & 0.9476 & 0.9524 & 0.9546 & 0.9572 & 0.9536 & 0.9498\tabularnewline
			\hline 
			$0.5$ & $n=500$ & 0.9824  & 0.9828  & 0.9858  & 0.9810  & 0.9722 & 0.9720 \tabularnewline
			& $n=1000$ & 0.9724  & 0.9786 & 0.9826  & 0.9810  & 0.9596  & 0.9594 \tabularnewline
			& $n=5000$ & 0.9552 & 0.9536 & 0.9596 & 0.9626 & 0.9596 & 0.9542\tabularnewline
			\hline 
			& \multicolumn{1}{c}{} &  &  &  &  &  & \tabularnewline
			\multicolumn{8}{c}{Average Length of CI}\tabularnewline
			\hline 
			\hline 
			& \multicolumn{1}{c}{} & \multicolumn{3}{c}{Specification 1} & \multicolumn{3}{c}{Specification 2}\tabularnewline
			\cline{3-8} \cline{4-8} \cline{5-8} \cline{6-8} \cline{7-8} \cline{8-8} 
			$\beta_{0}$ & \multicolumn{1}{c|}{} & $m=1$ & $m=2$ & $m=3$ & $\lambda=1$ & $\lambda=2$ & $\lambda=3$\tabularnewline
			\hline 
			$-0.5$ & $n=500$ & 5.4643  & 10.2562 & 17.4488  & 7.0549  & 5.6285  & 6.0639\tabularnewline
			& $n=1000$ & 3.3501  & 5.9311  & 8.1165  & 4.2270  & 3.4243  & 3.8800 \tabularnewline
			& $n=5000$ & 0.7489 & 1.8254 & 2.4970 & 1.1165 & 0.6337 & 0.6588\tabularnewline
			\hline 
			$-0.3$ & $n=500$ & 4.2511 & 6.8326  & 9.5995  & 5.6856  & 4.9154  & 5.4091 \tabularnewline
			& $n=1000$ & 2.5915  & 4.3335  & 5.6775 & 3.4711  & 3.0685  & 3.5331 \tabularnewline
			& $n=5000$ & 0.5297 & 1.3514 & 1.9123 & 0.9505 & 0.7346 & 0.8560\tabularnewline
			\hline 
			$0$ & $n=500$ & 3.5812  & 5.3587  & 6.8508 & 4.7774  & 4.3760  & 4.8399 \tabularnewline
			& $n=1000$ & 2.1797 & 3.4445  & 4.4160  & 2.9651 & 2.7944  & 3.2136\tabularnewline
			& $n=5000$ & 0.4238 & 1.0993 & 1.5392 & 0.8365 & 0.7964 & 0.9882\tabularnewline
			\hline 
			$0.3$ & $n=500$ & 3.3664  & 4.5527  & 5.7822 & 4.4438  & 3.2458  & 3.1962 \tabularnewline
			& $n=1000$ & 2.0559  & 2.8904 & 3.6607  & 2.6872 & 2.0100  & 2.1074 \tabularnewline
			& $n=5000$ & 0.4072 & 0.9592 & 1.2989 & 0.6961 & 0.4399 & 0.4713\tabularnewline
			\hline 
			$0.5$ & $n=500$ & 3.0230  & 4.1201  & 5.8780  & 3.7624  & 2.1350  & 1.9826 \tabularnewline
			& $n=1000$ & 1.7684  & 2.1613  & 3.4957  & 2.1027  & 1.1718  & 1.2075 \tabularnewline
			& $n=5000$ & 0.3576 & 0.6764 & 1.0002 & 0.3995 & 0.3723 & 0.4081\tabularnewline
			\hline 
			& \multicolumn{1}{c}{} &  &  &  &  &  & \tabularnewline
		\end{tabular}
		\par\end{centering}
	\parbox[c]{6.2in}{\footnotesize
		Notes: The true $a'\rho_{0}$ is equal to 14. The first half of the
		table reports the empirical coverage probability of the asymptotic
		confidence interval and the second half its average length for $a'\rho_{0}$.
		The empirical coverage probability of the confidence interval for
		$a'\rho_{0}$ is generally conservative which is expected from the
		use of the Bonferroni approach. Nevertheless, the length of the confidence
		interval is reasonably small. The simulation number, $R$, is 5000.%
	}
\label{table: rho MC}
\end{table}
	
	\subsection{Results}
	The finite sample performance of the asymptotic inference is shown in Tables \ref{table: beta MC} and \ref{table: rho MC}. Overall, the simulation results illustrate good power and size
	properties for the asymptotic inference on $\beta_{0}$ and $a'\rho_{0}$.
	
	As for the size properties, the coverage probabilities of confidence intervals for $\beta_0$ are close to 95\% nominal level, as shown in Table \ref{table: beta MC}. The size properties are already good with $n=500$, and thus, show little improvement as $n$ increase to $5000$. The coverage probabilities of confidence intervals for $a'\rho_0$ (as shown in Table \ref{table: rho MC}) are a little conservative. This conservativeness
	is expected, given the fact that the interval is constructed using
	a Bonferroni approach. The conservativeness is alleviated as we increase the sample size.
	
	As for the power properties, we consider the average length of the confidence intervals. The confidence intervals for $\beta_0$ are very short, with average length around 0.1 - 0.2 when $n = 500$ and 0 - 0.03 when $n =5000$. As for $a'\rho_0$, the average length of the confidence intervals is around 2 - 17 when $n=500$ and 0.4 - 2.5 when $n = 5000$. Since $a'\rho_0 = 14$, the average length shows good power properties of the inference.
	
	\section{Empirical Application: State Presence across Municipalities}
	\subsection{Motivation and Background}
	State capacity (i.e., the capacity of a country to provide public goods, basic services, and the rule of law) can be limited for various reasons. (See e.g. \citemain{Besley/Persson:09:AER} and \citemain{Gennaioli/Voth:15:REStud}).\footnote{See also an early work by \citemain{Brett/Pinkse:00:CJE} for an empirical study on the spatial effects on municipal governments' decisions on business property tax rates.} A ``weak state" may arise due to political corruption and clientelism, and result in spending inadequately on public goods (\citemain{Acemoglu:05:JME}), accommodating armed opponents of the government (\citemain{Powell:13:QJE}), and war (\citemain{McBride/Milante/Skaperdas:11:JCE}). Empirical evidence has shown how these weak states can persist from precolonial times, with higher state capacities apparently related to current level prosperity at the ethnic and national levels (\citemain{Gennaioli/Rainer:07:JEG} and \citemain{Michalopoulos/Papaioannou:13:Eca}).
	
	Our empirical application is based on a recent study by \citemain{Acemoglu/GarciaJimeno/Robinson:15:AER} who investigate the local choices of state capacity in Colombia, using a model of a complete information game on an exogenously formed network. In their set-up, municipalities choose a level of spending on public goods and state presence (as measured by either the number of state employees or state agencies). Network externalities in a municipality's choice exist because municipalities that are adjacent to one another can benefit from their neighbors' choices of public goods provisions, such as increased security, infrastructure and bureaucratic connections. Thus, a municipality's choice of state capacity can be thought of as a strategic decision on a geographic network.   
	
	It is not obvious that public good provision in one municipality leads to higher spending on public goods in neighboring municipalities.  Some neighbors may free-ride and under-invest in state presence if they anticipate others will invest highly. Rent-seeking by municipal politicians would also limit the provision of public goods.  On the other hand, economies of scale could lend to complementarities in state presence across neighboring municipalities.
	
	In our study, we extend the model in \citemain{Acemoglu/GarciaJimeno/Robinson:15:AER} to an incomplete information game where information may be shared across municipalities. In particular, we do not assume that all municipalities know and observe all characteristics and decisions of the others. It seems reasonable that the decisions made across the country may not be observed or well known by those municipalities that are geographically remote. 
	
	\subsection{Empirical Set-up}
	
	Let $y_i$ denote the state capacity in municipality $i$ (as measured by the log number of public employees in municipality $i$) and $G_P$ denote the geographic network, where an edge is defined on two municipalities that are geographically adjacent.\footnote{This corresponds to the case in of $\delta_1=\delta_2 = 0$ in \citemain{Acemoglu/GarciaJimeno/Robinson:15:AER}.} We assume that $G_P$ is exogenously formed.
	\begin{figure}[t]
		\centering
		\caption{\small Degree Distribution of $G_P$}
		\label{degree_dist}
		\includegraphics[scale=0.3]{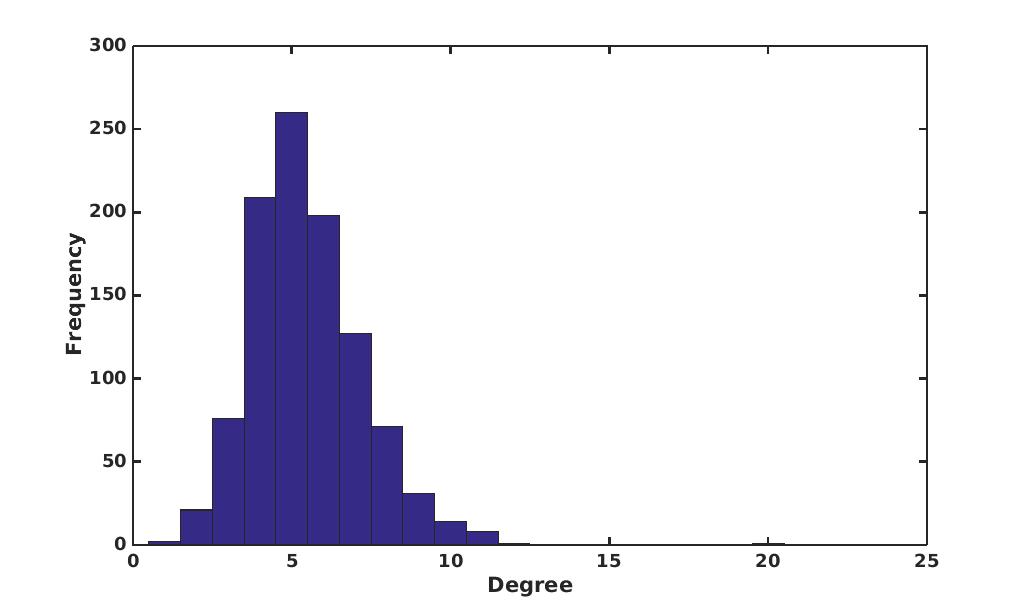}
		\parbox{6.2in}{\footnotesize
			Notes: The figure presents the degree distribution of the graph $G_P$ used in the empirical specification. The average degree is 5.48, the maximum degree is 20, and the minimum degree is 1. \bigskip}
	\end{figure}
	The degree distribution of $G_P$ is shown in Figure \ref{degree_dist}. We study the optimal choice of $y_i$, where $y_i$ leads to a larger prosperity $p_i$. Prosperity in municipality $i$ is modeled as:
	\begin{equation}
	\label{prosperity}
	p_i = \left(\beta \bar{y}_i + x_{1,i}' \gamma +\eta_i + \varepsilon_i + \varsigma^D_i\right) y_i,
	\end{equation}
	where $\varsigma^D_i$ is a district specific dummy variable, $\varepsilon_i$ and $\eta_i$ are our sharable and non-sharable private information, and $\overline{y}_i=\frac{1}{n_P(i)}\sum_{j \in N_P(i)} y_j.$  The term $x_{1,i}$ represents municipality characteristics.  These include geographic characteristics, such as land quality, altitude, latitude, rainfall; and municipal characteristics, such as distance to highways, distance to royal roads and Colonial State Presence.\footnote{Note that $p_i$ is only a function of terms that are multiplied by $y_i$.  This is a simplification from their specification.  We do so because we will focus on the best response equation.  The best response equation, derived from the first order condition to this problem, would not include any term that is not a function of $y_i$ itself.}
	
	The welfare of a municipality is given by
	\begin{equation}
	\label{empiricalu}
	u_i(y_i,y_{-i},\tau,\eta_i) = p_i(y_i,\bar{y}_i,\tau, \eta_i)  -\frac{1}{2}y_i^2,
	\end{equation}
	where the second term refers to the cost of higher state presence, and the first term is the prosperity $p_i$.
	
	We can rewrite the welfare of the municipality by substituting (\ref{prosperity}) into (\ref{empiricalu}):
	\begin{equation}
	\label{empiricalu2}
	u_i(y_i,y_{-i},\tau,\eta_i) = \left(\beta \bar{y}_i + x_{1,i}' \gamma +\eta_i + \varepsilon_i + \varsigma^D_i\right) y_i  -\frac{1}{2}y_i^2.
	\end{equation}
	We assume that municipalities (or the mayor in charge), wishes to maximize welfare by choosing state presence, given their beliefs about the types of the other municipalities.
	
	In our specification, we allow for incomplete information.  This is reflected in the terms $\varepsilon_i$, $\eta_i$, which will be present in the best response function.  The municipality, when choosing state presence $y_i$, will be able to observe $\varepsilon_i$ of its neighbors and will use its beliefs over the types of the others to generate its best response.  The best response will follow the results from Theorem \ref{thm: best response}.
	
	\subsection{Model Specification}
	
	We closely follow Table 3 in \citemain{Acemoglu/GarciaJimeno/Robinson:15:AER} for the choice of specifications and variables. First, we will consider the model with simple types.\footnote{In the supplemental note, we consider the empirical application with first order sophisticated types (game $\Gamma_1$), as well as the model selection test between simple types and first order sophisticated types from Appendix F. Since the simple type model is not rejected in the data and it is more parsimonious, we present it in the main text. The results for the first order sophisticated case are more or less similar except that the confidence intervals of $\beta_0$ are wider. At 5\%, the model selection procedure did not reject either of the sets of the moment conditions from the simple type and the first order sophisticated players.}
	
	Throughout the specifications, we include longitude, latitude, surface area, elevation, annual rainfall, department fixed effects and a department capital dummy (all in $X_1$). We further consider the effect of variables distance to current highways, land quality and presence of rivers in the municipality.
	
	For the choice of instruments, we consider two separate types of instruments. The first is the sum of neighbor values (across $G_P$) of the historical variables (denoted as $C_i$).\footnote{For this, we assume the exclusion restriction in \citemain{Acemoglu/GarciaJimeno/Robinson:15:AER}, namely that historical variables only affect prosperity in the same municipality. This means that although one's historical variables (Total Crown Employees, Distance to Royal Roads, Colonial State Agencies and Historical Population, as well as functions thereof) can affect the same municipality's prosperity, it can only affect those of the neighbors by impacting the choice of state capacity in the first, which then impacts the choice of the state capacity in the neighbors.} The historical variables used are Total Crown Employees (also called Colonial State Officials), Distance to Royal Roads, Colonial State Agencies and Historical Population, as well as Colonial State Presence Index squared and Distance to Royal Roads squared.  Using the latter two additionally sharpens inference. We also use the variable $\tilde{Z}_{i}=n_P(i)^{-1} \sum_{j \in N_P(i)} \lambda_{ij} X_{j,1}$ as part of the instrumental variables, which was shown to perform well in the Monte Carlo Simulations in Section 4. This variable captures cross sectional dependence as a crucial source of variation for inference on the strategic interactions.  We use downweighting of our instruments as explained in a preceding section and rescale instruments by multiplying them by $S_{\varphi \varphi}^{-1/2}$.
	
	\subsection{Results}
	\begin{table}[t]
		\begin{centering}
			\small
			\caption{\small State Presence and Networks Effects across Colombian Municipalities}
			\label{statecapacitytable}
			\resizebox{\columnwidth}{!}{%
				\begin{tabular}{cccccc}
					\hline 
					\hline 
					& &\multicolumn{4}{c}{Outcome: The Number of State Employees}
					\\
					\cline{2-6}
					&  & Baseline &  Distance to Highway & Land Quality & Rivers\\
					&  & (1) & (2) & (3) & (4) \\
					\hline
					\multicolumn{1}{c}{} & &  &  &  & \\
					$\beta_0$ &  &$[0.16,0.31]$ & $[0.16,0.32]$ & $[0.17,0.39]$ & $[0.07,0.38]$ \\
					\tabularnewline
					%Confidence Set for $\beta_0$
					$dy_i/d(\text{\footnotesize colonial state}$  & & $[-0.060,0.003]$ &$[-0.048,-0.001]$&$[-0.051,0.003]$ &$[-0.034,0.009]$\\
					$~\text{\footnotesize officials})$ & & & & & \\
					\tabularnewline
					\footnotesize Average & & & & & \\
					$dy_i/d(\text{\footnotesize colonial state}$  & & $[-1.323,4.051]$ &$[-1.249,2.793]$&$[-0.972,3.545]$ &$[-4.186,2.719]$\\
					$~\text{\footnotesize agencies})$& & & & & \\
					\tabularnewline
					\footnotesize Average & & & & & \\
					$dy_i/d(\text{\footnotesize distance to}$  & & $[-0.010,0.011]$ &$[-0.009,0.011]$&$[-0.008,0.018]$ &$[-0.010,0.013]$ \\
					$~\text{\footnotesize Royal Roads})$ & & & & & \\
					&  &  &  &  & \tabularnewline
					\hline 
					$n$ & & 1018 & 1018 & 1003 & 1003\\
					\hline 
					\multicolumn{1}{c}{} & &  &  &  &\\
				\end{tabular}
			}
			\par\end{centering}
		\parbox{6.2in}{\footnotesize
			Notes: Confidence sets for $\beta$ are presented in the table, obtained from inverting the test statistic $T(\beta)$ from Section \ref{sec: moment restrictions} for First Order Sophisticated types, with confidence level of 95\%.  The critical values in the first row come from the asymptotic statistic.  Downweighting is used.  The average marginal effects for historical variables upon state capacity are also shown.  The marginal effect of Colonial State Officials is equal to its $\gamma$ coefficient. The marginal effect for Distance to Royal Roads for municipality $i$ equals $\gamma_{Royal~Roads} + 2*\gamma_{Royal~Roads^2}(Royal~Roads)_i$, where $\gamma_{Royal~Roads}$ is the $\gamma$ coefficient of its linear term, and $\gamma_{Royal~Roads^2}$ is the coefficient of its quadratic term, as this variable enters $X_1$as a quadratic form. The analogous expression holds for the variable Colonial State Agencies.  We show the average marginal effect for these two variables. We then present the confidence set for these marginal effects, computed by the inference procedure on $a'\gamma$ developed in Section \ref{sec: moment restrictions}.
			All specifications include controls of latitude, longitude, surface area, elevation, rainfall, as well as Department and Department capital dummies.  Instruments are constructed from payoff neighbors' sum of the $G_P$ neighbors values of the historical variables Total Crown Employees, Colonial State Agencies, Colonial State Agencies squared, population in 1843, distance to Royal Roads, distance to Royal Roads squared, together with the non-linear function $\tilde{Z}_{i}=n_P(i)^{-1} \sum_{j \in N_P(i)} \lambda_{ij} X_{j,1}$.  Column (2) includes distance to current highway in $X_1$, Column (3) expands the specification of Column (2) by also including controls for land quality (share in each quality level). 	Column (4) controls for rivers in the municipality and land quality, in addition to those controls from Column (1).  
			One can see that the results are very stable across specifications.\bigskip}
	\end{table}
	
	The results across a range of specifications are presented in Table \ref{statecapacitytable}. In these results, we see that the effect is statistically different than 0 and stable across specifications.  It indicates that there is complementarity in the provision of public goods and state presence ($\beta>0$).
	
	Let us compare our results to those in \citemain{Acemoglu/GarciaJimeno/Robinson:15:AER}.  There, the authors report the average marginal effects over their weighted graph.  The (weighted) average degree is 0.0329, so our results can be compared in an approximation, by considering 0.0329 $\hat{\beta}$.  
	
	In general, our estimates have the same sign and significance as those of \citemain{Acemoglu/GarciaJimeno/Robinson:15:AER}.  Our estimates are in the range of [0.002, 0.013], after reweighting as mentioned before, somewhat comparable to theirs of [0.016, 0.022] (in the case of the outcome of the number of public employees, in Table 3).  Hence, we find similar qualitative effects, although a smaller magnitude.  Recall that our confidence set is built without assuming that $\beta_0$ is consistently estimable.
	
\begin{figure}[t]
		\centering
		\caption{\small Average Network Externality from being a Department Capital}
		\label{networkext}
		\includegraphics[scale = 0.65, trim={7cm 5cm 7cm 5.2cm},clip]{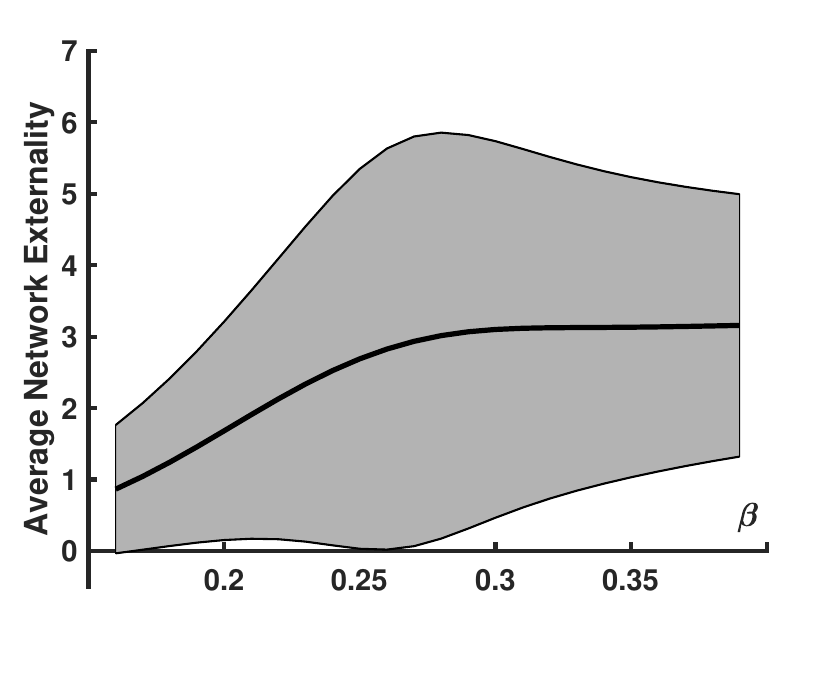}
		\parbox{6.2in}{\footnotesize
			Notes: The figure presents the average network externalities from being a department capital.  We use the estimated results from Column (3) in Table \ref{statecapacitytable}. This captures the externality for a municipality from being a department capital, which involves higher state presence and centralization of resources.  This effect is not only the direct effect, but it also quantifies a reflection effect: neighbors of department capitals also benefit from it. The grey shaded area represents the 95\% confidence interval for this ANE following the inference procedure developed in Section 3.\bigskip}
	\end{figure}

	In Figure \ref{networkext}, we show the results of our estimated network externalities for the estimates from Table \ref{statecapacitytable}, for the importance of being a department capital. The average network externality (ANE) is computed as
	\begin{eqnarray*}
		\frac{1}{N}\sum_{i \in N} \sum_{j \in N_P(i)}\frac{\beta_0 \hat{\gamma}_{dc}}{n_P(i)(1-\beta_0 c_{ij})} \left(1+\frac{\beta_0^2 \overline \lambda_i}{n_P(i) - \beta_0^2 \overline \lambda_i}\right),
	\end{eqnarray*}
	where $\hat{\gamma}_{dc}$ is the estimated parameter of the $X_1$ variable department capital. The parameter is defined in Section \ref{subsubsec: est and inf}, and captures the average effect of a neighbor being a department capital. We construct a confidence interval as in (\ref{CI network extern}).
	
	The figure shows that there is a strong and increasing network externality from being a department capital over the range of the confidence set of $\beta$.  This indicates that the effect of being a capital has spillovers on other municipalities: since $\beta>0$, and one expects that department capitals have more state presence and resources, being a department capital yields increasing returns the stronger the complementarity.
	
	\section{Conclusion}
	
	This paper proposes a new approach of empirical modeling for interactions among many agents when the agents observe the types of their neighbors in a single large network. The main challenge arises from the fact that the information sharing relations are typically connected among a large number of players whereas the econometrician observes only a fraction of those agents. Using a behavioral model of belief formation, this paper produces an explicit form of best responses from which an asymptotic inference procedure for the payoff parameters is developed. As we showed in our paper, this explicit form gives a reduced form for the observed actions, and exhibits various intuitive features. For example, the best responses show that network externality is heterogeneous across agents depending on the relations of their payoff neighbors.
	
	The main advantage of our paper's approach is two-fold. First, the empirical modeling according to our approach accommodates a wide range of sampling processes. Such a feature is crucial because the econometrician rarely has precise knowledge about the actual sampling process through which data are generated. Second, the model can be used when only a fraction of the players are observed from a large connected network of agents. This can be quite useful as the econometrician typically does not observe the entire set of agents who interact with each other. 
	
	An interesting extension from this research is to consider the situation where the network formation is endogenous, i.e., the randomness of $G_P$ is correlated with the unobserved payoff heterogeneities of agents even after conditioning on the covariates. It is not trivial to extend our framework to this situation, because we need to explicitly consider the randomness of $G_P$ in relation to other variables instead of conditioning our inference on $G_P$ as we do in this paper. Such consideration should be properly made in the construction of moment conditions and the proposal of appropriate instrumental variables. We relegate this extension to future research.
	
	\bibliographystylemain{econometrica}
	\bibliographymain{matching_networks_A3}

\newpage
\appendix

\vspace*{5ex minus 1ex}
\begin{center}
	\large \textsc{Supplemental Note to ``Estimating Local Interactions Among Many Agents Who Observe Their Neighbors"}
	\bigskip
\end{center}

\date{%
	%TCIMACRO{\TeXButton{Today}{\today}}%
	%BeginExpansion
	\today%
	%EndExpansion
}

\vspace*{3ex minus 1ex}
\begin{center}
	Nathan Canen, Jacob Schwartz, and Kyungchul Song\\
	\textit{University of Houston, University of Haifa, and University of British Columbia}\\
	\smallskip
	
\end{center}
\maketitle

\bigskip
\bigskip

This supplemental note provides further details on \citetsupp{Canen/Schwartz/Song:19:WP}. This note consists of nine appendices. In Appendix A, we formally present a model of information sharing over time among many agents where the econometrician is interested in the estimation of local interactions in a particular decision problem. We explain how this extended model maps to the static model of the main paper. In Appendix B, we give the mathematical proofs of Theorems \ref{thm: best response} and \ref{thm: best response2}. We also provide details on the best response of a Bayesian decision maker and the result of coincidence of equilibrium strategies and behavioral strategies when the payoff graph has multiple disjoint complete subgraphs. In Appendix C, we explain how estimation of the asymptotic covariance matrix is motivated.  Appendix D gives the proof of local identification (Theorem \ref{thm: cons estimability}) and the asymptotic results (Theorem \ref{thm: theta hat exog}). Appendix E explains inference based on the model with first order sophisticated agents. The secion also provides the proof of the asymptotic validity of the proposed inference, and results from Monte Carlo simulation studies. Appendix F provides details on the model selection procedure between different games $\Gamma_0$ and $\Gamma_1$. Appendix G presents a proposal on testing for information sharing on unobservables. Appendix H gives the proofs of the results on the convergence of behavioral strategies to equilibrium strategies from game $\Gamma_\infty$ in Section \ref{subsubsec: convergence of behav str}. Appendix I gives the results from the empirical application using the game with first-order sophisticated agents.

\section{Information Sharing Among Many Agents Over Time}

Information sharing among people takes place over time, and the econometrician usually observes part of these people as a snapshot in the process. The process involves information sharing, network formation and decision making. Agents can form a network and share information for various purposes. There is no reason to believe that each agent's particular decision problem which is of interest to the econometrician is the single ultimate concern of the agents when they form a network at an earlier time.\footnote{For example, it is highly implausible to assume that friendship formation among the students is done for the sole purpose of achieving maximal performance in a math exam observed by the econometrician.} In this section, we provide an extended model of information sharing which fits the main set-up of the paper. The main idea is that people receive signals, form networks, and share information with their neighbors repeatedly. Then there is a decision making stage. The network formation (of either a payoff graph or an information graph) can be made, if not exclusively, in anticipation of the decision making later. However, as we will explain later in detail, our model assumes that when the agents form an  information and a payoff graph, they do not observe other agents'  payoff relevant signals that are not observed by the econometrician. This is the precise sense in which the information and the payoff graph are exogenously formed.

Let us present a formal model of information sharing over time. Let $N$ be the set of a finite yet large number of players who share their type information over time recursively, where at each stage, players go through three steps sequentially: information graph formation, type realization and information sharing. Then at the final stage, players make a decision, maximizing their expected utilities.

At Stage 0, each agent $i \in N$ is endowed with signal $\mathcal{C}_{i,0}$. The signals can be correlated across agents in an arbitrary way. Then information sharing among agents happens recursively over time as follows starting from Stage $s=1$.
\smallskip

\noindent \textbf{Stage s-1:} (\textsc{Information Graph Formation}) Each player $i \in N$ receives signal $\mathcal{C}_{i,s-1}$. Using these signals, the players in $N$ form an information sharing network $G_{I,s-1}=(N,E_{I,s-1})$ among themselves, where $G_{I,s-1}=(N,E_{I,s-1})$ is a directed graph on $N$.\footnote{The formation of an information sharing network is tantamount to each agent making a (unilateral) binary decision to share his type information with others. Details of this decision making process are not of focus in the empirical model and hence are not elaborated further. What suffices for us is that the strategy of each agent is measurable with respect to the information the agent has. This latter condition is satisfied typically when the agent chooses a pure strategy given his information.}
\smallskip

(\textsc{Type Realization}) Each player $i \in N$ is given his type vector $(\tau_{i,s-1}',\eta_{i,s-1})'$, where $\eta_{i,s-1}$ is a private type which player $i$ keeps to himself and $\tau_{i,s-1}$ a sharable type which is potentially observed by other agents.
\smallskip

(\textsc{Information Sharing}) Each player $i$ observes the sharable type $\tau_{j,s-1}$ of each player $j$ in his neighborhood in the information sharing network.
\smallskip

\noindent \textbf{Stage s:} (\textsc{Information Graph Formation}) Each player $i \in N$ receives signal $\mathcal{C}_{i,s}$ which contains part of $\mathcal{C}_{i,s-1}$ and part of the information about $G_{I,s-1}$ and $\tau_{j,s-1}$ with $j \in N_{I,s-1}(i)$.  Using these signals, the players in $N$ form an information sharing network $G_{I,s}=(N,E_{I,s})$ among themselves, where $G_{I,s}=(N,E_{I,s})$ is a directed graph on $N$, and receives signal $\mathcal{C}_{i,s}$.
\smallskip

(\textsc{Type Realization}) Each player $i \in N$ is given his type vector $(\tau_{i,s}',\eta_{i,s})'$, where $\eta_{i,s}$ is a private type which player $i$ keeps to himself and $\tau_{i,s}$ a sharable type which is potentially observed by other agents.
\smallskip

(\textsc{Information Sharing}) Each player $i$ observes the sharable type $\tau_{j,s}$ of each player $j$ in his neighborhood in the information sharing network $G_{I,s}$.
\smallskip

The information sharing activities proceed up to Stage $S-1$. Then each agent faces a decision making problem.
\smallskip

\noindent \textbf{Decision Stage:} (\textsc{Payoff Graph Formation}) Each player $i \in N$ receives signal $\mathcal{C}_{i,S}$ which contains part of $\mathcal{C}_{i,S-1}$ and part of information about $G_{I,S-1}$ and $\tau_{j,S-1}$ with $j \in N_{I,S-1}(i)$.  Using these signals, the players in $N$ form a payoff graph $G_P=(N,E_P)$ among themselves, where $G_{I,S}=(N,E_{I,S})$ is a directed graph on $N$, and receives signal $\mathcal{C}_{i,S}$.
\smallskip

(\textsc{Type Realization}) Each player $i \in N$ is given his type vector $(\tau_{i,S}',\eta_{i,S})'$, where $\eta_{i,S}$ is a private type which player $i$ keeps to himself and $\tau_{i,S}$ a sharable type which is potentially observed by other agents.
\smallskip

(\textsc{Information Sharing}) Each player $i$ observes the sharable type $\tau_{j,S}$ of each player $j$ in his  neighborhood in the information sharing network $G_{I,S}$.
\smallskip

(\textsc{Decision}) Each player $i$ makes a decision which maximizes his expected utility given his beliefs about other agents' strategies using the information accumulated so far.
\smallskip

Now let us consider how this model of information sharing over time maps to our static local interactions model in our main paper. Our local interactions model captures the state where each player in $N$ faces the Decision Stage as follows. The information graph $G_I$ corresponds to $G_{I,S}$ and the payoff graph $G_P$ as described in Decision Stage above. The type vector $\tau_i$ and $\eta_i$ for each player $i \in N$ correspond to $\tau_{i,S}$ and $\eta_{i,S}$ as above. The signal $\mathcal{C}_i$ for each agent $i$ is defined to be
\begin{eqnarray*}
	\mathcal{C}_i = \bigvee_{s=1}^S \mathcal{C}_{i,s},
\end{eqnarray*}
where $\mathcal{A}_1 \vee \mathcal{A}_2$ denotes the smallest $\sigma$-field that contains both $\mathcal{A}_1$ and $\mathcal{A}_2$. Therefore, each agent $i$'s signal $\mathcal{C}_i$ contains the information that has been accumulated so far. This information contains past information sharing experiences that have happened over time.

Observe that here we deliberately separate the information sharing stage and the decision stage. This is because people typically share information without necessarily anticipating the particular decision problem that the econometrician happens to later investigate.

\section{Best Responses}

\subsection{Proofs of Theorems \ref{thm: best response}- \ref{thm: best response2}}

\begin{lemma}
	\label{lemm: exist BR}
	Suppose that $\beta_0 \in (-1,1)$. Then, for any $i \in N$ such that $n_P(i) \ge 1$,
	\begin{eqnarray*}
		n_P(i) - \beta_0^2 \overline \lambda_i \ge \frac{n_P(i)(n_P(i) + |\beta_0|)(1 - |\beta_0|)}{n_P(i)(1 - |\beta_0|) + |\beta_0|},
	\end{eqnarray*}
and
\begin{eqnarray}
\label{bd wij}
	0 \le w_{ii}^{[0]} \le 1 + \frac{\beta_0^2}{1 - \beta_0^2}, \text{ and }
	\left| w_{ij}^{[0]} \right| \le \frac{|\beta_0|}{n_P(i)(1 - |\beta_0|)}\left(1 + \frac{\beta_0^2}{1 - \beta_0^2} \right).
\end{eqnarray}
\end{lemma}

\noindent \textbf{Proof: } First, note that for any $j \in N_P(i)$,
\begin{eqnarray}
\label{cij}
	c_{ij} \le \frac{n_P(i)-1}{n_P(i)}.	
\end{eqnarray}
Hence
\begin{eqnarray}
\label{bd lam2}
	\lambda_{ij} \le \frac{n_P(i)}{n_P(i) - |\beta_0| (n_P(i)-1)} = \frac{n_P(i)}{n_P(i) (1 - |\beta_0|) + |\beta_0|}.
\end{eqnarray}
Therefore,
\begin{eqnarray}
\label{bd lam}
	\overline \lambda_i \le \frac{n_P(i)}{n_P(i) (1 - |\beta_0|) + |\beta_0|}.
\end{eqnarray}
Now, we find that
\begin{eqnarray*}
	n_P(i) - \beta_0^2 \overline \lambda_i &\ge& n_P(i) - \frac{\beta_0^2 n_P(i)}{n_P(i) (1 - |\beta_0|) + |\beta_0|}\\
	&=& \frac{n_P(i)(n_P(i) + |\beta_0|)(1 - |\beta_0|)}{n_P(i)(1 - |\beta_0|) + |\beta_0|}.
\end{eqnarray*}
This gives the desired lower bound for $n_P(i) - \beta_0^2 \overline \lambda_i$.

Let us turn to $w_{ii}^{[0]}$. Since $\lambda_{ij} \ge 0$, we have $w_{ii}^{[0]} \ge 0$ by the previous bound. As for its upper bound, from (\ref{bd lam}), we have 
\begin{eqnarray*}
	w_{ii}^{[0]} \le 1 + \frac{\beta_0^2}{(n_P(i)+ |\beta_0|)(1 - |\beta_0|)} \le 1 + \frac{\beta_0^2}{1 - \beta_0^2},
\end{eqnarray*}
because $n_P(i) \ge 1$. Finally, as for $w_{ij}^{[0]}$, it suffices to note that by (\ref{cij}),
\begin{eqnarray*}
	0 \le \lambda_{ij} \le \frac{1}{1 - |\beta_0|}.
\end{eqnarray*}
$\blacksquare$\medskip

\noindent \textbf{Proof of Theorem \ref{thm: best response}:} The case of $n_P(i) = 0$ is trivial. Let us assume that $n_P(i) \ge 1$. From the optimization of agent $i$, 
\begin{eqnarray}
s_{i}^{[0]}(\mathcal{I}_{i,0})=\tau_{i}+\beta_{0}\left(\frac{1}{n_{P}(i)}\sum_{k\in N_{P}(i)}\sum_{j\in\overline{N}_{P}(k)}w_{kj}^{i}\tau_{j}\right)+\eta_{i}.\label{FOC}
\end{eqnarray}
Reorganizing the terms, we have 
\begin{eqnarray}
\label{si0}
	s_{i}^{[0]}(\mathcal{I}_{i,0}) &=& \left(1+\frac{\beta_{0}}{n_{P}(i)}\sum_{k\in N_{P}(i)}w_{ki}^{i}\right)\tau_{i}\\ \notag
	&& +\beta_{0}\sum_{j\in N_{I}(i)}\frac{1}{n_{P}(i)}\sum_{k\in N_{P}(i)}w_{kj}^{i}1\{j\in\overline{N}_{P}(k)\} \tau_{j} + \eta_i.
\end{eqnarray}
By setting the coefficient of $\tau_{j}$ to be $w_{ij}$, we obtain
\begin{eqnarray}
w_{ii} & = & 1+\frac{\beta_{0}}{n_{P}(i)}\sum_{k\in N_{P}(i)}w_{ki}^{i},\label{wii}
\end{eqnarray}
and for all $j\in N_{I}(i)$, 
\begin{eqnarray}
w_{ij} & = & \frac{\beta_{0}}{n_{P}(i)}\sum_{k\in N_{P}(i)}w_{kj}^{i}1\{j\in \overline N_{P}(k)\}\label{wij}\\ \notag
& = & \frac{\beta_{0}}{n_{P}(i)}\sum_{k\in N_{P}(i)}w_{kj}^{i}1\{j\in N_{P}(k)\}+\beta_{0}\frac{w_{jj}^{i}1\{j\in N_{P}(i)\}}{n_{P}(i)}, 
\end{eqnarray}
where the last term corresponds to the case $j=k\in N_{P}(i)$.
We now apply our behavioural assumptions to (\ref{wii}) and (\ref{wij}). First, we apply the behavioural assumption to (\ref{wii}) to write
\begin{eqnarray}
w_{ii} & = & 1+\frac{\beta_{0}}{n_{P}(i)}\sum_{k\in N_{P}(i)}w_{ki}^{i}1\{i\in \overline N_{P}(k)\}\nonumber \\
& = & 1+\frac{\beta_{0}}{n_{P}(i)}\sum_{k\in N_{P}(i)}w_{ik}1\{i\in\overline N_{P}(k)\}\nonumber \\
& = & 1+\frac{\beta_{0}}{n_{P}(i)}\sum_{k\in N_{P}(i)}w_{ik},\label{eq:wii}
\end{eqnarray}
where the last line follows by the undirectedness of $G_{P}$. Turning
to (\ref{wij}), we have 
\begin{eqnarray}
w_{ij} & = & w_{ij}\frac{\beta_{0}}{n_{P}(i)}\sum_{k\in N_{P}(i)}1\{j\in N_{P}(k)\}+\beta_{0}\frac{w_{ii}1\{j\in N_{P}(i)\}}{n_{P}(i)}\nonumber \\
& = & w_{ij}\beta_{0}c_{ij}+\beta_{0}\frac{w_{ii}1\{j\in N_{P}(i)\}}{n_{P}(i)}.\label{eq:wij_2}
\end{eqnarray}
Rearranging (\ref{eq:wij_2}) for $w_{ij}$, we obtain
\begin{eqnarray}
w_{ij} & = & \beta_{0}w_{ii}\frac{\lambda_{ij}1\{j\in N_{P}(i)\}}{n_{P}(i)}.\label{wij_3}
\end{eqnarray}
We first solve for $w_{ii}$. By plugging (\ref{eq:wii}) into (\ref{wij_3})
and averaging over $N_{P}(i)$ we obtain
\begin{align*}
\frac{1}{n_P(i)}\sum_{j\in N_{p}(i)}w_{ij} & =\beta_{0}\left(1+\frac{\beta_{0}}{n_{P}(i)}\sum_{k\in N_{P}(i)}w_{ik}\right)\frac{\overline \lambda_{i}}{n_{P}(i)},
\end{align*}
which, after rearranging terms, becomes
\begin{align}
\frac{1}{n_P(i)}\sum_{j\in N_{p}(i)}w_{ij} & =\frac{\beta_{0}\overline \lambda_{i}}{n_P(i)-\beta_{0}^{2}\overline \lambda_{i}}.\label{eq:wijav}
\end{align}
Hence, by plugging (\ref{eq:wijav}) into (\ref{eq:wii}), we obtain 
\begin{eqnarray*}
	w_{ii} & = & 1+\frac{\beta_{0}^{2}\overline \lambda_{i}}{n_P(i)-\beta_{0}^{2}\overline \lambda_{i}}.
\end{eqnarray*}
(Note that $n_P(i) - \beta_0 \overline \lambda_i >0$ by Lemma \ref{lemm: exist BR}.) By plugging this back into (\ref{wij_3}), we have 
\begin{eqnarray}
w_{ij} & = & \frac{\beta_{0}w_{ii}\lambda_{ij}1\{j\in \overline N_{P}(i)\}}{n_{P}(i)}.\label{iter wij-1-1}
\end{eqnarray}
Thus relevant $j$'s that appear in the best responses are only those
$j$'s such that $j\in N_{P}(i)$. Taking $w_{ii}^{[0]} = w_{ii}$ and $w_{ij}^{[0]} = w_{ij}$, we obain the desired result. $\blacksquare$ \medskip

\noindent \textbf{Proof of Theorem \ref{thm: best response2}:}  We prove the result by induction. We begin
by showing the result holds for $m=1$. Suppose each agent is the first order sophisticated type ($m=1$), i.e., each $i\in N$ believes that each $k\neq i$ is a simple type ($m=0$) and chooses strategies according to: 
\[
s_{k}^{i}(\mathcal{I}_{i,0})=\sum_{j\in\overline{N}_{P}(k)}\tau_{j}w_{kj}^{[0]}+\eta_{k}.
\]
The best responses of the first order sophisticated types are linear because the payoff
is quadratic in the player's own actions, and they believe simple
types play according to linear strategies. Hence the best response of the first order sophisticated type takes the form 
\begin{align*}
s_{i}^{[1]}(\mathcal{I}_{i,1})= & \tau_{i}+\frac{\beta_{0}}{n_{P}(i)}\sum_{k\in N_{P}(i)}\left(\sum_{j\in\overline{N}_{P}(k)}\tau_{j}w_{kj}^{[0]}\right)+\eta_{i}\\
= & \left(1+\frac{\beta_{0}}{n_{P}(i)}\sum_{k\in N_{P}(i)}w_{ki}^{[0]}\right)\tau_{i}+\frac{\beta_{0}}{n_{P}(i)}\sum_{j\in N_{I}(i)}\sum_{k\in N_{P}(i)}\tau_{j}w_{kj}^{[0]}1\{j\in\overline{N}_{P}(k)\}+\eta_{i}.
\end{align*}
By setting the coefficient of $\tau_{j}$ to $w_{ij}^{[1]}$, we obtain
the weights
\begin{align}
w_{ii}^{[1]} & =1+\frac{\beta_{0}}{n_{P}(i)}\sum_{k\in N_{P}(i)}w_{ki}^{[0]},\label{eq:wiisup1}
\end{align}
and for each $j\in N_{I}(i)$
\begin{align}
& w_{ij}^{[1]}=\frac{\beta_{0}}{n_{P}(i)}\sum_{k\in N_{P}(i)}w_{kj}^{[0]}1\{j\in\overline N_{P}(k)\}\label{eq:wij_sup1}\\
& =\frac{\beta_{0}}{n_{P}(i)}\sum_{k\in N_{P}(i)}w_{kj}^{[0]}1\{j\in N_{P}(k)\}+\beta_{0}\frac{w_{jj}^{[0]}1\{j\in N_{P}(i)\}}{n_{P}(i)}.\nonumber 
\end{align}
Therefore, the weights for $j\in N_{I}(i)$ involve agents up to $N_{P,2}(i)$.
Since 
\begin{align*}
s_{i}^{[1]} & =\tau_{i}w_{ii}^{[1]}+\sum_{j\in N_{P,2}(i)}w_{ij}^{[1]}\tau_{j}+\eta_{i},
\end{align*}
with the weights for $w_{ii}^{[1]}$ and $w_{ij}^{[1]}$ given in
(\ref{eq:wiisup1}) and (\ref{eq:wij_sup1}), we have shown that the result
holds for $m=1$. Now suppose that for some $m$, the best responses
are given by
\begin{eqnarray*}
	s_{i}^{[m]}(\mathcal{I}_{i,m})=w_{ii}^{[m]}\tau_{i}+\sum_{j\in N_{P,m+1}(i)}w_{ij}^{[m]}\tau_{j}+\eta_{i},
\end{eqnarray*}
where $w_{ii}^{[m]}$ and $w_{ij}^{[m]}$ are as defined in the statement
of the result. From the optimization problem of $m+1$ types, we can
write the best response for these types as 

\noindent 
\begin{align*}
s_{i}^{[m+1]}(\mathcal{I}_{i,1})= & \tau_{i}+\beta_{0}\left(\frac{1}{n_{P}(i)}\sum_{k\in N_{P}(i)}\sum_{j\in\overline{N}_{P,m+1}(k)}\tau_{j}w_{kj}^{[m]}\right)+\eta_{i}\\
= & \left(1+\frac{\beta_{0}}{n_{P}(i)}\sum_{k\in N_{P}(i)}w_{ki}^{[m]}\right)\tau_{i}+\frac{\beta_{0}}{n_{P}(i)}\sum_{j\in N_{I}(i)}\sum_{k\in N_{P}(i)}\tau_{j}w_{kj}^{[m]}1\{j\in\overline N_{P,m+1}(k)\}+\eta_{i}.
\end{align*}
Note that this expression involves only those $j\in N_{I}(i)$ up
to $N_{P,m+2}(i)$. Hence we conclude that
\begin{eqnarray*}
	s_{i}^{[m+1]}(\mathcal{I}_{i,m})=w_{ii}^{[m+1]}\tau_{i}+\sum_{j\in N_{P,m+2}(i)}w_{ij}^{[m+1]}\tau_{j}+\eta_{i},
\end{eqnarray*}

\noindent with 
\begin{align*}
w_{ii}^{[m+1]} & =1+\frac{\beta_{0}}{n_{P}(i)}\sum_{k\in N_{P}(i)}w_{ki}^{[m]}
\end{align*}
and

\begin{align*}
w_{ij}^{[m+1]} & =\frac{\beta_{0}}{n_{P}(i)}\sum_{k\in N_{P}(i)}w_{kj}^{[m]}1\{j\in\overline{N}_{P,m+1}(k)\}
\end{align*}
as required. Therefore, we have shown that the result holds for all
$m\geq1$ by mathematical induction. $\blacksquare$

\subsection{The Best Response of a Bayesian Decision Maker}
In this section, we elaborate on the points made in Section \ref{subsubsec:Bayesian DM}, and prove the claim that under the prior $Q_i$ satisfying Conditions (a)-(c), the best response of the Bayesian decision maker coincides with quadratic utility with that of a simple type player with belief projection.

To see this, consider the following problem of Bayesian decision maker $i$:
\begin{eqnarray*}
	\sup_{y_i \in \mathcal{Y}_i} \mathbf{E}_{Q_i}\left[u_i(y_i,s_{-i}(\mathcal{I}_{i,0};W_{-i}),\tau,\eta_i)|W_i = w_i, \mathcal{I}_{i,0} \right].
\end{eqnarray*}
The solution $y_i$ takes the following form:
\begin{eqnarray}
\label{best response 3}
	s_i(\mathcal{I}_{i,0};w_i) = \tau_i + \frac{\beta_0}{n_P(i)}\sum_{k \in N_P(i)} \mathbf{E}_{Q_i}\left[ \sum_{j \in \overline N_I(k)} W_{kj}' \tau_j|W_i = w_i, \mathcal{I}_{i,0} \right].
\end{eqnarray}
By condition (c), we write
\begin{eqnarray*}
	&& \frac{1}{n_P(i)}\sum_{k \in N_P(i)} \mathbf{E}_{Q_i}\left[ \sum_{j \in \overline N_I(k)} W_{kj}' \tau_j|W_i = w_i, \mathcal{I}_{i,0} \right] \\
	&=& \frac{1}{n_P(i)}\sum_{k \in N_P(i)} \sum_{j \in \overline N_P(k)} \mathbf{E}_{Q_i}\left[  W_{kj}'|W_i = w_i, \mathcal{I}_{i,0} \right]\tau_j.
\end{eqnarray*}
We plug this back into (\ref{best response 3}). Using restrictions (a) and (b), we can rewrite
\begin{eqnarray}
\label{s0}
	s_i(\mathcal{I}_{i,0};w_i) = w_i' h(\tau_i) + \eta_i,
\end{eqnarray}
for some function $h$ of $\tau_i$. We equate this to 
\begin{eqnarray*}
	s_i(\mathcal{I}_{i,0};w_i) = w_i' \tau_i + \eta_i,
\end{eqnarray*}
as $s_i(\mathcal{I}_{i,0};w_i)$ is a linear strategy, and find out the weight vector $w_i$. However, this is precisely how the best response in Theorem \ref{thm: best response} was derived. Indeed, By restrictions (a) and (b), the right hand side of (\ref{si0}) coincides with that of (\ref{s0}).

\subsection{The Coincidence of the Equilibrium Strategies and the Behavioral Strategies in the Case of  Complete Payoff Subgraphs}
Here we show that the equilibrium strategies and the behavioral strategies coincide when the payoff graphs $G_P$ have disjoint multiple subgraphs and each subgraph is a complete graph as in Section \ref{subsubsec: complete payoff subgraphs}. For this, it suffices to derive the Bayesian Nash Equilibrium (BNE) as in (\ref{BNE2}). Suppose that $\mathbf{y}(i)$ is the $n_P(i)+1$ dimensional column vector of actions that are realized from the BNE of the game. Let $\boldsymbol{\eta}(i)$ and $\boldsymbol{\tau}(i)$ be an $n_P(i)+1$ dimensional column vector whose entries are  $\eta_j$ and $\tau_j$ respectively for $j \in \overline N_P(i)$. Then, from the first order condition of the expected payoff, we find that
\begin{eqnarray}
\label{eq42}
	\mathbf{y}(i) = \mathbf{y}^*(i) + \boldsymbol{\eta}(i),
\end{eqnarray}
where $\mathbf{y}^*(i)$ is a vector that satisfies:
\begin{eqnarray*}
	\mathbf{y}^*(i) = \beta_0 A(i) \mathbf{y}^*(i) + \boldsymbol{\tau}(i),
\end{eqnarray*}
with
\begin{eqnarray*}
	A(i) = \frac{1}{n_P(i)} \left(\mathbf{1}(i)\mathbf{1}(i)' - I_{n_P(i)+1} \right).
\end{eqnarray*}
Hence
\begin{eqnarray}
\label{eq32}
	\mathbf{y}^*(i) = (I_{n_P(i)+1} - \beta_0 A(i))^{-1} \boldsymbol{\tau}(i),
\end{eqnarray}
where $ \mathbf{1}(i)$ is the $n_P(i)+1$ dimensional column vector of ones and $I_{n_P(i)+1}$ is the $n_P(i)+1$ dimensional identity matrix. Using the Woodbury formula (see \citesupp{Hager:89:SIAM}), we obtain that
\begin{eqnarray*}
	(I_{n_P(i)+1} - \beta_0 A(i))^{-1} &=& \left( 1 + \beta_0/n_P(i) \right)^{-1} I_{n_P(i)+1}\\
	&+& \frac{1}{(1 + \beta_0/n_P(i))^2 - \beta_0 (1 + \beta_0/n_P(i) )(n_P(i)+1)/n_P(i)} \cdot \frac{\beta_0 \mathbf{1}(i) \mathbf{1}(i)'}{n_P(i)}.
\end{eqnarray*}
Hence plugging this back to (\ref{eq32}), we obtain that the entry $y_i^*$ of $\mathbf{y}^*(i)$ that corresponds to agent $i$ takes the following form:
\begin{eqnarray*}
	y_i^* &=& \left( 1 + \frac{\beta_0}{n_P(i)} \right)^{-1} \tau_i \\
	&+& \left( 1 + \frac{\beta_0}{n_P(i)} \right)^{-1} \frac{\beta_0 n_P(i)}{n_P(i) (1 - \beta_0)} \cdot \overline \tau(i) \cdot \frac{n_P(i)+1}{n_P(i)},
\end{eqnarray*}
where
\begin{eqnarray*}
	\overline \tau(i) = \frac{\mathbf{1}(i)' \boldsymbol{\tau}(i)}{n_P(i) + 1}.
\end{eqnarray*}
By rearranging the terms, we obtain that
\begin{eqnarray*}
	y_i^* &=& \left( 1 + \frac{\beta_0^2}{(n_P(i) + \beta_0)(1 - \beta_0)}\right) \tau_i \\
	&& + \frac{\beta_0}{(n_P(i) + \beta_0)(1 - \beta_0)} \sum_{j \in N_P(i)} \tau_j.
\end{eqnarray*}
Thus (\ref{eq42}) coincides with (\ref{BNE2}).

\section{Estimation of the Asymptotic Covariance Matrix}
We now explain our proposal to estimate the asymptotic covariance matrix, given in equation (\ref{hat Omega and hat V}) for the model with agents of simple type.

We first explain our proposal to estimate $\Lambda$ consistently for the case of $\beta_0 \ne 0$. Then, we later show how the estimator works even for the case of $\beta_0 = 0$. We first write
\begin{eqnarray}
\label{decomp v}
v_i = R_i(\varepsilon) + \eta_i,
\end{eqnarray}
where
\begin{eqnarray*}
	R_i(\varepsilon) = w_{ii}^{[0]} \varepsilon_i +  \frac{\beta_0 w_{ii}^{[0]}}{n_P(i)} \sum_{j \in N_P(i)} \lambda_{ij} \varepsilon_j.
\end{eqnarray*}
Define for $i,j \in N$,
\begin{eqnarray}
\label{eij}
	e_{ij} &=& \mathbf{E}\left[R_i(\varepsilon) R_j(\varepsilon) |\mathcal{F}\right]/\sigma_{\varepsilon}^2,
\end{eqnarray}
where $\sigma_{\varepsilon}^2 = \text{Var}(\varepsilon_i^2|\mathcal{F})$  denotes the variance of $\varepsilon_i$. It is not hard to see that for all $i \in N$,
\begin{eqnarray}
\label{eii}
	e_{ii} = (w_{ii}^{[0]})^2 +  \frac{\beta_0^2 (w_{ii}^{[0]})^2}{n_P^2(i)}\sum_{j \in N_P(i)} \lambda_{ij}^2,
\end{eqnarray}
and for $i \ne j$ such that $N_P(i) \cap N_P(j) \ne \varnothing$, $e_{ij} = \beta_0 q_{\varepsilon,ij}$, where
\begin{eqnarray*}
	q_{\varepsilon,ij} =  w_{ii}^{[0]}w_{jj}^{[0]}\left(\frac{\lambda_{ji} 1\{i \in N_P(j)\}}{n_P(j)} 
	+ \frac{\lambda_{ij} 1\{j \in N_P(i)\}}{n_P(i)}
	+ \frac{\beta_0}{n_P(i)n_P(j)} \sum_{k \in N_P(i) \cap N_P(j)} \lambda_{ik} \lambda_{jk}\right).
\end{eqnarray*}
Thus, we write
\begin{eqnarray}
\label{2 eqs}
\frac{1}{n^*} \sum_{i \in N^*} \mathbf{E}[v_i^2|\mathcal{F}] &=& a_\varepsilon \sigma_\varepsilon^2 + \sigma_\eta^2, \text{ and } \\ \notag
\frac{1}{n^*} \sum_{i \in N^*} \sum_{j \in N_P(i)\cap N^*} \mathbf{E}[v_i v_j|\mathcal{F}] &=& \beta_0 b_\varepsilon \sigma_\varepsilon^2,
\end{eqnarray}
where $\sigma_\eta^2$ denotes the variance of $\eta_i$,
\begin{eqnarray*}
	a_\varepsilon = \frac{1}{n^*}\sum_{i \in N^*} e_{ii}, \text{ and }
	b_\varepsilon = \frac{1}{n^*}\sum_{i \in N^*} \sum_{j \in N_P(i)\cap N^*} q_{\varepsilon,ij}.
\end{eqnarray*}
(Note that since not all agents in $N_P(i)$ are in $N^*$ for all $i \in N^*$, the set $N_P(i)\cap N^*$ does not necessarily coincide with $N_P(i)$.) When $\beta_0 \ne 0$, the solution takes the following form:
\begin{eqnarray}
\label{sigmas}
\sigma_\varepsilon^2 &=& \frac{1}{n^* \beta_0 b_\varepsilon} \sum_{i \in N^*} \sum_{j \in N_P(i)\cap N^*} \mathbf{E}[v_i v_j|\mathcal{F}] \text{ and } \\ \notag
\sigma_\eta^2 &=& \frac{1}{n^*} \sum_{i \in N^*} \mathbf{E}[v_i^2|\mathcal{F}] - \frac{a_\varepsilon}{n^* \beta_0 b_\varepsilon} \sum_{i \in N^*} \sum_{j \in N_P(i)\cap N^*} \mathbf{E}[v_i v_j|\mathcal{F}].
\end{eqnarray}
In other words, when $\beta_0 \ne 0$, i.e., when there is strategic interaction among the players, we can ``identify" $\sigma_\varepsilon^2$ and $\sigma_\eta^2$ by using the variances and covariances of residuals $v_i$'s. The intuition is as follows. Since the source of cross-sectional dependence of $v_i$'s is due to the presence of $\varepsilon_i$'s, we can identify first $\sigma_{\varepsilon}^2$ using covariance between $v_i$ and $v_j$ for linked pairs $i,j$, and then identify $\sigma_\eta^2$ by subtracting from the variance of $v_i$ the contribution from $\varepsilon_i$. 

In order to obtain a consistent estimator of $\Lambda$ which does not require that $\beta_0 \ne 0$, we derive its alternative expression. Let us first write
\begin{eqnarray}
\label{lambda 2}
\Lambda = \Lambda_1 + \Lambda_2,
\end{eqnarray}
where
\begin{eqnarray*}
	\Lambda_1 &=& \frac{1}{n^*} \sum_{i \in N^*} \mathbf{E}[v_i^2|\mathcal{F}] \tilde \varphi_i \tilde \varphi_i', \text{ and }\\
	\Lambda_2 &=& \frac{1}{n^*} \sum_{i \in N^*} \sum_{j \in N^*_{-i}} \mathbf{E}[v_i v_j|\mathcal{F}] \tilde \varphi_i \tilde \varphi_j',
\end{eqnarray*}
where $N^*_{-i} = N^*\backslash\{i\}$. Using (\ref{decomp v}) and (\ref{sigmas}), we can rewrite
\begin{eqnarray*}
	\Lambda_2 &=& \frac{1}{n^*} \sum_{i \in N^*} \sum_{j \in N^*_{-i}: N_P(i) \cap N_P(j) \ne \varnothing} e_{ij} \sigma_\varepsilon^2 \tilde \varphi_i \tilde \varphi_j'\\
	&=& \frac{\beta_0}{n^*} \sum_{i \in N^*} \sum_{j \in N^*_{-i}: N_P(i) \cap N_P(j) \ne \varnothing} q_{\varepsilon,ij} \sigma_\varepsilon^2 \tilde \varphi_i \tilde \varphi_j'\\
	&=& \frac{s_\varepsilon}{n^*} \sum_{i \in N^*} \sum_{j \in N^*_{-i}: N_P(i) \cap N_P(j) \ne \varnothing} q_{\varepsilon,ij} \tilde \varphi_i \tilde \varphi_j',
\end{eqnarray*} 
where 
\begin{eqnarray*}
	s_\varepsilon = \frac{\displaystyle \sum_{i \in N^*} \sum_{j \in N_P(i)\cap N^*} \mathbf{E}[v_i v_j|\mathcal{F}]}{\displaystyle  \sum_{i \in N^*} \sum_{j \in N_P(i)\cap N^*} q_{\varepsilon,ij}}.
\end{eqnarray*}

Now, it is clear that with this expression for $\Lambda_2$, the definition of $\Lambda$ is well defined regardless of whether $\beta_0=0$ or $\beta_0 \ne 0$.  We can then find the estimator of $\Lambda$, $\hat{\Lambda}$, by using the empirical analogues to the above, as shown in the main text.  %Plugging $\hat{\Lambda}$ into (\ref{hat Omega and hat V}), yields the estimator for our asymptotic covariance matrix.

\section{Local Identification and Estimation}
\subsection{Preliminary Results for Local Identification}
In this section, we give computations of derivatives that are used for the proof of local identification. Define
\begin{equation}
v_{i}(\beta,\rho)=Y_{i}-q_{i}(\beta)\left(X_{i}+\frac{\beta}{n_{P}(i)}\sum_{j\in N_{P}(i)}\lambda_{ij}(\beta)X_{j}\right)'\rho,\label{eq:v-1}
\end{equation}
where 
\begin{align*}
q_{i}(\beta) &=\frac{n_{P}(i)}{n_{P}(i)-\beta^{2}\overline \lambda_{i}}\\
\lambda_{ij}(\beta) & =\frac{1}{1-\beta c_{ij}},\text{ and }\\
\overline \lambda_{i}(\beta) & =\frac{1}{n_{P}(i)}\sum_{j\in N_{P}(i)}\lambda_{ij}(\beta).
\end{align*}
The relevant derivatives of (\ref{eq:v-1}) are given by
\begin{eqnarray}
\label{eq:first_deriv}
\frac{\partial v_{i}(\beta,\rho)}{\partial\beta} 
&=& -q_{i}(\beta)\frac{1}{n_{P}(i)}\sum_{j\in N_{P}(i)}\lambda_{ij}^{2}(\beta)X_{j}'\rho\\ \notag
&& -q_{i}'(\beta)\left(X_{i}'+\frac{\beta}{n_{P}(i)}\sum_{j\in N_{P}(i)}\lambda_{ij}(\beta)X_{j}'\right)\rho.
\end{eqnarray}
\begin{eqnarray}
	\frac{\partial^{2}v_{i}(\beta,\rho)}{\partial\beta^{2}}  &=& -\frac{2q_{i}(\beta)}{n_{P}(i)}\sum_{j\in N_{P}(i)}\lambda_{ij}^{3}(\beta)c_{ij}X_{j}'\rho-\frac{2 q_{i}'(\beta)}{n_{P}(i)}\sum_{j\in N_{P}(i)}\lambda_{ij}^{2}(\beta)X_{j}'\rho\label{eq:second_deriv}\\
	&& -q_{i}''(\beta)\left(X_{i}'+\frac{\beta}{n_{P}(i)}\sum_{j\in N_{P}(i)}\lambda_{ij}(\beta)X_{j}'\right)\rho,\nonumber
\end{eqnarray}
\begin{eqnarray}
\label{eq:third_deriv}
	\frac{\partial^{3}v_{i}(\beta,\rho)}{\partial\beta^{3}} &=&-\frac{6 q_{i}(\beta)}{n_{P}(i)}\sum_{j\in N_{P}(i)}\lambda_{ij}^{4}(\beta)c_{ij}^{2}X_{j}'\rho\\ \notag
	&& -\frac{6q_{i}'(\beta)}{n_{P}(i)}\sum_{j\in N_{P}(i)}\lambda_{ij}^{3}(\beta)c_{ij}X_{j}'\rho\\ \notag
	&& -\frac{3q_{i}''(\beta)}{n_{P}(i)}\sum_{j\in N_{P}(i)}\lambda_{ij}^{2}(\beta)X_{j}'\rho\\ \notag
	&& -q_{i}'''(\beta)\left(X_{i}'+\frac{\beta}{n_{P}(i)}\sum_{j\in N_{P}(i)}\lambda_{ij}(\beta)X_{j}'\right)\rho,\nonumber
\end{eqnarray}
and
\begin{eqnarray}
	\frac{\partial^{3}v_{i}(\beta,\rho)}{\partial\beta^{2}\partial\rho} 
	=\left(\frac{\partial^{3}v_{i}(\beta,\rho)}{\partial\beta^{2}\partial\rho_{1}},...,\frac{\partial^{3}v_{i}(\beta,\rho)}{\partial\beta^{2}\partial\rho_{d}}\right)',\label{eq:cross_deriv}
\end{eqnarray}
where for each $s=1,...,d$, we have 
\begin{align*}
\frac{\partial^{3}v_{i}(\beta,\rho)}{\partial\beta^{2}\partial\rho_{s}} & =-\frac{2q_{i}(\beta)}{n_{P}(i)}\sum_{j\in N_{P}(i)}\lambda_{ij}^{3}(\beta)c_{ij}X_{j,s}-\frac{2q_{i}'(\beta)}{n_{P}(i)}\sum_{j\in N_{P}(i)}\lambda_{ij}^{2}(\beta)X_{j,s}\\
& -q_{i}''(\beta)\left(X_{i,s}+\frac{\beta}{n_{P}(i)}\sum_{j\in N_{P}(i)}\lambda_{ij}(\beta)X_{j,s}\right).
\end{align*}
We will focus on showing (\ref{eq:first_deriv}), (\ref{eq:second_deriv}),
and (\ref{eq:third_deriv}), since (\ref{eq:cross_deriv}) follows immediately from (\ref{eq:second_deriv}). We provide expressions
for $q_{i}'(\beta)$, $q_{i}''(\beta)$, and $q_{i}'''(\beta)$ in Section \ref{subsec:derivatives_of_qi} later.

For the derivations, we will repeatedly make use of the following
result.
\begin{lemma} For any integer $m\geq1$,
\begin{eqnarray}
	\label{eq:deriv_lamb}
	\frac{\partial\lambda_{ij}^{m}}{\partial\beta}=m\lambda_{ij}^{m+1}c_{ij}.
\end{eqnarray}
\end{lemma}

\noindent \textbf{Proof:} First observe that 
	\begin{eqnarray*}
		\frac{\partial\lambda_{ij}}{\partial\beta}  =  \frac{\partial}{\partial\beta}(1-\beta c_{ij})^{-1}
		& = & (-1)(1-\beta c_{ij})^{-2}(-c_{ij})\\
		& = & \frac{c_{ij}}{(1-\beta c_{ij})^{2}}
		 =  \lambda_{ij}^{2}c_{ij},
	\end{eqnarray*}
	Then we note that equation (\ref{eq:deriv_lamb}) with $m$ implies
	that equation (\ref{eq:deriv_lamb}) also holds for $m+1$: 
	\begin{align*}
	\frac{\partial\lambda_{ij}^{m+1}}{\partial\beta} & =\frac{\partial\lambda_{ij}\lambda_{ij}^{m}}{\partial\beta}\\
	& =\lambda_{ij}\frac{\partial\lambda_{ij}^{m}}{\partial\beta}+\lambda_{ij}^{m}\frac{\partial\lambda_{ij}}{\partial\beta}
	 =\lambda_{ij}\left(m\lambda_{ij}^{m+1}c_{ij}\right)+\lambda_{ij}^{m}\lambda_{ij}^{2}c_{ij}
	 =(m+1)\lambda_{ij}^{m+2}c_{ij}.
	\end{align*}
	
	It follows from the claim above that for each integer $m\geq1$:
	\begin{equation}
	\frac{\partial\bar{\lambda}_{ij}^{m}}{\partial\beta}=\frac{1}{n_P(i)} \sum_{j \in N_P(i)} m\lambda_{ij}^{m+1}c_{ij}.\label{eq:deriv_lambprime}
	\end{equation}
	$\blacksquare$

\subsubsection{First Derivative, $\partial v_{i}(\beta,\rho)/\partial\beta$}

Recall that 
\[
v_{i}(\beta,\rho)=Y_{i}-q_{i}(\beta)\left(X_{i}'+\frac{\beta}{n_{P}(i)}\sum_{j\in N_{P}(i)}\lambda_{ij}(\beta)X_{j}'\right)\rho.
\]
Therefore, by the chain rule:
\begin{align}
\label{eq:deriv_1_proof}
\frac{\partial v_{i}(\beta,\rho)}{\partial\beta} &=-\frac{\partial}{\partial\beta}\left(q_{i}(\beta)\left(X_{i}'+\frac{\beta}{n_{P}(i)}\sum_{j\in N_{P}(i)}\lambda_{ij}(\beta)X_{j}'\right)\rho\right) \\ \notag &=-q_{i}(\beta)\left[\frac{\partial}{\partial\beta}\left(X_{i}'+\frac{\beta}{n_{P}(i)}\sum_{j\in N_{P}(i)}\lambda_{ij}(\beta)X_{j}'\right)\right]\rho\\ \notag
& -\left(\frac{\partial}{\partial\beta}q_{i}(\beta)\right)\left(X_{i}'+\frac{\beta}{n_{P}(i)}\sum_{j\in N_{P}(i)}\lambda_{ij}(\beta)X_{j}'\right)\rho,\nonumber 
\end{align}
where
\begin{align}
\label{eq:term_deriv_2}
\frac{\partial}{\partial\beta}\left(X_{i}'+\frac{\beta}{n_{P}(i)}\sum_{j\in N_{P}(i)}\lambda_{ij}(\beta)X_{j}'\right)  \rho
&=\beta\left(\frac{1}{n_{P}(i)}\sum_{j\in N_{P}(i)}\lambda_{ij}^{2}(\beta)c_{ij}X_{j}'\right)\rho\\ \notag
&+\frac{1}{n_{P}(i)}\sum_{j\in N_{P}(i)}\lambda_{ij}(\beta)X_{j}'\rho\\ \notag
& =\left(\sum_{j\in N_{P}(i)}\frac{1}{n_{P}(i)}\lambda_{ij}(\beta)\left(\beta\lambda_{ij}(\beta)c_{ij}+1\right)X_{j}'\right)\rho. \notag
\end{align}
The first equality follows from chain rule and using (\ref{eq:deriv_lambprime}),
with $m=1$. Since we have $(1-\beta c_{ij})\lambda_{ij}(\beta)=1$, we find that
\begin{eqnarray*}
	\frac{\partial}{\partial\beta}\left(X_{i}'+\frac{\beta}{n_{P}(i)}\sum_{j\in N_{P}(i)}\lambda_{ij}(\beta)X_{j}'\right)  \rho
	= \left(\frac{1}{n_{P}(i)}\sum_{j\in N_{P}(i)}\lambda_{ij}^{2}(\beta)X_{j}'\right)\rho.
\end{eqnarray*}
Therefore, plugging (\ref{eq:term_deriv_2}) into (\ref{eq:deriv_1_proof})
yields the desired expression:
\begin{align*}
\frac{\partial v_{i}(\beta,\rho)}{\partial\beta}= & -q_{i}(\beta)\frac{1}{n_{P}(i)}\sum_{j\in N_{P}(i)}\lambda_{ij}^{2}(\beta)X_{j}'\rho-q_{i}'(\beta)\left(X_{i}'+\frac{\beta}{n_{P}(i)}\sum_{j\in N_{P}(i)}\lambda_{ij}(\beta)X_{j}'\right)\rho.
\end{align*}

\subsubsection{Second Derivative, $\partial^{2}v_{i}(\beta,\rho)/\partial\beta^{2}$.}
Note that
\begin{eqnarray}
\label{q2}
\frac{\partial^{2}v_{i}(\beta,\rho)}{\partial\beta^{2}}
&=& -q_{i}(\beta)\frac{2}{n_{P}(i)}\sum_{j\in N_{P}(i)}\lambda_{ij}^{3}(\beta)c_{ij}X_{j}'\rho-q_{i}'(\beta)\frac{1}{n_{P}(i)}\sum_{j\in N_{P}(i)}\lambda_{ij}^{2}(\beta)X_{j}'\rho \nonumber\\
&& -q_{i}'(\beta)\frac{1}{n_{P}(i)}\sum_{j\in N_{P}(i)}\lambda_{ij}^{2}(\beta)X_{j}'\rho-q_{i}''(\beta)\left(X_{i}+\frac{\beta}{n_{P}(i)}\sum_{j\in N_{P}(i)}\lambda_{ij}(\beta)X_{j}'\right)\rho.
\end{eqnarray}
By rearranging the terms, we can rewrite
\begin{eqnarray}
\label{q21}
\frac{\partial^{2}v_{i}(\beta,\rho)}{\partial\beta^{2}}
&=& -q_{i}(\beta)\frac{2}{n_{P}(i)}\sum_{j\in N_{P}(i)}\lambda_{ij}^{3}(\beta)c_{ij}X_{j}'\rho-q_{i}'(\beta)\frac{2}{n_{P}(i)}\sum_{j\in N_{P}(i)}\lambda_{ij}^{2}(\beta)X_{j}'\rho  \nonumber\\
&& -q_{i}''(\beta)\left(X_{i}'+\frac{\beta}{n_{P}(i)}\sum_{j\in N_{P}(i)}\lambda_{ij}(\beta)X_{j}'\right)\rho.  \notag
\end{eqnarray}

\subsubsection{Third Derivative, $\partial^{3}v_{i}(\beta,\rho)/\partial\beta^{3}$.}

We take derivatives of each term in (\ref{q2}) in turn. Observe that 
\begin{align*}
-\frac{\partial}{\partial\beta}\left(q_{i}(\beta)\frac{2}{n_{P}(i)}\sum_{j\in N_{P}(i)}\lambda_{ij}^{3}(\beta)c_{ij}X_{j}'\rho\right) & =-q_{i}(\beta)\frac{6}{n_{P}(i)}\sum_{j\in N_{P}(i)}\lambda_{ij}^{4}(\beta)c_{ij}^{2}X_{j}'\rho\\
& -q_{i}'(\beta)\frac{2}{n_{P}(i)}\sum_{j\in N_{P}(i)}\lambda_{ij}^{3}(\beta)c_{ij}X_{j}'\rho\\
-\frac{\partial}{\partial\beta}\left(2q_{i}'(\beta)\frac{1}{n_{P}(i)}\sum_{j\in N_{P}(i)}\lambda_{ij}^{2}(\beta)X_{j}'\rho\right) & =-q_{i}'(\beta)\frac{4}{n_{P}(i)}\sum_{j\in N_{P}(i)}\lambda_{ij}^{3}(\beta)c_{ij}X_{j}'\rho\\
& -q_{i}''(\beta)\frac{2}{n_{P}(i)}\sum_{j\in N_{P}(i)}\lambda_{ij}^{2}(\beta)X_{j}'\rho.
\end{align*}
Also, note that
\begin{align*}
-\frac{\partial}{\partial\beta}\left(q_{i}''(\beta)\left(X_{i}'+\frac{\beta}{n_{P}(i)}\sum_{j\in N_{P}(i)}\lambda_{ij}(\beta)X_{j}'\right)\rho\right) & =-q_{i}''(\beta)\frac{1}{n_{P}(i)}\sum_{j\in N_{P}(i)}\lambda_{ij}^{2}(\beta)X_{j}'\rho\\
& -q_{i}'''(\beta)\left(X_{i}'+\frac{\beta}{n_{P}(i)}\sum_{j\in N_{P}(i)}\lambda_{ij}(\beta)X_{j}'\right)\rho.
\end{align*}
Therefore, the sum of the three preceding terms yields the required
derivative
\begin{eqnarray*}
 \frac{\partial^{3}v_{i}(\beta,\rho)}{\partial\beta^{3}}
&=&-q_{i}(\beta)\frac{6}{n_{P}(i)}\sum_{j\in N_{P}(i)}\lambda_{ij}^{4}(\beta)c_{ij}^{2}X_{j}'\rho-q_{i}'(\beta)\frac{2}{n_{P}(i)}\sum_{j\in N_{P}(i)}\lambda_{ij}^{3}(\beta)c_{ij}X_{j}'\rho\\
&& -q_{i}'(\beta)\frac{4}{n_{P}(i)}\sum_{j\in N_{P}(i)}\lambda_{ij}^{3}(\beta)c_{ij}X_{j}'\rho\\
&& -q_{i}''(\beta)\frac{2}{n_{P}(i)}\sum_{j\in N_{P}(i)}\lambda_{ij}^{2}(\beta)X_{j}'\rho-q_{i}''(\beta)\frac{1}{n_{P}(i)}\sum_{j\in N_{P}(i)}\lambda_{ij}^{2}(\beta)X_{j}'\rho\\
&& -q_{i}'''(\beta)\left(X_{i}'+\frac{\beta}{n_{P}(i)}\sum_{j\in N_{P}(i)}\lambda_{ij}(\beta)X_{j}'\right)\rho\\
&=& -q_{i}(\beta)\frac{6}{n_{P}(i)}\sum_{j\in N_{P}(i)}\lambda_{ij}^{4}(\beta)c_{ij}^{2}X_{j}'\rho-q_{i}'(\beta)\frac{6}{n_{P}(i)}\sum_{j\in N_{P}(i)}\lambda_{ij}^{3}(\beta)c_{ij}X_{j}'\rho\\
&& -q_{i}''(\beta)\frac{3}{n_{P}(i)}\sum_{j\in N_{P}(i)}\lambda_{ij}^{2}(\beta)X_{j}'\rho-q_{i}'''(\beta)\left(X_{i}'+\frac{\beta}{n_{P}(i)}\sum_{j\in N_{P}(i)}\lambda_{ij}(\beta)X_{j}'\right)\rho.
\end{eqnarray*}

\subsubsection{Expressions for derivatives of $q_{i}(\beta)$\label{subsec:derivatives_of_qi}}

Recall that 
\[
q_{i}(\beta)=\frac{n_{P}(i)}{n_{P}(i)-\beta^{2}\overline \lambda_{i}}.
\]
We provide expressions for the first three derivatives of $q_{i}(\beta)$. Namely, we show that:
\begin{align}
q_{i}'(\beta) & =\frac{\beta}{n_P(i)}q_{i}^{2}(\beta)(2\overline \lambda_i + \beta \overline \lambda_i'),\label{eq:q_first}\\ 
q_{i}''(\beta) & =\frac{\beta}{n_P(i)} q_i(\beta)^2(3 \overline \lambda_i'+ \beta \overline \lambda_i'') + \frac{q_i(\beta)}{n_P(i)}(2 \overline \lambda_i + \beta \overline \lambda_i')\left(q_i(\beta)+ 2\beta q_i'(\beta)\right),\text{ and}\label{eq:q_second}\\ 
\small
q_{i}'''(\beta) & =\frac{\beta}{n_P(i)} q_i(\beta)^2(4 \overline \lambda_i''+ \beta \overline \lambda_i''')+\frac{q_i(\beta)}{n_P(i)}(6 \overline \lambda_i + 2\beta \overline \lambda_i'')\left(q_i(\beta) +  2\beta q_i'(\beta) \right) \label{eq:q_third} \\ 
& \quad + \frac{2 (2 \overline \lambda_i + \beta \overline \lambda_i')}{n_P(i)}\left(q_i(\beta) q_i'(\beta) + \beta q_i(\beta) q_i''(\beta) + q_i'(\beta) (q_i(\beta) + \beta q_i'(\beta))\right) ,\nonumber 
\end{align}
where we have from (\ref{eq:deriv_lambprime}) that:
\begin{align}
\overline \lambda_{i}'(\beta) & =\frac{1}{n_{P}(i)}\sum_{j\in N_{P}(i)} c_{ij} \lambda_{ij}^{2},\text{ and}\label{eq:r_first}\\
\overline \lambda_{i}''(\beta) & = \frac{2}{n_{P}(i)}\sum_{j\in N_{P}(i)} c_{ij}^2 \lambda_{ij}^3.\label{eq:r_second} \\ \notag
\overline \lambda_{i}'''(\beta) & = \frac{6}{n_{P}(i)}\sum_{j\in N_{P}(i)} c_{ij}^3 \lambda_{ij}^4.\label{eq:r_third}
\end{align}

First, (\ref{eq:q_first}) holds from applying chain rule to $q_{i}(\beta)$
as follows
\begin{align*}
q_{i}'(\beta) & =(-1)\left[1-\beta^{2}\overline \lambda_{i}(\beta)/n_{P}(i)\right]^{-2}(-)\frac{1}{n_{P}(i)}\left[2\beta\overline \lambda_{i}(\beta)+\beta^{2}\overline \lambda_{i}'(\beta)\right]\\
& =\frac{\beta\left(\beta\overline \lambda_{i}'(\beta)+2\overline \lambda_{i}(\beta)\right)}{n_{P}(i)\left(1-\beta^{2}\overline \lambda_{i}/n_{P}(i)\right)^{2}}\\
& = \frac{\beta}{n_P(i)}q_{i}(\beta)^2(2\overline \lambda_i + \beta \overline \lambda_i'),
\end{align*}
which is our first equation. For the second derivative of $q_i(\beta)$, we find that:
\begin{eqnarray*}
	q_i''(\beta) &=&\frac{\beta}{n_P(i)} q_i(\beta)^2(2 \overline \lambda_i' + \overline \lambda_i' + \beta \overline \lambda_i'') + (2 \overline \lambda_i + \beta \overline \lambda_i')\left(\frac{q_i(\beta)^2}{n_P(i)} + \frac{2\beta}{n_P(i)} q_i'(\beta) q_i(\beta)\right)\\
	&=&\frac{\beta}{n_P(i)} q_i(\beta)^2(3 \overline \lambda_i'+ \beta \overline \lambda_i'') + \frac{(2 \overline \lambda_i + \beta \overline \lambda_i')}{n_P(i)}\left(q_i(\beta)^2+ 2\beta q_i(\beta) q_i'(\beta)\right).
\end{eqnarray*}

For the final equation, we take derivatives of the above to find:

\begin{eqnarray*}
	q_i'''(\beta)  &=&\frac{\beta}{n_P(i)} q_i(\beta)^2(3 \overline \lambda_i''+\overline \lambda_i'' + \beta \overline \lambda_i''') + (3 \overline \lambda_i' + \beta \overline \lambda_i'')\left(\frac{q_i(\beta)^2}{n_P(i)} + \frac{2\beta}{n_P(i)} q_i'(\beta) q_i(\beta)\right)\\
	&\quad &+ \frac{(2 \overline \lambda_i' + \overline \lambda_i' + \beta \overline \lambda_i'')}{n_P(i)} q_i(\beta)(q_i(\beta)+2\beta q_i'(\beta)) \\
	&\quad &+ \frac{(2 \overline \lambda_i + \beta \overline \lambda_i')}{n_P(i)}\left(2 q_i(\beta) q_i'(\beta) + 2\beta q_i(\beta) q_i''(\beta) + 2q_i'(\beta) (q_i(\beta) + \beta q_i'(\beta))\right) \\
	&=& \frac{\beta}{n_P(i)} q_i(\beta)^2(4 \overline \lambda_i''+ \beta \overline \lambda_i''')+\frac{q_i(\beta)}{n_P(i)}(6 \overline \lambda_i + 2\beta \overline \lambda_i'')\left(q_i(\beta) +  2\beta q_i'(\beta) \right) \\
	&\quad& + \frac{2 (2 \overline \lambda_i + \beta \overline \lambda_i')}{n_P(i)}\left(q_i(\beta) q_i'(\beta) + \beta q_i(\beta) q_i''(\beta) + q_i'(\beta) (q_i(\beta) + \beta q_i'(\beta))\right).
\end{eqnarray*}
\subsection{Proof of Theorem \ref{thm: cons estimability}}

Let us recall our notation first. Let $\theta = [\beta,\rho']'$ and $\theta_0 = [\beta_0,\rho_0']'$. As in Theorem \ref{thm: cons estimability}, we assume that $\varphi_i$ does not depend on $\theta$. Let $\lambda_{ij}(\beta) = 1/(1 - \beta c_{ij})$, and
\begin{eqnarray*}
	\overline \lambda_i(\beta) = \frac{1}{n_P(i)}\sum_{j \in N_P(i)} \lambda_{ij}(\beta).
\end{eqnarray*}
Let
\begin{eqnarray*}
	Z_i(\beta) = \left(1 + \frac{\beta^2 \overline \lambda_i(\beta)}{n_P(i) - \beta^2 \overline \lambda_i(\beta)} \right)\left( X_i + \frac{\beta}{n_P(i)}\sum_{j \in N_P(i)} \lambda_{ij}(\beta) X_j\right).
\end{eqnarray*}
Then, we define
\begin{eqnarray*}
	v_i(\theta) = Y_i - Z_i(\beta)' \rho.
\end{eqnarray*}
We let
\begin{eqnarray*}
	G_n(\theta) &\equiv& \frac{1}{n}\sum_{i=1}^n \mathbf{E}\left[v_i(\theta) \varphi_i|G_P\right] \text{ and }\\
	\hat G_n(\theta) &\equiv& \frac{1}{n}\sum_{i=1}^n v_i(\theta) \varphi_i. 
\end{eqnarray*}
We also let for $m=1,...,M$
\begin{eqnarray*}
	G_{n,m}(\theta) \equiv \frac{1}{n}\sum_{i=1}^n \mathbf{E}[v_i(\theta)\varphi_{i,m}|G_P],
\end{eqnarray*} 
where we recall that $\varphi_{i,m}$ is the $m$-th entry of $\varphi_i$. 

\begin{lemma}
	\label{lemm: der bound}
	Suppose that Assumption \ref{assump: ident}(ii) holds and that there exists $\varepsilon>0$ such that $\Theta = \overline B(\theta_0;\varepsilon)$. Then, there exists a constant $C>0$ such that for each $m=1,...,M$ and for all $n \ge 2$,
	\begin{eqnarray*}
		\frac{1}{n}\sum_{i \in N} \mathbf{E}\left[ \left. \sup_{\theta \in \Theta}  \left\| \frac{\partial v_i(\theta)}{\partial \theta}\right\|^2 \varphi_{i,m}^2 \right\vert G_P\right] &\le& C,
	\end{eqnarray*}
	and  for all $k_1,k_2,k_3 =1,...,d$,
	\begin{eqnarray*}
		\frac{1}{n}\sum_{i \in N} \mathbf{E}\left[ \left. \sup_{\theta \in \Theta}  \left|\frac{\partial^3 v_i(\theta) }{\partial \theta_{k_1}\partial \theta_{k_2}\partial \theta_{k_2}}\right| |\varphi_{i,m}| \right\vert G_P\right] \le C.
	\end{eqnarray*}
\end{lemma}	

\noindent \textbf{Proof: } By the assumption that $\beta \in [-1+ \nu, 1 - \nu]$ for some $\nu>0$, whenever $\beta$ is such that $(\beta,\rho) \in \Theta$, and that $\Theta$ is compact, and $0\le c_{ij} \le 1$, we have
\begin{eqnarray*}
	\frac{1}{2 - \nu} \le \lambda_{ij}(\beta) \le \frac{1}{\nu}.
\end{eqnarray*}
for all $\beta \in [-1+\nu, 1- \nu]$ such that $(\beta,\rho) \in \Theta$. Hence the results immediately follow from Assumption \ref{assump: ident}(ii) and the derivatives that we computed previously. $\blacksquare$\medskip 

\noindent \textbf{Proof of Theorem \ref{thm: cons estimability}: } Fix a small $\varepsilon>0$ and take $\Theta = \overline B(\theta_0,\varepsilon)$.  Note that $\|\hat G_n(\theta)\|$ is continuous in $\theta \in \Theta$ for every realization of the payoff graph $G_P$ and every realization of $(Y_i,X_{i,1},X_{i,2})_{i \in N}$. Since $\Theta$ is compact, the minimizer of $\|\hat G_n(\theta)\|$ over $\Theta$ exists in $\Theta$. Let us take
\begin{eqnarray*}
	\hat \theta \in \text{argmin}_{\theta \in \Theta} \|\hat G_n(\theta)\|.
\end{eqnarray*}
It suffices to show that $\hat \theta$ is consistent for $\theta_0$. For this we prove the following two claims:\medskip

\noindent \textbf{Claim 1: } There exists $\bar \delta>0$ such that for any $\delta \in (0,\bar \delta]$, there exists $\varepsilon_{\delta}>0$ such that for all $n \ge 1$, 
\begin{eqnarray*}
	\inf_{\theta \in \Theta: \|\theta - \theta_0\| > \delta} \left\| G_n(\theta) \right\| > \varepsilon_\delta.
\end{eqnarray*}

\noindent \textbf{Claim 2: } $\sup_{\theta \in \Theta} \left\| \hat G_n(\theta) - G_n(\theta) \right\| = o_P(1).$\medskip
 
Then, the consistency of $\hat \theta$ over $\Theta = \overline B(\theta_0,\varepsilon)$ follows as in the proof of Corollary 3.2 of \citesupp{Pakes/Pollard:89:Eca}. 

Let us first prove Claim 1. By $\beta \in [-1+ \nu, 1 - \nu]$ with $(\beta,\rho) \in \Theta$ and $0 \le c_{ij} \le 1$, $G_n(\theta)$ is infinite times differentiable over $\theta \in \Theta$. For each $m=1,...,M$ and $\theta,\theta_0 \in \Theta$, there exists a point $\theta_m^*$ on the line segment between $\theta$ and $\theta_0$ such that
\begin{eqnarray*}
	G_{n,m}^2(\theta) &=& \frac{\partial G_{n,m}^2(\theta_0)}{\partial \theta'}(\theta - \theta_0) 
	 + \frac{1}{2}(\theta - \theta_0)'\frac{\partial^2 G_{n,m}^2(\theta_m^*)}{\partial \theta \partial \theta'}(\theta - \theta_0)\\
	&=&\frac{1}{2}(\theta - \theta_0)'\frac{\partial^2 G_{n,m}^2(\theta_m^*)}{\partial \theta \partial \theta'}(\theta - \theta_0),
\end{eqnarray*}
because $G_{n,m}(\theta_0) = 0$. As for the last term, by Lemma \ref{lemm: der bound}, there exists a constant $C_1>0$ such that for all $n \ge 2$,
\begin{eqnarray*}
	&& \frac{1}{2}(\theta - \theta_0)'\frac{\partial^2 G_{n,m}^2(\theta^*)}{\partial \theta \partial \theta'}(\theta - \theta_0)\\
	&\ge& \frac{1}{2}(\theta - \theta_0)'\frac{\partial^2 G_{n,m}^2(\theta_0)}{\partial \theta \partial \theta'}(\theta - \theta_0)  - C_1 \|\theta - \theta_0\|^3.
\end{eqnarray*}
Note that (again, from $G_{n,m}(\theta_0) = 0$)
\begin{eqnarray*}
	\frac{\partial^2 G_{n,m}^2(\theta_0)}{\partial \theta \partial \theta'} =  2\left(\frac{1}{n}\sum_{i=1}^n H_{i,m}(\theta_0) \right) \left(\frac{1}{n}\sum_{i=1}^n H_{i,m}(\theta_0) \right)'.
\end{eqnarray*}
Hence
\begin{eqnarray*}
	\|G_n(\theta)\|^2 &=& G_n(\theta)' G_n(\theta) = \sum_{m=1}^M G_{n,m}^2(\theta) \\
	&\ge& \sum_{m=1}^M (\theta - \theta_0)'\left(\frac{1}{n}\sum_{i=1}^n H_{i,m}(\theta_0) \right) \left(\frac{1}{n}\sum_{i=1}^n H_{i,m}(\theta_0) \right)'(\theta - \theta_0) - C_1 M \|\theta - \theta_0\|^3.
\end{eqnarray*}
By Assumption \ref{assump: ident}(iii), with the constant $c>0$ there, we obtain that for all $n \ge 1$,
\begin{eqnarray*}
	\left\| G_n(\theta) \right\| \ge c \| \theta - \theta_0\|^2 - C_1\|\theta - \theta_0\|^3.
\end{eqnarray*}
We can find $C_2>0$ and $\bar \delta>0$ such that for all $0< \delta \le \bar \delta$, $c \delta^2 - C_1 \delta^3 > C_2 \delta^2$. Thus, we obtain Claim 1.

Let us turn to the proof of Claim 2. Let $G_P^*=(N,E_P^*)$ be a graph on $N$ such that $ij \in E_P^*$ if and only if $N_P(i) \cap N_P(j) \ne \varnothing$. Then the maximum degree of $G_P^*$ is bounded by $\max_{i \in N} n_P^2(i)$. Note that $v_i(\theta)\varphi_i$ is a function of $Y_i$ and $X_j$'s, and that $Y_i$'s have $G_P^*$ as a conditional dependency graph given $\mathcal{F}$.\footnote{Random variables $(W_i)_{i \in N}$ having $G_P^*$ as a conditional dependency graph given $\mathcal{F}$ means that for any set $A \subset N$,  $(W_i)_{i \in A}$ and $(W_i)_{i \in N \setminus \overline N_P(A)}$ are conditionally independent given $\mathcal{F}$, where $\overline N_P(A)$ is the union of $\overline N_P(j)$ over $j \in A$.} Since $\mathcal{F}$ involves the $\sigma$-field of $X = (X_i)_{i \in N}$, $v_i(\theta)\varphi_i$ has $G_P^*$ as a conditional dependency graph given $\mathcal{F}$. For the proof, we use Lemma 3.4 of \citesupp{Lee/Song:19:BJ}. We first show that there exists $C>0$ such that for all $n \ge 1$ and all $\theta, \tilde \theta \in \Theta$,
\begin{eqnarray}
\label{cond}
	\sqrt{\frac{1}{n}\sum_{i \in N}\mathbf{E}\left[ \left. (v_i(\theta) - v_i(\tilde \theta))^2 \varphi_{i,m}^2 \right\vert G_P\right]} \le C \| \theta - \tilde \theta\|.
\end{eqnarray}
By Assumption \ref{assump: ident}(ii) and Lemma \ref{lemm: der bound}, there exist constants $C_1,C_2>0$ such that for all $m=1,...,M$, and for all $n \ge 1$,
\begin{eqnarray}
\label{statements}
	\frac{1}{n}\sum_{i \in N} \mathbf{E}\left[ \left. \sup_{\theta \in \Theta}  \left\| \frac{\partial v_i(\theta)}{\partial \theta}\right\|^2 \varphi_{i,m}^2 \right\vert G_P\right] &\le& C_1, \text{ and }\\ \notag
	\frac{1}{n}\sum_{i \in N} \mathbf{E}\left[ \left. \sup_{\theta \in \Theta} v_i^2(\theta) \varphi_{i,m}^2 \right\vert G_P\right] &\le& C_2.
\end{eqnarray}
The first statement of (\ref{statements}) immediately yields (\ref{cond}) by the first order Taylor expansion. Combining this with the second statement of (\ref{statements}), and noting that $\Theta$ is compact in a finite dimensional Euclidean space, we find from Lemma 3.4 of \citesupp{Lee/Song:19:BJ} that there exists $C>0$ such that for all $n \ge 1$,
\begin{eqnarray*}
	\mathbf{E}\left[ \left. \sup_{\theta \in \Theta} \left| \frac{1}{n}\sum_{i=1}^n \left(v_i(\theta) \varphi_{i,m} - \mathbf{E}[v_i(\theta) \varphi_{i,m}|G_P] \right) \right| \right\vert G_P \right] \le C \left(1 + \max_{i \in N} n_P^2(i)\right) / \sqrt{n}.
\end{eqnarray*}
By Assumption \ref{assump: ident}(iv), we obtain Claim 2, which completes the proof. $\blacksquare$

\subsection{Proof of Theorem \ref{thm: theta hat exog}}
Throughout the proofs, we use the notation $C_1$ and $C_2$ to represent a constant which does not depend on $n$ or $n^*$. Without loss of generality, we also assume that $N^*$ is $\mathcal{F}$-measurable. This loses no generality because due to Condition A of the sampling process in the paper, the same proof goes through if we redefine $\mathcal{F}$ to be the $\sigma$-field generated by both $\mathcal{F}$ and $N^*$.

We introduce auxiliary lemmas which are used for proving Theorem \ref{thm: theta hat exog}.
\begin{lemma}
	\label{lemma: aux}
	For any array of numbers $\{a_{ij}\}_{i,j \in N}$ and a sequence $\{b_i\}_{i \in N}$ of numbers, we have for any subsets $A,B \subset N$ and for any undirected graph $G=(N,E)$,
	\begin{eqnarray*}
		\sum_{i \in B} \sum_{j \in N(i) \cap A} a_{ij} b_j = \sum_{i \in A} \left(\sum_{j \in N(i) \cap B} a_{ji}\right) b_i,
	\end{eqnarray*}
	where $N(i) = \{i \in N: ij \in E\}$.
\end{lemma}
\noindent \textbf{Proof: } Since the graph $G$ is undirected, i.e, $1\{j \in N(i)\} = 1\{i \in N(j)\}$, we write the left hand side sum as
\begin{eqnarray*}
	\sum_{i \in B} \sum_{j \in A} 1\{j \in N(i)\} a_{ij} b_j = \sum_{j \in A} \sum_{i \in B} 1\{i \in N(j)\} a_{ij} b_j.
\end{eqnarray*}
Interchanging the index notation $i$ and $j$ gives the desired result. $\blacksquare$

\begin{lemma}
	\label{lemma: residual CLT}
	Suppose that the conditions of Theorem \ref{thm: theta hat exog} hold. Then,
	\begin{eqnarray*}
		\Lambda^{-1/2} \frac{1}{\sqrt{n^*}} \sum_{i \in N^*} \tilde \varphi_i v_i \rightarrow_d N(0,I_M).
	\end{eqnarray*}
\end{lemma}

\noindent \textbf{Proof:} Choose any vector $b \in \mathbf{R}^M$ such that $||b||=1$ and let $\tilde \varphi_{i,b} = b'\tilde \varphi_i$. Recall that
\begin{eqnarray*}
	v_i = w_{ii}^{[0]} \varepsilon_i + \sum_{j \in N_P(i)} w_{ij}^{[0]} \varepsilon_j + \eta_i.
\end{eqnarray*}
Define
\begin{eqnarray*}
	a_{i} = w_{ii}^{[0]} \tilde \varphi_{i,b} 1\{i \in N^*\} + \sum_{j \in N_P(i) \cap N^*} \tilde \varphi_{j,b} w_{ji}^{[0]}.
\end{eqnarray*}
Then, from (\ref{bd wij}),
\begin{eqnarray*}
	|a_{i}| \le \left(1 + \frac{\beta_0^2}{1 - \beta_0^2} \right) \left( |\tilde \varphi_{i,b}| 1\{i \in N^*\} + |\beta_0| \sum_{j \in N_P(i) \cap N^*} \frac{|\tilde \varphi_{j,b}|}{n_P(j)( 1 - |\beta_0|)}\right).
\end{eqnarray*}
Using Lemma \ref{lemma: aux}, we can write
\begin{eqnarray}
\label{decomp}
\frac{1}{\sqrt{n^*}}\sum_{i \in N^*} \tilde \varphi_{i,b} v_i = \sum_{i \in N^\circ} \xi_i,
\end{eqnarray}
where we recall $N^\circ = \bigcup_{ i \in N^*} \overline N_P(i)$, and
\begin{eqnarray*}
	\xi_i = ( a_i \varepsilon_i + \tilde \varphi_{i,b} \eta_i 1\{i \in N^*\})/\sqrt{n^*}.
\end{eqnarray*}
By the Berry-Esseen Lemma for independent random variables (see, e.g., \citesupp{Shorack:00:ProbStat}, p.259),
\begin{eqnarray}
\label{berry-esseen}
\sup_{t \in \mathbf{R}} \Big \vert P\left\{ \sum_{i \in N^\circ} \frac{\xi_i}{\sigma_{\xi,i}} \le t |\mathcal{F}\right\} - \Phi(t) \Big \vert
\le \frac{\displaystyle 9 \mathbf{E}\left[\sum_{i \in N^\circ} |\xi_i|^3|\mathcal{F}\right]}{\displaystyle \left(\sum_{i \in N^\circ} \sigma_{\xi,i}^2 \right)^{3/2}},
\end{eqnarray}
where $\sigma_{\xi,i}^2 = \text{Var}(\xi_i|\mathcal{F})$. It suffices to show that the last bound vanishes in probability as $n^* \rightarrow \infty$. First, observe that
\begin{eqnarray*}
	\sum_{i \in N^\circ} \sigma_{\xi,i}^2 = \frac{1}{n^*}\sum_{i \in N^\circ} (a_i^2 \sigma_\varepsilon^2 + \tilde \varphi_{i,b}^2 \sigma_\eta^2 1\{i \in N^*\})
	&\ge& \frac{\sigma_\eta^2}{n^*} \sum_{i \in N^*} \tilde \varphi_{i,b}^2 = \sigma_\eta^2 >0,
\end{eqnarray*}
because $\frac{1}{n^*} \sum_{i \in N^*} \tilde \varphi_{i,b}^2 = 1$. Observe that
\begin{eqnarray}
\label{eq 5}
\quad \quad \mathbf{E}\left[\sum_{i \in N^\circ} |\xi_i|^3 |\mathcal{F}\right]
&\le& \frac{4\max_{i \in N} \mathbf{E}[|\varepsilon_i|^3 |\mathcal{F}]}{(n^*)^{3/2}}\sum_{i \in N^\circ} |\tilde \varphi_{i,b}|^3 |a_i|^3\\ \notag
&& + \frac{4\max_{i \in N}\mathbf{E}[|\eta_i|^3 |\mathcal{F}]}{(n^*)^{3/2}}\sum_{i \in N^\circ} |\tilde \varphi_{i,b}|^3\\ \notag
&\le& \frac{C_1\max_{i \in N} \mathbf{E}[|\varepsilon_i|^3 |\mathcal{F}]}{(n^*)^{3/2}}\sum_{i \in N^\circ} |a_i|^3
+ \frac{C_1 n^\circ \max_{i \in N}\mathbf{E}[|\eta_i|^3 |\mathcal{F}]}{(n^*)^{3/2}},
\end{eqnarray}
for some constant $C_1>0$, by Assumption \ref{assump: moment bound}. Now, using the fact that $|\tilde \varphi_{i,b}| \le C$, for some constant $C>0$, we bound the leading term as (for some constants $C_2,C_3>0$)
\begin{eqnarray*}
	\frac{C_2}{n^*}\sum_{i \in N^\circ} |a_i|^3 &\le& C \left(1 + \frac{\beta_0^2}{1 - \beta_0^2} \right)^3 \frac{1}{n^*}\sum_{i \in N^\circ} \left( |\tilde \varphi_{i,b}| 1\{i \in N^*\} + |\beta_0| \sum_{j \in N_P(i) \cap N^*} \frac{|\tilde \varphi_{j,b}|}{n_P(j)( 1 - |\beta_0|)}\right)^3\\
	&\le& \frac{C_3}{(1-|\beta_0|)^3} \left(1 + \frac{\beta_0^2}{1 - \beta_0^2} \right)^3,
\end{eqnarray*}
by Assumption \ref{assump: degree and moment}. Therefore, for some constant $C_2>0$,
\begin{eqnarray*}
	\mathbf{E}\left[\sum_{i \in N^\circ} |\xi_i|^3 |\mathcal{F}\right] \le \frac{C_2}{\sqrt{n^*}(1-|\beta_0|)^6}\max_{i \in N^*} \mathbf{E}[|\varepsilon_i|^3 |\mathcal{F}] + \frac{C_2 n^\circ}{(n^*)^{3/2}} \max_{i \in N^*}\mathbf{E}[|\eta_i|^3 |\mathcal{F}].
\end{eqnarray*}
Thus we conclude that the bound in (\ref{berry-esseen}) is $O_P((n^*)^{-1/2}+n^\circ (n^*)^{-3/2})$, where $n^\circ = |N^\circ|$. However, for some constant $C>0$,
\begin{eqnarray*}
	n^\circ \le \sum_{i \in N^*} |\overline{N}_P(i)| \le C n^*,
\end{eqnarray*}
by Assumption \ref{assump: degree and moment}. Hence we obtain the desired result. $\blacksquare$

\begin{lemma}
	\label{lemma: Lambda}
	Suppose that the conditions of Theorem \ref{thm: theta hat exog} hold. Then,
	\begin{eqnarray*}
		\mathbf{E}\left[||S_{\tilde \varphi v}||^2 |\mathcal{F}\right] = O((n^*)^{-1}), \text{ and }
		\mathbf{E}\left[||S_{Z^* v}||^2 |\mathcal{F}\right] = O((n^*)^{-1}),
	\end{eqnarray*}
	where
	\begin{eqnarray*}
		Z_i^* = \sum_{j \in N_P(i)\cap N^*} Z_j, S_{\tilde \varphi v} = \frac{1}{n^*} \sum_{i \in N^*} \tilde \varphi_i v_i, \text{ and } S_{Z^* v} = \frac{1}{n^*} \sum_{i \in N^*} Z_i^* v_i.
	\end{eqnarray*}
\end{lemma}
\noindent \textbf{Proof: } Recall the definitions of $e_{ij}$ and $e_{ii}$ in (\ref{eij}) and (\ref{eii}). First, observe that
\begin{eqnarray}
\label{bd eij}
	e_{ii} &\le& \left(1 + \frac{\beta_0^2}{1 - \beta_0^2}\right)^2\left( 1 + \frac{\beta_0^2}{(1 - |\beta_0|)^2}\right), \text{ and }\\ \notag
	|e_{ij}| &\le& 
	\frac{2 + |\beta_0|}{(1- |\beta_0|)^2} \left(1 + \frac{\beta_0^2}{1 - \beta_0^2}\right)^2.
\end{eqnarray}
Note that
\begin{eqnarray*}
	\mathbf{E}\left[ ||S_{\tilde \varphi v}||^2 |\mathcal{F}\right] &\le& \frac{\sigma_{\varepsilon}^2}{(n^*)^2} \sum_{i \in N^*} \sum_{j \in N_{-i}^*: N_P(i)\cap N_P(j) \ne \varnothing} |e_{ij}| ||\tilde \varphi_i|| ||\tilde \varphi_j||  \\
	&&+ \frac{1}{(n^*)^2} \sum_{i \in N^*} (|e_{ii}|\sigma_\varepsilon^2 + \sigma_\eta^2) ||\tilde \varphi_i||^2.
\end{eqnarray*}
However, since $||\tilde \varphi_i|| \le C$ by Assumption \ref{assump: moment bound}, we use (\ref{bd eij}) to obtain that $\mathbf{E}\left[ ||S_{\tilde \varphi v}||^2 |\mathcal{F}\right] = O((n^*)^{-1})$. 

Let us turn to the second bound. Observe that by Assumption \ref{assump: moment bound}, we have some $C>0$ such that for all $i \in N^*$, $||Z_i^*|| \le C$. Following the same proof as before, we obtain the desired result for $\mathbf{E}\left[ ||S_{Z^* v}||^2 |\mathcal{F}\right]$ as well. $\blacksquare$

\begin{lemma}
	\label{lemma: v tildes}
	Suppose that the conditions of Theorem \ref{thm: theta hat exog} hold. Then the following holds.
	\smallskip
	
	\noindent (i) $\frac{1}{n^*} \sum_{i \in N^*} (\tilde v_i^2 - v_i^2) \tilde \varphi_i \tilde \varphi_i' = O_P(1/\sqrt{n^*}).$
	\smallskip
	
	\noindent (ii) $\frac{1}{n^*} \sum_{i \in N^*} \sum_{j \in N_P(i) \cap N^*} (\tilde v_i \tilde v_j - v_i v_j) \tilde \varphi_i \tilde \varphi_j'= O_P(1/n^*).$
	\smallskip
	
	\noindent (iii) $\frac{1}{n^*} \sum_{i \in N^*} (v_i^2 - \mathbf{E}[v_i^2|\mathcal{F}]) \tilde \varphi_i \tilde \varphi_i' = O_P(1/\sqrt{n^*})$.
	\smallskip
	
	\noindent (iv) $\frac{1}{n^*} \sum_{i \in N^*} \sum_{j \in N_P(i)\cap N^*} (v_i v_j - \mathbf{E}[v_i v_j|\mathcal{F}]) \tilde \varphi_i \tilde \varphi_j'= O_P(1/\sqrt{n^*})$.
\end{lemma}

\noindent \textbf{Proof: } (i) First, write $\tilde v - v = - Z (\tilde \rho - \rho_0)$, where
$\tilde \rho - \rho_0 = \left[S_{Z\tilde \varphi} S_{Z \tilde \varphi}'\right]^{-1} S_{Z\tilde \varphi} S_{\tilde \varphi v}$.
Hence
\begin{eqnarray*}
	\left\|\frac{1}{n^*}\sum_{i \in N^*} (\tilde v_i - v_i)^2 \tilde \varphi_i \tilde \varphi_i'\right\|
	\le \frac{C_1}{n^*}\sum_{i \in N^*} (\tilde v_i - v_i)^2,
\end{eqnarray*}
for some constant $C_1>0$. As for the last term, note that
\begin{eqnarray}
\label{eq}
&& \frac{1}{n^*}\sum_{i \in N^*} \mathbf{E}\left[ (\tilde v_i - v_i)^2 |\mathcal{F}\right]\\ \notag
&=& \frac{1}{n^*} \text{tr} \left( S_{Z \tilde \varphi}' \left[S_{Z\tilde \varphi} S_{Z \tilde \varphi}'\right]^{-1} S_{ZZ} \left[S_{Z\tilde \varphi} S_{Z \tilde \varphi}'\right]^{-1} S_{Z \tilde \varphi} \Lambda \right) = O_P\left(\frac{1}{n^*}\right),
\end{eqnarray}
by the definition of $\Lambda$ in (\ref{lambda}) and by Lemma \ref{lemma: Lambda}. However, we need to deal with
\begin{eqnarray}
\label{eq6}
\Big \vert \frac{1}{n^*} \sum_{i \in N^*} (\tilde v_i^2 - v_i^2) \Big \vert
\le \sqrt{\frac{1}{n^*} \sum_{i \in N^*} (\tilde v_i - v_i)^2}
\sqrt{\frac{1}{n^*} \sum_{i \in N^*} (\tilde v_i + v_i)^2}.
\end{eqnarray}
Note that
\begin{eqnarray*}
	\frac{1}{n^*} \sum_{i \in N^*} (\tilde v_i + v_i)^2 &\le& \frac{2}{n^*} \sum_{i \in N^*} (\tilde v_i - v_i)^2 + \frac{8}{n^*} \sum_{i \in N^*} v_i^2\\
	&=& O_P\left(\frac{1}{n^*}\right) + \frac{8}{n^*} \sum_{i \in N^*} v_i^2,
\end{eqnarray*}
by (\ref{eq}). As for the last term,
\begin{eqnarray*}
	\frac{1}{n^*} \sum_{i \in N^*} \mathbf{E}[v_i^2 |\mathcal{F}]
	\le \frac{2}{n^*} \sum_{i \in N^*} \mathbf{E}[R_i^2(\varepsilon) |\mathcal{F}]
	+ \frac{2}{n^*} \sum_{i \in N^*} \mathbf{E}[\eta_i^2 |\mathcal{F}].
\end{eqnarray*}
The last term is bounded by $2 \sigma_{\eta}^2$, and the first term on the right hand side is bounded by
\begin{eqnarray*}
	\frac{2 \sigma_{\varepsilon}^2}{n^*} \sum_{i \in N^*} e_{ii} \le C,
\end{eqnarray*}
by (\ref{bd eij}). Combining this with (\ref{eq}) and (\ref{eq6}), we obtain the desired result.\smallskip

\noindent (ii) Let us first write
\begin{eqnarray*}
	&& \frac{1}{n^*} \sum_{i \in N^*} \sum_{j \in N_P(i) \cap N^*} (\tilde v_i \tilde v_j - v_i v_j) \\
	&=& \frac{1}{n^*} \sum_{i \in N^*} \sum_{j \in N_P(i)\cap N^*}(\tilde v_i - v_i)(\tilde v_j - v_j)\\
	&+& \frac{1}{n^*} \sum_{i \in N^*} \sum_{j \in N_P(i)\cap N^*}(\tilde v_i - v_i) v_j\\
	&+& \frac{1}{n^*} \sum_{i \in N^*} \sum_{j \in N_P(i)\cap N^*} v_i (\tilde v_j - v_j)
	= A_{n,1} + A_{n,2} + A_{n,3}, \text{ say.}
\end{eqnarray*}
As for the leading term, by Cauchy-Schwarz inequality,
\begin{eqnarray*}
	|A_{n,1}| = \sqrt{\frac{1}{n^*}\sum_{i \in N^*} (\tilde v_i - v_i)^2}\sqrt{\frac{1}{n^*} \sum_{i \in N^*} \left(\sum_{j \in N_P(i)\cap N^*} (\tilde v_j - v_j) \right)^2}.
\end{eqnarray*}
Note that
\begin{eqnarray*}
	&& \frac{1}{n^*} \sum_{i \in N^*} \mathbf{E}\left[\left(\sum_{j \in N_P(i)\cap N^*} (\tilde v_j - v_j) \right)^2 |\mathcal{F} \right]\\
	&\le&
	\frac{1}{n^*} \sum_{i \in N^*} |N_P(i)\cap N^*| \sum_{j \in N_P(i)\cap N^*} \mathbf{E}\left[\left(\tilde v_j - v_j\right)^2 |\mathcal{F} \right]\\
	&=&\frac{1}{n^*} \sum_{i \in N^*} \left(\sum_{j \in N_P(i)\cap N^*} |N_P(j)\cap N^*| \right)\mathbf{E}\left[\left(\tilde v_i - v_i\right)^2 |\mathcal{F} \right],
\end{eqnarray*}
where the inequality above uses Jensen's inequality and the equality above uses Lemma \ref{lemma: aux}. Hence the last term is bounded by
\begin{eqnarray*}
	\frac{\max_{i \in N^*}|N_P(i)\cap N^*|^2}{n^*} \sum_{i \in N^*} \mathbf{E}\left[\left(\tilde v_i - v_i\right)^2 |\mathcal{F} \right] 
	\le O_P\left(\frac{1}{n^*}\right).
\end{eqnarray*}
by (\ref{eq}). Thus we conclude that
\begin{eqnarray*}
	|A_{n,1}| = O_P\left(\frac{1}{n^*}\right).
\end{eqnarray*}
Now, let us turn to $A_{n,2}$. Observe that
\begin{eqnarray*}
	A_{n,2} &=& -\frac{1}{n^*}\sum_{i \in N^*} Z_i' \sum_{j \in N_P(i) \cap N^*} v_j (\tilde \rho - \rho_0)\\
	&=& - \left(\frac{1}{n^*}\sum_{i \in N^*} Z_i^{*'} v_i \right) (\tilde \rho - \rho_0) = - S_{Z^* v}(\tilde \rho - \rho_0)
\end{eqnarray*}
using Lemma \ref{lemma: aux}. From the proof of (i), we obtain that
\begin{eqnarray*}
	\tilde \rho - \rho_0 = O_P\left(\frac{1}{\sqrt{n^*}} \right).
\end{eqnarray*}
Hence combined with Lemma \ref{lemma: Lambda}, we have 
\begin{eqnarray*}
	|A_{n,2}| = O_P\left(\frac{1}{n^*}\right).
\end{eqnarray*}

Since by Lemma \ref{lemma: aux}, $A_{n,2} = A_{n,3}$, the proof of (ii) is complete.\\

\noindent (iii) Note that 
\begin{eqnarray*}
	\text{Var} \left(\frac{1}{n^*}\sum_{i \in N^*} R_i^2(\varepsilon) |\mathcal{F} \right)
	&\le& \frac{2}{(n^*)^2}\sum_{i \in N^*} \text{Var} \left((w_{ii}^{[0]})^2 \varepsilon_i^2|\mathcal{F}\right)\\
	&+& \frac{2}{(n^*)^2}\sum_{i \in N^*} \text{Var} \left(\left(\frac{\beta_0 w_{ii}^{[0]}}{n_P(i)} \sum_{j \in N_P(i)}  \lambda_{ij} \varepsilon_j \right)^2|\mathcal{F} \right).
\end{eqnarray*}
The leading term is $O_P((n^*)^{-1})$. The last term is bounded by
\begin{eqnarray*}
	\frac{2}{(n^*)^2}\sum_{i \in N^*}  \frac{\beta_0^4 (w_{ii}^{[0]})^4}{n_P(i)} \sum_{j \in N_P(i)}  \lambda_{ij}^4 \mathbf{E}[\varepsilon_j^4|\mathcal{F}] = O_P((n^*)^{-1}).
\end{eqnarray*}
Since $v_i = R_i(\varepsilon) + \eta_i$ and $\varepsilon_i$'s and $\eta_i$'s are independent, we obtain the desired rate.
\smallskip

\noindent (iv) For simplicity of notation, define
\begin{eqnarray*}
	V_{ij} = (v_i v_j - \mathbf{E}[v_i v_j|\mathcal{F}])\tilde \varphi_i \tilde \varphi_j'.
\end{eqnarray*}
Then we write
\begin{eqnarray*}
	&& \mathbf{E}\left[\left(\frac{1}{n^*} \sum_{i \in N^*} \sum_{j \in N_P(i)\cap N^*} V_{ij} \right)^2 |\mathcal{F}\right]\\
	&=& \frac{1}{(n^*)^2} \sum_{i_1 \in N^*} \sum_{j_1 \in N_P(i)\cap N^*}  \sum_{i_2 \in N^*} \sum_{j_2 \in N_P(i)\cap N^*} \mathbf{E}\left[V_{i_1 j_1} V_{i_2 j_2}|\mathcal{F}\right].
\end{eqnarray*}
The last expection is zero, whenever $(i_2,j_2)$ is away from $(i_1,j_1)$ by more than two edges. Hence we can bound the last term by (using Assumption \ref{assump: degree and moment}))
\begin{eqnarray*}
	\frac{C_1}{n^*} \max_{i \in N} \mathbf{E}[v_i^2|\mathcal{F}] \le \frac{C_2}{n^*}
\end{eqnarray*}
for some constants $C_1,C_2$ which do not depend on $n$. $\blacksquare$

\begin{lemma}
	\label{lemma: Consistency of Lambda hat}
	Suppose that the conditions of Theorem \ref{thm: theta hat exog} hold. Then,
	\begin{eqnarray*}
		\hat \Lambda - \Lambda = O_P\left(\frac{1}{\sqrt{n^*}}\right).
	\end{eqnarray*}
\end{lemma}

\noindent \textbf{Proof: } We write
\begin{eqnarray*}
	\hat \Lambda_1 - \Lambda_1 &=& \frac{1}{n^*}\sum_{i \in N^*} (\tilde v_i^2 - \mathbf{E}[v_i^2|\mathcal{F}])\tilde \varphi_i\tilde \varphi_i' \text{ and }\\
	\hat \Lambda_2 - \Lambda_2 &=& \frac{\hat s_{\varepsilon} - s_\varepsilon}{n^*}\sum_{i \in N^*}\sum_{j \in N_P(i) \cap N^*}  q_{\varepsilon,ij} \tilde \varphi_i\tilde \varphi_j'.
\end{eqnarray*}
By Assumption \ref{assump: nondeg} and Lemma \ref{lemma: v tildes}(ii)(iv), we have
\begin{eqnarray*}
	\hat s_{\varepsilon} - s_\varepsilon = O_P(1/\sqrt{n^*}).
\end{eqnarray*}
The desired result follows by using this and applying Lemma \ref{lemma: v tildes}(i) and (iii) to $\hat \Lambda_1 - \Lambda_1$. $\blacksquare$

\begin{lemma}
	\label{lemma: v_hats}
	Suppose that the conditions of Theorem \ref{thm: theta hat exog} hold. Then the following holds.
	
	\noindent (i) $\frac{1}{n^*} \sum_{i \in N^*} (\hat v_i^2 - v_i^2) \tilde \varphi_i \tilde \varphi_i' = O_P(1/\sqrt{n^*}).$
	
	\noindent (ii) $\frac{1}{n^*} \sum_{i \in N^*} \sum_{j \in N_P(i) \cap N^*} (\hat v_i \hat v_j - v_i v_j) \tilde \varphi_i \tilde \varphi_j'= O_P(1/n^*).$
\end{lemma}

\noindent \textbf{Proof: } First, write $\hat v - v = - Z (\hat \rho - \rho_0)$, where
\begin{eqnarray}
\label{expr}
\hat \rho - \rho_0 = \left[S_{Z\tilde \varphi} \hat \Lambda^{-1} S_{Z \tilde \varphi}'\right]^{-1} S_{Z\tilde \varphi} \hat \Lambda^{-1} S_{\tilde \varphi v}.
\end{eqnarray}
Following the same arguments as in the proof of Lemma \ref{lemma: v tildes}(i) and (ii) and Lemma \ref{lemma: Consistency of Lambda hat}, we obtain the desired result. $\blacksquare$
\smallskip

\noindent \textbf{Proof of Theorem \ref{thm: theta hat exog}:} Let us consider the first statement. We write
\begin{eqnarray*}
	\frac{1}{\sqrt{n^*}} \hat \Lambda^{-1/2} \tilde \varphi' \hat v &=&  \frac{1}{\sqrt{n^*}} \hat \Lambda^{-1/2} \tilde \varphi' (\hat v - v)
	+ \frac{1}{\sqrt{n^*}} \hat \Lambda^{-1/2} \tilde \varphi' v\\
	&=&  - \frac{1}{\sqrt{n^*}} \hat \Lambda^{-1/2} \tilde \varphi' Z (\hat \rho - \rho_0)
	+ \frac{1}{\sqrt{n^*}} \hat \Lambda^{-1/2} \tilde \varphi' v = \sqrt{n^*}(I - P)\hat \Lambda^{-1/2} S_{\tilde \varphi v},
\end{eqnarray*}
using (\ref{expr}), where
\begin{eqnarray*}
	P = \hat \Lambda^{-1/2}S_{Z \tilde \varphi}'\left[S_{Z \tilde \varphi} \hat \Lambda^{-1} S_{Z \tilde \varphi}' \right]^{-1} S_{Z \tilde \varphi} \hat \Lambda^{-1/2}.
\end{eqnarray*}
Note that $P$ is a projection matrix from $\mathbf{R}^M$ onto the range space of $\hat \Lambda^{-1/2}S_{Z \tilde \varphi}'$. Hence combining Lemmas \ref{lemma: residual CLT} and \ref{lemma: Consistency of Lambda hat}. We obtain the desired result. The second result follows from Lemma \ref{lemma: residual CLT} and equation (\ref{expr}). $\blacksquare$

\section{Inference for the Model with First Order Sophisticated Agents}
\label{App sec:Inf FOS}
\subsection{Overview}
Let us consider inference on payoff parameters using a model that assumes all the agents to be of first-order sophisticated type. The network externality is more extensive than when the agents are of simple type, and best responses involve more extensive network externality, and we require more data accordingly. In particular, we strengthen Conditions B and C as follows:
\medskip

\noindent \textbf{Condition B1:} For each $i \in N^*$, the econometrician observes $N_{P,2}(i)$ and $(Y_i,X_i)$ and for any $j \in N_P(i)$ and any $k \in N_{P,2}(i)\backslash N_P(i)$, the econometrician observes $n_P(j)$, $n_P(k)$, $|N_P(i)\cap N_P(j)|$ and $|N_P(j)\cap N_P(k)|$, $X_j$ and $X_k$.
\smallskip

\noindent \textbf{Condition C1:} Either of the following two conditions is satisfied.

(a) For any $i,j \in N^*$ such that $i \ne j$, $N_{P,2}(i) \cap N_{P,2}(j) = \varnothing$.

(b) For each agent $i \in N^*$, and for any agent $j \in N^*$ such that $N_{P,2}(i) \cap N_{P,2}(j) \ne \varnothing$, the econometrician observes $Y_j$, $|N_{P,2}(j)\cap N_{P,2}(k)|$, $n_P(k)$ and $X_k$ for all $k \in N_P(j)$.
\medskip

Condition B1 requires that the data contain many agents such that $N_{P,2}(i)$ for each $i$ of such agents is available together with the number of common $G_P$-neighbors between each agent $k \in N_{P,2}(i)$ and agent $i$ and between each agent $j \in N_P(i)$ and agent $i$. Condition C1 is again trivially satisfied if data contain many agents such that $G_P$-neighbors of $G_P$-neighbors do not overlap. In this case, we can select $N^*$ to include only those agents.

The inference is similar as in the case with agents of simple type, except that we redefine $Z_i$ and $v_i$ into $Z_i^{\mathsf{FS}}$, and $v_i^{\mathsf{FS}}$ as we explain below. Define
\begin{eqnarray}
\label{Z_i*}
Z_i^{\textsf{FS}} &=& \left(1 + \frac{\beta_0}{n_P(i)}\sum_{j \in N_P(i)} w_{ji}^{[0]} \right) X_i\\ \notag
&& + \sum_{j \in N_{P,2}(i)} \left( \frac{\beta_0}{n_P(i)} \sum_{k \in N_P(i)} w_{kj}^{[0]} 1\{j \in \overline N_P(k)\}\right) X_j.
\end{eqnarray}
Then, by the previous results (see (\ref{FOS BR})), we can write
\begin{eqnarray}
\label{first-order sophisticated}
Y_i = Z_i^{{\textsf{FS}}'}\rho_0 + v_i^{\textsf{FS}},
\end{eqnarray}
where
\begin{eqnarray*}
	v_i^{\textsf{FS}} = R_i^{\textsf{FS}}(\varepsilon) + \eta_i,
\end{eqnarray*}
with
\begin{eqnarray*}
	R_i^{\textsf{FS}}(\varepsilon) &=& \left(1 + \frac{\beta_0}{n_P(i)}\sum_{j \in N_P(i)} w_{ji}^{[0]} \right) \varepsilon_i\\ \notag
	&& + \sum_{j \in N_{P,2}(i)} \left( \frac{\beta_0}{n_P(i)} \sum_{k \in N_P(i)} w_{kj}^{[0]} 1\{j \in \overline N_P(k)\}\right) \varepsilon_j.
\end{eqnarray*}
Using this reformulation, we can develop inference similarly as before. More specifically, let us define
\begin{eqnarray}
\label{lambda FS}
\Lambda^{\textsf{FS}} = \frac{1}{n^*}\sum_{i \in N^*} \sum_{j \in N^*} \mathbf{E}[v_i^{\textsf{FS}} v_j^{\textsf{FS}}|\mathcal{F}] \tilde \varphi_i \tilde \varphi_j',
\end{eqnarray}
and let $\hat \Lambda^{\textsf{FS}}$ be a consistent estimator of $\Lambda^{\textsf{FS}}$. (See the next subsection of the construction of the estimator.) Define
\begin{eqnarray}
\label{rho hat 2}
\hat \rho^{\textsf{FS}} = \left[S_{Z \tilde \varphi}^{\textsf{FS}} (\hat \Lambda^{\textsf{FS}})^{-1} (S_{Z \tilde \varphi}^{\textsf{FS}})'\right]^{-1} S_{Z \tilde \varphi}^{\textsf{FS}} (\hat \Lambda^{\textsf{FS}})^{-1} S_{\tilde \varphi y}^{\textsf{FS}},
\end{eqnarray}
where $S_{Z \tilde \varphi}^{\textsf{FS}}$ and $S_{\tilde \varphi y}^{\textsf{FS}}$ are the same as $S_{Z \tilde \varphi}^{\textsf{FS}}$ and $S_{\tilde \varphi y}^{\textsf{FS}}$ except that we use $Z^{\textsf{FS}}$ in place of $Z$. Using this, we construct the estimator
\begin{eqnarray}
\label{Omega and V hat}
\hat V^{\textsf{FS}}  =  \left[S_{Z \tilde \varphi }^{\textsf{FS}} (\hat \Lambda^{\textsf{FS}})^{-1} S_{\tilde \varphi Z}^{\textsf{FS}}\right]^{-1}.
\end{eqnarray}
We construct a vector of residuals $\hat v^{\textsf{FS}} = [\hat v_i^{\textsf{FS}}]_{i \in N^*}$, where
\begin{eqnarray}
\label{tilde v beta2}
\hat v_i^{\textsf{FS}} = Y_i - Z_i^{\textsf{FS} \prime} \hat \rho^{\textsf{FS}}.
\end{eqnarray}
Finally, we form a profiled test statistic as follows:
\begin{eqnarray}
\label{Pop Obj2} \quad \quad \quad
T^{\mathsf{FS}}(\beta_0) = \frac{(\hat v^{\textsf{FS}})' \tilde \varphi (\hat \Lambda^{\textsf{FS}})^{-1} \tilde \varphi' \hat v^{\textsf{FS}}}{n^*}.
\end{eqnarray}
Then, we construct confidence intervals
\begin{eqnarray*}
	C_{1 - \alpha}^{\beta,\mathsf{FS}} \equiv \left\{\beta \in (-1,1): T^{\textsf{FS}}(\beta) \le c_{1-\alpha} \right\},
\end{eqnarray*}
where $c_{1-\alpha}$ is the $(1-\alpha)$-quantile of $\chi_{M-d}^2$.

The confidence intervals for $a'\rho$ can be similarly constructed as in Section \ref{subsubsec: est and inf}. More specifically, let
\begin{eqnarray*}
	\hat V^{\mathsf{FS}} = \left[S_{Z\tilde \varphi}^{\mathsf{FS}} (\hat \Lambda^{\mathsf{FS}})^{-1} S_{Z\tilde \varphi}^{\mathsf{FS} \prime} \right]^{-1},
\end{eqnarray*}
and define
\begin{eqnarray*}
	\hat \sigma^{\mathsf{FS}}(a) = \sqrt{a'\hat V^{\mathsf{FS}} a}.
\end{eqnarray*}
Then, the confidence interval for $a'\rho$ is found as
\begin{eqnarray*}
	C_{1-\alpha}^{\rho,\mathsf{FS}}(a) = \bigcup_{\beta \in C_{1-(\alpha/2)}^{\rho,\mathsf{FS}}} C_{1-(\alpha/2)}^{\beta,\mathsf{FS}}(\beta,a),
\end{eqnarray*}
where
\begin{eqnarray*}
	C_{1-(\alpha/2)}^{\rho,\mathsf{FS}}(\beta_0,a) = \left[a'\hat \rho^{\mathsf{FS}} - \frac{z_{1-(\alpha/4)} \hat \sigma^{\mathsf{FS}}(a)}{\sqrt{n}}, a'\hat \rho^{\mathsf{FS}} + \frac{z_{1-(\alpha/4)} \hat \sigma^{\mathsf{FS}}(a)}{\sqrt{n}}\right].
\end{eqnarray*}
and $z_{1-(\alpha/4)}$ is the $(1 - (\alpha/4))$-percentile of $N(0,1)$.

\subsection{Estimation of the Asymptotic Covariance Matrix}

We first construct a consistent estimator $\hat \Lambda^{\textsf{FS}}$ of $\hat \Lambda$. Define for $i,j \in N$,
\begin{eqnarray*}
	e_{ij}^{\textsf{FS}} = \mathbf{E}[R_i^{\textsf{FS}}(\varepsilon)R_j^{\textsf{FS}}(\varepsilon)|\mathcal{F}] /\sigma_\varepsilon^2.
\end{eqnarray*}
If we let
\begin{eqnarray}
\label{w bar}
\overline w_{ij}^{[0]} = \frac{1}{n_P(i)} \sum_{k \in N_P(i)} w_{kj}^{[0]} 1\{j \in \overline N_P(k)\},
\end{eqnarray}
we can rewrite
\begin{eqnarray}
\label{e star ii}
e_{ii}^{\textsf{FS}} = \left( 1 + \beta_0 \overline w_{ii}^{[0]} \right)^2 + \beta_0^2 \sum_{j \in N_{P,2}(i)} \left(\overline w_{ij}^{[0]} \right)^2,
\end{eqnarray}
and for $i \ne j$, $e_{ij}^{\textsf{FS}} = \beta_0 q_{\varepsilon,ij}^{\textsf{FS}}$, where
\begin{eqnarray}
\label{q star ij 2}
\quad \quad
q_{\varepsilon,ij}^{\textsf{FS}} &=& \overline w_{ji}^{[0]} \left( 1 + \beta_0 \overline w_{ii}^{[0]} \right)1\{i \in N_{P,2}(j)\} + \overline w_{ij}^{[0]} \left( 1 + \beta_0 \overline w_{jj}^{[0]} \right)1\{j \in N_{P,2}(i)\}\\ \notag
&& + \beta_0 \sum_{s \in N_{P,2}(i) \cap N_{P,2}(j)} \overline w_{is}^{[0]} \overline w_{js}^{[0]},
\end{eqnarray}
where the last term is zero if $N_{P,2}(i) \cap N_{P,2}(j)$ is empty. Similarly, sums over empty sets in any of the terms above are zero. Let us now write
\begin{eqnarray*}
	\Lambda^{\textsf{FS}} = \Lambda_1^{\textsf{FS}} + \Lambda_2^{\textsf{FS}},
\end{eqnarray*}
where
\begin{eqnarray*}
	\Lambda_1^{\textsf{FS}} &=& \frac{1}{n^*}\sum_{i \in N^*} \mathbf{E}[(v_i^{\textsf{FS}})^2|\mathcal{F}] \tilde \varphi_i \tilde \varphi_i', \text{ and }\\
	\Lambda_2^{\textsf{FS}} &=& \frac{1}{n^*}\sum_{i \in N^*} \sum_{j \in N^*_{-i}} \mathbf{E}[v_i^{\textsf{FS}} v_j^{\textsf{FS}}|\mathcal{F}] \tilde \varphi_i \tilde \varphi_j'.
\end{eqnarray*}
To motivate estimation of $\Lambda_2^{\textsf{FS}}$, we rewrite
\begin{eqnarray}
\label{eq44}
\Lambda_2^{\textsf{FS}} = \frac{1}{n^*} \sum_{i \in N^*} \sum_{j \in N^*_{-i}} (e_{ij}^{\textsf{FS}}) \sigma_\varepsilon^2 \tilde \varphi_i \tilde \varphi_j'
= \frac{\beta_0}{n^*} \sum_{i \in N^*} \sum_{j \in N^*_{-i}} q_{\varepsilon,ij}^{\textsf{FS}} \sigma_\varepsilon^2 \tilde \varphi_i \tilde \varphi_j'.
\end{eqnarray} 
Let us find an expression for $\sigma_\varepsilon^2$. Note that
\begin{eqnarray*}
	\frac{1}{n^*}\sum_{i \in N^*} \sum_{j \in N_P(i) \cap N^*}\mathbf{E}[v_i^{\mathsf{FS}}v_j^{\mathsf{FS}}|\mathcal{F}] = \beta_0 b_\varepsilon^{\mathsf{FS}} \sigma_\varepsilon^2, 
\end{eqnarray*}
where
\begin{eqnarray*}
	b_\varepsilon^{\mathsf{FS}} = \frac{1}{n^*}\sum_{i \in N^*} \sum_{j \in N_P(i) \cap N^*} q_{\varepsilon,ij}^{\textsf{FS}}.
\end{eqnarray*}
Hence if we let
\begin{eqnarray*}
	s_\varepsilon^{\textsf{FS}} = \frac{\displaystyle \sum_{i \in N^*} \sum_{j \in N_P(i) \cap N^*} \mathbf{E}[v_i^{\textsf{FS}} v_j^{\textsf{FS}}|\mathcal{F}]}{\displaystyle \sum_{i \in N^*} \sum_{j \in N_P(i) \cap N^*} q_{\varepsilon,ij}^{\textsf{FS}}},
\end{eqnarray*}
we have
\begin{eqnarray*}
	\sigma_\varepsilon^2 \beta_0 = s_\varepsilon^{\textsf{FS}}.
\end{eqnarray*}
Plugging this in the last term in (\ref{eq44}), we obtain that
\begin{eqnarray*}
	\Lambda_2^{\textsf{FS}} = \frac{s_\varepsilon^{\textsf{FS}}}{n^*} \sum_{i \in N^*} \sum_{j \in N^*_{-i}} q_{\varepsilon,ij}^{\textsf{FS}} \tilde \varphi_i \tilde \varphi_j'.
\end{eqnarray*}
Our estimator then uses the empirical analogues to find $\hat{\Lambda}^{\textsf{FS}}$. 

First define
\begin{eqnarray}
\label{rho tilde 2}
\tilde \rho^{\textsf{FS}} = \left[(S_{Z \tilde \varphi}^{\textsf{FS}})(S_{Z \tilde \varphi}^{\textsf{FS}})'\right]^{-1} S_{Z \tilde \varphi}^{\textsf{FS}} S_{\tilde \varphi y}^{\textsf{FS}},
\end{eqnarray}
and let
\begin{eqnarray}
	\label{v tilde FS}
	\tilde v_i^{\textsf{FS}} = Y_i - Z_i^{\textsf{FS} \prime} \tilde \rho^{\textsf{FS}}.
\end{eqnarray}
We now present a consistent estimator $\hat \Lambda^{\textsf{FS}}$:
\begin{eqnarray*}
	\hat \Lambda^{\textsf{FS}} = \hat \Lambda_1^{\textsf{FS}} + \hat \Lambda_2^{\textsf{FS}},  
\end{eqnarray*}
where
\begin{eqnarray*}
	\hat \Lambda_1^{\textsf{FS}} &=& \frac{1}{n^*}\sum_{i \in N^*} (\tilde v_i^{\textsf{FS}})^2 \tilde \varphi_i \tilde \varphi_i', \text{ and }\\
	\hat \Lambda_2^{\textsf{FS}} &=& \frac{\hat s_\varepsilon^{\textsf{FS}}}{n^*} \sum_{i \in N^*} \sum_{j \in N^*_{-i}:N_{P,2}(i) \cap N_{P,2}(j) \ne \varnothing} q_{\varepsilon,ij}^{\textsf{FS}} \tilde \varphi_i \tilde \varphi_j',
\end{eqnarray*}
and
\begin{eqnarray*}
	\hat s_\varepsilon^{\textsf{FS}}
	= \frac{\displaystyle \sum_{i \in N^*} \sum_{j \in N_P(i)\cap N^*} \tilde v_i^{\textsf{FS}} \tilde v_j^{\textsf{FS}}}{\displaystyle \sum_{i \in N^*} \sum_{j \in N_P(i) \cap N^*} q_{\varepsilon,ij}^{\textsf{FS}}},
\end{eqnarray*}
with $q_{\varepsilon,ij}^{\textsf{FS}}$ as defined in (\ref{q star ij 2}). For this, we construct $\tilde v_i^\mathsf{FS}$ as we constructed $\tilde v_i$ using $Z^\mathsf{FS}$ in place of $Z$.

\subsection{Asymptotic Theory for Inference from the Model with First-Order Sophisticated Agents}
In this section, we develop asymptotic theory for the game with first-order sophisticated agents. Recall from (\ref{FOS BR}) and (\ref{tau spec}) that each player $i$'s best response $s_i^{[1]}$ takes the following form: (recall the definition of $\overline w_{ij}^{[0]}$ in (\ref{w bar}))
\begin{align*}
s_{i}^{[1]}(\mathcal{I}_{i,1})  = \left( 1 + \beta_0 \overline w_{ii}^{[0]}\right) X_i'\rho_0
& + \beta_0 \sum_{j \in N_{P,2}(i)} \overline w_{ij}^{[0]} X_j'\rho_0 + R_i^{\mathsf{FS}}(\varepsilon) + \eta_i,
\end{align*}
where,
\begin{eqnarray*}
	R_i^{\mathsf{FS}}(\varepsilon) = \left( 1 + \beta_0 \overline w_{ii}^{[0]}\right) \varepsilon_i + \beta_0 \sum_{j \in N_{P,2}(i)} \overline w_{ij}^{[0]} \varepsilon_j.
\end{eqnarray*}

In place of Assumptions \ref{assump: nondeg} - \ref{assump: moment bound}, we make the following assumptions.

\begin{assumption}
	\label{assump: nondeg 2} There exists $c>0$ such that for all $n^* \ge 1$,
	\begin{eqnarray*}
		&& \lambda_{\min}(S_{\varphi \varphi}) \ge c, \lambda_{\min} (\Lambda^{\mathsf{FS}}) \ge  c, \\
		&& \lambda_{\min} ((S_{Z \tilde \varphi}^{\mathsf{FS}}) (S_{Z \tilde \varphi}^{\mathsf{FS}})') \ge  c, \text{ and  }\\
		&& \lambda_{\min} ((S_{Z \tilde \varphi}^{\mathsf{FS}}) (\Lambda^{\mathsf{FS}})^{-1} (S_{Z \tilde \varphi}^{\mathsf{FS}})') \ge  c.
	\end{eqnarray*}
\end{assumption}

\begin{assumption}
	\label{assump: moment bound 2}
	\noindent There exists a constant $C>0$ such that for all $n^* \ge 1$,
	\begin{eqnarray*}
		\max_{i \in N_2^\circ} ||X_i|| +  \max_{i \in N_2^\circ} ||\tilde \varphi_i|| \le C
	\end{eqnarray*}
	and $\mathbf{E}[\varepsilon_i^4|\mathcal{F}] + \mathbf{E}[\eta_i^4|\mathcal{F}] <C$,
	where $n_2^\circ = |N_2^\circ|$ and
	\begin{eqnarray*}
		N_2^\circ = \bigcup_{ i \in N^*} \overline N_{P,2}(i).
	\end{eqnarray*}
\end{assumption}

Then the asymptotic results are summarized in the following theorem.
\begin{theorem}
	\label{thm: theta hat exog 2}
	Suppose that the conditions of Theorem \ref{thm: best response2}, Assumption \ref{assump: degree and moment}, and Assumptions \ref{assump: nondeg 2} - \ref{assump: moment bound 2} hold. Then,
	\begin{eqnarray*}
		T^{\mathsf{FS}}(\beta_0) \rightarrow_d \chi^2_{M-d}, \text{ and } (\hat V^{\mathsf{FS}})^{-1/2} \sqrt{n^*} ( \hat \rho^{\mathsf{FS}}  - \rho_0 ) \rightarrow_d N(0,I_d),
	\end{eqnarray*}
	as $n^* \rightarrow \infty$.
\end{theorem}

The proofs follow similar steps as in the proof of Theorem \ref{thm: theta hat exog}. For the sake of transparency, we provide complete proofs here.

\begin{lemma}
	\label{lemma: residual CLT 2}
	Suppose that the conditions of Theorem \ref{thm: theta hat exog 2} hold. Then, as $n^* \rightarrow \infty$,
	\begin{eqnarray*}
		(\Lambda^{\mathsf{FS}})^{-1/2} \frac{1}{\sqrt{n^*}} \sum_{i \in N^*} \tilde \varphi_i v_i^{\mathsf{FS}} \rightarrow_d N(0,I_M).
	\end{eqnarray*}
\end{lemma}

\noindent \textbf{Proof:} Choose any vector $b \in \mathbf{R}^M$ such that $||b||=1$ and let $\tilde \varphi_{i,b} = b'\tilde \varphi_i$. Define
\begin{eqnarray*}
	a_i^{\textsf{FS}} =  \left(1 + \beta_0 \overline w_{ii}^{[0]} \right) \tilde \varphi_{i,b} 1\{i \in N^*\} + \beta_0 \sum_{j \in N_{P,2}(i) \cap N^*} \tilde \varphi_{j,b} \overline w_{ji}^{[0]}.
\end{eqnarray*}
By (\ref{bd wij}), we have
\begin{eqnarray}
\label{bd wij2}
0 \le \overline w_{ii}^{[0]} \le 1 + \frac{\beta_0^2}{1 - \beta_0^2}, \text{ and }
\left| \overline w_{ij}^{[0]} \right| \le \frac{|\beta_0|}{n_P(i)(1 - |\beta_0|)}\left(1 + \frac{\beta_0^2}{1 - \beta_0^2} \right).
\end{eqnarray}
Then we can write
\begin{eqnarray}
\label{decomp 2}
\frac{1}{\sqrt{n^*}}\sum_{i \in N^*} \tilde \varphi_{i,b} v_i^{\textsf{FS}} = \sum_{i \in N_2^\circ} \xi_i^{\textsf{FS}},
\end{eqnarray}
where $\xi_i^{\textsf{FS}} =  (a_i^{\textsf{FS}} \varepsilon_i + \tilde \varphi_{i,b} \eta_i 1\{i \in N^*\} )/\sqrt{n^*}$. By the  Berry-Esseen Lemma (e.g., \citesupp{Shorack:00:ProbStat}, p.259),
\begin{eqnarray}
\label{berry-esseen 2}
\sup_{t \in \mathbf{R}} \Big \vert P\left\{ \sum_{i \in N_2^\circ} \frac{\xi_i^{\textsf{FS}}}{\sigma_{\xi,i}^{\textsf{FS}}} \le t |\mathcal{F}\right\} - \Phi(t) \Big \vert
\le \frac{\displaystyle 9 \mathbf{E}\left[\sum_{i \in N_2^\circ} |\xi_i^{\textsf{FS}}|^3|\mathcal{F}\right]}{\displaystyle \left(\sum_{i \in N_2^\circ} (\sigma_{\xi,i}^{\textsf{FS}})^2\right)^{3/2}},
\end{eqnarray}
where $(\sigma_{\xi,i}^{\textsf{FS}})^2 = \text{Var}(\xi_i^{\textsf{FS}}|\mathcal{F})$.
It suffices to show that the last bound vanishes in probability as $n^* \rightarrow \infty$. Again, since $\varepsilon_i$'s and $\eta_i$'s are independent,
\begin{eqnarray*}
	\sum_{i \in N_2 ^\circ} \sigma_{\xi,i}^2 \ge \sigma_\eta^2 >0.
\end{eqnarray*}
Observe that as in (\ref{eq 5}), for some constant $C_1>0$,
\begin{eqnarray*}
	\mathbf{E}\left[\sum_{i \in N_2^\circ} |\xi_i^{\textsf{FS}}|^3 |\mathcal{F}\right]
	\le \frac{C_1\max_{i \in N} \mathbf{E}[|\varepsilon_i|^3 |\mathcal{F}]}{(n^*)^{3/2}}\sum_{i \in N_2^\circ} |a_i^{\textsf{FS}}|^3
	+ \frac{C_1 n_2^\circ \max_{i \in N}\mathbf{E}[|\eta_i|^3 |\mathcal{F}]}{(n^*)^{3/2}}.
\end{eqnarray*}
Now, as for the leading term, note that by (\ref{bd wij2}), Assumption \ref{assump: limited spillover} and the assumption that $|\tilde \varphi_{i,b}|<C$ for all $i \in N^*$ for some $C>0$, we have for some constant $C_4>0$ that does not depend on $n$, 
\begin{eqnarray*}
	\frac{1}{n^*}\sum_{i \in N_2^\circ} |a_i^{\textsf{FS}}|^3 \le C.
\end{eqnarray*}
Hence, we find that for some $C_1>0$,
\begin{eqnarray*}
	\mathbf{E}\left[\sum_{i \in N^*} |\xi_i^{\textsf{FS}}|^3 |\mathcal{F}\right] \le \frac{C_1}{\sqrt{n^*}} \max_{i \in N} \mathbf{E}[|\varepsilon_i|^3 |\mathcal{F}]
	+  \frac{C_1 n_2^\circ}{(n^*)^{3/2}} \max_{i \in N} \mathbf{E}[|\eta_i|^3 |\mathcal{F}].
\end{eqnarray*}
Thus we conclude that the bound in (\ref{berry-esseen}) is $O_P(1/\sqrt{n^*} + n_2^\circ/(n^*\sqrt{n^*}))$. Since $n^\circ_2 \le C n^*$, we obtain the desired result. $\blacksquare$

\begin{lemma}
	\label{lemma: Lambda 2}
	Suppose that the conditions of Theorem \ref{thm: theta hat exog 2} hold. Then,
	\begin{eqnarray*}
		||S_{\tilde \varphi v}^{\textsf{FS}}||^2 \le \frac{C}{n^*},
	\end{eqnarray*}
	for some constant $C$ that does not depend on $n$.
\end{lemma}
\noindent \textbf{Proof: } Recall the definitions of $e_{ii}^{\mathsf{FS}}$ and $e_{ij}^{\mathsf{FS}}$ in (\ref{e star ii}) and below. Note that
\begin{eqnarray}
\label{bd}
\quad \quad ||\Lambda^{\textsf{FS}}|| &\le& \frac{\sigma_{\varepsilon}^2}{n^*} \sum_{i \in N^*} \sum_{j \in N_{-i}^*: N_{P,2}(i)\cap N_{P,2}(j) \ne \varnothing} |e_{ij}^\textsf{FS}| ||\tilde \varphi_i|| ||\tilde \varphi_j|| \\ \notag
&& + \frac{1}{n^*} \sum_{i \in N^*} (|e_{ii}^\textsf{FS}|\sigma_\varepsilon^2 + \sigma_\eta^2) ||\tilde \varphi_i||^2.
\end{eqnarray}
By (\ref{bd wij2}), we have
\begin{eqnarray}
\label{bd32}
\max_{i \in N^*} |e_{ii}^\textsf{FS}| \le C, \text{ and } \max_{i,j \in N^*: i \ne j} |e_{ij}^\textsf{FS}| \le C,
\end{eqnarray}
for constant $C>0$. Thus, we find that $|e_{ij}^\textsf{FS}| \le C.$ Therefore, both terms on the right hand side of (\ref{bd}) is bounded by $C/n^*$. $\blacksquare$\medskip

Recall the definition of $\tilde v_i^{\textsf{FS}}$ in (\ref{v tilde FS}).

\begin{lemma}
	\label{lemma: v tildes 2}
	Suppose that the conditions of Theorem \ref{thm: theta hat exog 2} hold. Then the following holds.
	\smallskip
	
	\noindent (i) $\frac{1}{n^*} \sum_{i \in N^*} ((\tilde v_i^\mathsf{FS})^2 - (v_i^\mathsf{FS})^2) \tilde \varphi_i \tilde \varphi_i' = O_P(1/\sqrt{n^*}).$
	\smallskip
	
	\noindent (ii) $\frac{1}{n^*} \sum_{i \in N^*} \sum_{j \in N_P(i) \cap N^*} (\tilde v_i^\mathsf{FS} \tilde v_j^\mathsf{FS} - v_i^\mathsf{FS} v_j^\mathsf{FS}) \tilde \varphi_i \tilde \varphi_j'= O_P(1/n^*).$
	\smallskip
	
	\noindent (iii) $\frac{1}{n^*} \sum_{i \in N^*} ((v_i^\mathsf{FS})^2 - \mathbf{E}[(v_i^\mathsf{FS})^2|\mathcal{F}]) \tilde \varphi_i \tilde \varphi_i' = O_P(1/\sqrt{n^*})$.
	\smallskip
	
	\noindent (iv) $\frac{1}{n^*} \sum_{i \in N^*} \sum_{j \in N_P(i)\cap N^*} ( v_i^\mathsf{FS} v_j^\mathsf{FS} - \mathbf{E}[ v_i^\mathsf{FS} v_j^\mathsf{FS}|\mathcal{F}]) \tilde \varphi_i \tilde \varphi_j'= O_P(1/\sqrt{n^*})$.
\end{lemma}

\noindent \textbf{Proof: } (i) Note that
\begin{eqnarray*}
	\left\|\frac{1}{n^*}\sum_{i \in N^*} (\tilde v_i^\textsf{FS} - v_i^\textsf{FS})^2 \tilde \varphi_i \tilde \varphi_i'\right\|
	\le \frac{C}{n^*}\sum_{i \in N^*} (\tilde v_i^\textsf{FS} - v_i^\textsf{FS})^2,
\end{eqnarray*}
for some constant $C>0$. As for the last term, note that for some constant $C>0$,
\begin{eqnarray}
\label{eq 2}
&& \frac{1}{n^*}\sum_{i \in N^*} \mathbf{E}\left[ (\tilde v_i^\textsf{FS} - v_i^\textsf{FS})^2 |\mathcal{F}\right] \le \frac{C}{n^*}\text{tr}(\Lambda^\textsf{FS}) = O_P\left(\frac{1}{n^*}\right),
\end{eqnarray}
by Assumption \ref{assump: nondeg 2} and by Lemma \ref{lemma: Lambda 2}. However, we need to deal with
\begin{eqnarray*}
	\Big \vert \frac{1}{n^*} \sum_{i \in N^*} ((\tilde v_i^\textsf{FS})^2 - (v_i^\textsf{FS})^2) \Big \vert
	\le \sqrt{\frac{1}{n^*} \sum_{i \in N^*} (\tilde v_i^\textsf{FS} - v_i^\textsf{FS})^2}
	\sqrt{\frac{1}{n^*} \sum_{i \in N^*} (\tilde v_i^\textsf{FS} + v_i^\textsf{FS})^2}.
\end{eqnarray*}
Note that
\begin{eqnarray*}
	\frac{1}{n^*} \sum_{i \in N^*} (\tilde v_i^\textsf{FS} + v_i^\textsf{FS})^2 &\le& \frac{2}{n^*} \sum_{i \in N^*} (\tilde v_i^\textsf{FS} - v_i^\textsf{FS})^2 + \frac{8}{n^*} \sum_{i \in N^*} (v_i^\textsf{FS})^2\\
	&=& O_P\left(\frac{1}{n^*}\right) + \frac{8}{n^*} \sum_{i \in N^*} (v_i^\textsf{FS})^2.
\end{eqnarray*}
As for the last term,
\begin{eqnarray*}
	\frac{1}{n^*} \sum_{i \in N^*} \mathbf{E}[(v_i^\textsf{FS})^2 |\mathcal{F}]
	\le \frac{2}{n^*} \sum_{i \in N^*} \mathbf{E}[R_i^\textsf{FS}(\varepsilon)^2 |\mathcal{F}]
	+ \frac{2}{n^*} \sum_{i \in N^*} \mathbf{E}[\eta_i^2 |\mathcal{F}].
\end{eqnarray*}
The last term is bounded by $\sigma_{\eta}^2$, and the first term on the right hand side is bounded by
\begin{eqnarray*}
	\frac{2 \sigma_{\varepsilon}^2 }{n^*} \sum_{i \in N^*} e_{ii}^{\mathsf{FS}} \le C,
\end{eqnarray*}
for some constant $C>0$, by (\ref{bd32}). Combining this with (\ref{eq 2}), we obtain the desired result.
\smallskip

\noindent (ii) Define
\begin{eqnarray*}
	A_{n,1} &=& \frac{1}{n^*} \sum_{i \in N^*} \sum_{j \in N_P(i)\cap N^*}(\tilde v_i^\textsf{FS} - v_i^\textsf{FS})(\tilde v_j^\textsf{FS} - v_j^\textsf{FS})\\
	A_{n,2} &=& \frac{1}{n^*} \sum_{i \in N^*} \sum_{j \in N_P(i)\cap N^*}(\tilde v_i^\textsf{FS} - v_i^\textsf{FS}) v_j^\textsf{FS}, \text{ and }\\
	A_{n,3} &=& \frac{1}{n^*} \sum_{i \in N^*} \sum_{j \in N_P(i)\cap N^*} v_i^\textsf{FS} (\tilde v_j^\textsf{FS} - v_j^\textsf{FS}),
\end{eqnarray*}
and write
\begin{eqnarray*}
	\frac{1}{n^*} \sum_{i \in N^*} \sum_{j \in N_P(i) \cap N^*} (\tilde v_i^\mathsf{FS} \tilde v_j^\mathsf{FS} - v_i^\mathsf{FS} v_j^\mathsf{FS}) 
	= A_{n,1} + A_{n,2}+ A_{n,3}.
\end{eqnarray*}
As for the leading term, by Cauchy-Schwarz inequality,
\begin{eqnarray*}
	|A_{n,1}| = \sqrt{\frac{1}{n^*}\sum_{i \in N^*} (\tilde v_i^\textsf{FS} - v_i^\textsf{FS})^2}\sqrt{\frac{1}{n^*} \sum_{i \in N^*} \left(\sum_{j \in N_P(i)\cap N^*} (\tilde v_j^\textsf{FS} - v_j^\textsf{FS}) \right)^2}.
\end{eqnarray*}
Note that
\begin{eqnarray*}
	&& \frac{1}{n^*} \sum_{i \in N^*} \mathbf{E}\left[\left(\sum_{j \in N_P(i)\cap N^*} (\tilde v_j^\textsf{FS} - v_j^\textsf{FS}) \right)^2 |\mathcal{F} \right]\\
	&\le&
	\frac{1}{n^*} \sum_{i \in N^*} |N_P(i)\cap N^*| \sum_{j \in N_P(i)\cap N^*} \mathbf{E}\left[\left(\tilde v_j^\textsf{FS} - v_j^\textsf{FS}\right)^2 |\mathcal{F} \right]\\
	&=&\frac{1}{n^*} \sum_{i \in N^*} \left(\sum_{j \in N_P(i)\cap N^*} |N_P(j)\cap N^*| \right)\mathbf{E}\left[\left(\tilde v_i^\textsf{FS} - v_i^\textsf{FS}\right)^2 |\mathcal{F} \right],
\end{eqnarray*}
where the equality above uses Lemma \ref{lemma: aux}. Hence the last term is bounded by
\begin{eqnarray*}
	\frac{\max_{i \in N^*}|N_P(i)\cap N^*|^2}{n^*} \sum_{i \in N^*} \mathbf{E}\left[\left(\tilde v_i^\textsf{FS} - v_i^\textsf{FS}\right)^2 |\mathcal{F} \right] 
	\le O_P\left(\frac{1}{n^*}\right),
\end{eqnarray*}
by (\ref{eq 2}). Thus we conclude that $|A_{n,1}| = O_P(1/n^*)$.

Similarly, using Cauchy-Schwarz inequality and applying the same arguments, we have
\begin{eqnarray*}
	|A_{n,2}| = O_P\left(\frac{1}{n^*}\right) \text{ and } |A_{n,3}| = O_P\left(\frac{1}{n^*}\right),
\end{eqnarray*}
obtaining the desired result.\\

\noindent (iii) Note that
\begin{eqnarray*}
	\text{Var} \left(\frac{1}{n^*}\sum_{i \in N^*} R_i^\mathsf{FS}(\varepsilon) |\mathcal{F} \right) 
	\le \frac{1}{(n^*)^2}\sum_{i \in N^*} \mathbf{E}[(R_i^\mathsf{FS}(\varepsilon))^2 |\mathcal{F}] = O_P((n^*)^{-1}),
\end{eqnarray*}
from the proof of (i).
\smallskip

\noindent (iv) The proof is similar to (iii). Hence we omit the details. $\blacksquare$

\begin{lemma}
	\label{lemma: Consistency of Lambda hat 2}
	Suppose that the conditions of Theorem \ref{thm: theta hat exog 2} hold. Then,
	\begin{eqnarray*}
		\hat \Lambda^\mathsf{FS} - \Lambda^\mathsf{FS} = O_P\left(\frac{1}{\sqrt{n^*}}\right).
	\end{eqnarray*}
\end{lemma}

\noindent \textbf{Proof: } We write
\begin{eqnarray*}
	\hat \Lambda_1^\mathsf{FS} - \Lambda_1^\mathsf{FS} &=& \frac{1}{n^*}\sum_{i \in N^*} ((\tilde v_i^\mathsf{FS})^2 - \mathbf{E}[( v_i^\mathsf{FS})^2|\mathcal{F}])\tilde \varphi_i\tilde \varphi_i' \text{ and }\\
	\hat \Lambda_2^\mathsf{FS} - \Lambda_2^\mathsf{FS} &=& \frac{\hat s_{\varepsilon}^\mathsf{FS} - s_\varepsilon^\mathsf{FS}}{n^*}\sum_{i \in N^*}\sum_{j \in N_P(i) \cap N^*}  q_{\varepsilon,ij}^\mathsf{FS} \tilde \varphi_i\tilde \varphi_j'.
\end{eqnarray*}
Thus the desired result follows from Lemma \ref{lemma: v tildes 2}. $\blacksquare$

\begin{lemma}
	\label{lemma: v_hats 2}
	Suppose that the conditions of Theorem \ref{thm: theta hat exog 2} hold. Then the following holds.
	\smallskip
	
	\noindent (i) $\frac{1}{n^*} \sum_{i \in N^*} ((\hat v_i^\mathsf{FS})^2 - (v_i^\mathsf{FS})^2) \tilde \varphi_i \tilde \varphi_i' = O_P(1/n^*).$
	\smallskip
	
	\noindent (ii) $\frac{1}{n^*} \sum_{i \in N^*} \sum_{j \in N_P(i) \cap N^*} (\hat v_i^\mathsf{FS} \hat v_j^\mathsf{FS} - v_i^\mathsf{FS} v_j^\mathsf{FS}) \tilde \varphi_i \tilde \varphi_j'= O_P(1/\sqrt{n^*}).$
\end{lemma}

\noindent \textbf{Proof: } The proof is the same as that of Lemma \ref{lemma: v_hats}. $\blacksquare$
\smallskip

\noindent \textbf{Proof of Theorem \ref{thm: theta hat exog 2}:} The proof is precisely the same as that of Theorem \ref{thm: theta hat exog} except that we use the above auxiliary lemmas instead. Details are omitted. $\blacksquare$

\subsection{Monte Carlo Simulations for Games with the First Order Sophisticated
	Players}

\subsubsection{Simulation Design}
In this section, we investigate the finite sample properties of the
inference for first-order sophisticated types across various configurations
of the payoff graphs $G_{P}$. We generate graphs for the two specifications
models and check our inference under different parameters, described
in the following paragraph. Specification 1 uses an Erdös-Rényi (random
graph formation) payoff graph and Specification 2 uses Barabási-Albert
(preferential attachment) graphs seeded with an Erdös-Rényi graph
of the smallest integer larger than $5\sqrt{n}$. Some summary statistics
of the graphs used for the Monte Carlo study is given in Table \ref{table: FOS graph char}.

For the simulations, we also set the following: 
\begin{align*}
\tau_{i} & =X_{i}'\rho_{0}+\varepsilon_{i},
\end{align*}
where $\rho_{0}=(2,4,1,3,4)'$ and $X_{i}=(X_{i,1},\overline{X}_{i,2})'$,
and 
\begin{eqnarray*}
	\overline{X}_{i,2}=\frac{1}{n_{P}(i)}\sum_{j\in N_{P}(i)}X_{j,2}.
\end{eqnarray*}
We set and $a$ to be a column of ones so that $a'_{0}\rho=14.$ The
variables $\varepsilon$ and $\eta$ are drawn i.i.d. from $N(0,1)$.
The first column of $X_{i,1}$ is a column of ones, while remaining
columns of $X_{i,1}$ are drawn independently from $N(1,1)$. The
columns of $X_{i,2}$ are drawn independently from $N(3,1)$.

For instruments, we consider the following nonlinear transformations
of $X_{1}$ and $X_{2}$: 
\[
\varphi_{i}=[\tilde{Z}_{i,1},X_{i,1}^{2},\overline{X}_{i,2}^{2},\overline{X}_{i,2}^{3}]',
\]
where we define 
\[
\tilde{Z}_{i,1}\equiv\frac{1}{n_{P}(i)}\sum_{j\in N_{P,2}(i)}\lambda_{ij}X_{j,1}.
\]
We generate $Y_{i}$ from the best response function as in (\ref{first-order sophisticated}). We used the Monte Carlo simulation number equal to 5000.

\subsubsection{Results}
The results are found in Tables \ref{table 7}-\ref{table 10}. In Tables \ref{table 7} and \ref{table 8} we report
the finite sample coverage probabilities of the confidence intervals
for $\beta_{0}$ and for $a'\rho_{0}$ respectively. For $\beta_{0}$,
the coverage probabilities perform very well, whereas for $a'\rho_{0}$,
they are conservative. Overall,
for the range of the sample sizes $500-1000$, the finite sample properties
of the inference procedure seem reasonable.

In Tables \ref{table 9}-\ref{table 10}, we report the average length of the confidence intervals.
Clearly, as we increase the sample size from $500$ to $1000$, the
length of the confidence intervals tends to shrink substantially.
This suggests that accummulation of data leads to increased information
and improved accuracy in inference.

\begin{table}
	\caption{\small The Average and Maximum Degrees of Graphs in the Simulations}
	
	\begin{centering}
		
		\small
		\begin{tabular}{cc|cccccc}
			\hline 
			\hline
			&  & \multicolumn{3}{c}{Specification 1} & \multicolumn{3}{c}{Specification 2}\tabularnewline
			$n$ &  & $m=1$ & $m=2$ & $m=3$ & $\lambda=1$ & $\lambda=2$ & $\lambda=3$\tabularnewline
			\hline 
			$500$ & $d_{mx}$ & 17 & 21 & 30 & 5 & 8 & 11\tabularnewline
			& $d_{av}$ & 1.7600 & 3.2980 & 4.8340 & 0.9520 & 1.9360 & 2.9600\tabularnewline
			\hline 
			$1000$ & $d_{mx}$ & 18 & 29 & 34 & 6 & 7 & 9\tabularnewline
			& $d_{av}$ & 1.8460 & 3.5240 & 5.2050 & 0.9960 & 1.9620 & 3.0020\tabularnewline
			\hline 
			& \multicolumn{1}{c}{} &  &  &  &  &  & \tabularnewline
		\end{tabular}
		\par\end{centering}
	\parbox[c]{6.2in}{ \footnotesize
		Notes:  $d_{av}$ and $d_{mx}$ represent the average
		and maximum degrees of the networks respectively; that is, $d_{av}\equiv\frac{1}{n}\sum_{i\in N}n_{P}(i)$
		and $d_{mx}\equiv\max_{i\in N}n_{P}(i)$.}
	\label{table: FOS graph char}
\end{table}

\noindent \begin{center}
	\begin{table}
		\caption{\small The Empirical Coverage Probability of Confidence Intervals for $\beta_{0}$
			from First-Order Sophisticated Types}
		
		\begin{centering}
			\small
			\begin{tabular}{cc|cccccc}
				\hline 
				\hline
			 & \multicolumn{1}{c}{} & \multicolumn{3}{c}{Specification 1} & \multicolumn{3}{c}{Specification 2}\tabularnewline
			 \cline{3-8} \cline{4-8} \cline{5-8} \cline{6-8} \cline{7-8} \cline{8-8} 
			$\beta_{0}$ & \multicolumn{1}{c|}{} & $m=1$ & $m=2$ & $m=3$ & $\lambda=1$ & $\lambda=2$ & $\lambda=3$\tabularnewline
			\hline 
			$-0.5$ & $n=500$ & 0.9634 & 0.9566  & 0.9562  & 0.9576  & 0.9528 & 0.9586\tabularnewline
			& $n=1000$ & 0.9542  & 0.9552 & 0.9566  & 0.9566  & 0.9526 & 0.9558\tabularnewline
			\hline 
			$-0.3$ & $n=500$ & 0.9586  & 0.9542 & 0.9536  & 0.9526  & 0.9534  & 0.9570\tabularnewline
			& $n=1000$ & 0.9490  & 0.9518 & 0.9546 & 0.9508  & 0.9506  & 0.9542\tabularnewline
			\hline 
			$0$ & $n=500$ & 0.9576  & 0.9532  & 0.9548  & 0.9478 & 0.9530  & 0.9552\tabularnewline
			& $n=1000$ & 0.9502  & 0.9522  & 0.9530  & 0.9462  & 0.9474  & 0.9516 \tabularnewline
			\hline 
			$0.3$ & $n=500$ & 0.9652  & 0.9606 & 0.9582  & 0.9470 & 0.9514  & 0.9554 \tabularnewline
			& $n=1000$ & 0.9548 & 0.9520  & 0.9548 & 0.9454  & 0.9494  & 0.9530 \tabularnewline
			\hline 
			$0.5$ & $n=500$ & 0.9710 & 0.9658 & 0.9614 & 0.9502 & 0.9528 & 0.9584\tabularnewline
			& $n=1000$ & 0.9600 & 0.9570 & 0.9578 & 0.9474 & 0.9494 & 0.9566\tabularnewline
			\hline 
			& \multicolumn{1}{c}{} &  &  &  &  &  & \tabularnewline
			\end{tabular}
			\par\end{centering}
		\medskip{}
		
		\parbox[c]{6.2in}{ \footnotesize
			Notes: This table shows the empirical coverage probabilities
			$R=5000$ of the confidence intervals for $\beta_{0}$ under two models
			of graph formation. The nominal size is $\alpha=0.05$. As expected,
			the coverage probabilities are close to the nominal size.} 
		\label{table 7}
	\end{table}
	\par\end{center}

\begin{center}
	\begin{table}
		\caption{\small The Empirical Coverage Probability of Confidence Interval for $a'\rho_{0}$
			for First-Order Sophisticated Types}
		
		\begin{centering}
			\small
			\begin{tabular}{cc|cccccc}
				\hline 
				\hline
				& \multicolumn{1}{c}{} & \multicolumn{3}{c}{Specification 1} & \multicolumn{3}{c}{Specification 2}\tabularnewline
				\cline{3-8} \cline{4-8} \cline{5-8} \cline{6-8} \cline{7-8} \cline{8-8} 
				$\beta_{0}$ & \multicolumn{1}{c|}{} & $m=1$ & $m=2$ & $m=3$ & $\lambda=1$ & $\lambda=2$ & $\lambda=3$\tabularnewline
				\hline 
				$-0.5$ & $n=500$ & 0.9906  & 0.9928  & 0.9900  & 0.9896  & 0.9896  & 0.9910\tabularnewline
				& $n=1000$ & 0.9856 & 0.9872 & 0.9872 & 0.9836 & 0.9846 & 0.9896\tabularnewline
				\hline 
				$-0.3$ & $n=500$ & 0.9874 & 0.9878 & 0.9880 & 0.9830  & 0.9862  & 0.9874 \tabularnewline
				& $n=1000$ & 0.9802 & 0.9816 & 0.9862 & 0.9740 & 0.9786 & 0.9852\tabularnewline
				\hline 
				$0$ & $n=500$ & 0.9814  & 0.9848  & 0.9854 & 0.9760 & 0.9808  & 0.9848 \tabularnewline
				& $n=1000$ & 0.9714 & 0.9806 & 0.9812 & 0.9568 & 0.9720 & 0.9772\tabularnewline
				\hline 
				$0.3$ & $n=500$ & 0.9840 & 0.9856  & 0.9872 & 0.9644  & 0.9796  & 0.9828\tabularnewline
				& $n=1000$ & 0.9710 & 0.9810 & 0.9796 & 0.9488 & 0.9616 & 0.9766\tabularnewline
				\hline 
				$0.5$ & $n=500$ & 0.9842  & 0.9880 & 0.9886  & 0.9496  & 0.9750  & 0.9836\tabularnewline
				& $n=1000$ & 0.9650 & 0.9784 & 0.9820 & 0.9456 & 0.9526 & 0.9750\tabularnewline
				\hline 
				& \multicolumn{1}{c}{} &  &  &  &  &  & \tabularnewline
			\end{tabular}
			\par\end{centering}
		\parbox[c]{6.2in}{ \footnotesize
			Notes: This table shows the empirical coverage
			probabilities $R=5000$ of the confidence intervals for $a'\rho_{0}$
			under two models of graph formation. The nominal size is $\alpha=0.05$.
			The procedure is conservative, as expected from the Bonferroni procedure.}
		\label{table 8}
	\end{table}
	\par\end{center}

\begin{table}
	\caption{\small The Average Length of Confidence Intervals for $\beta_{0}$ for First-Order
		Sophisticated Types}
	
	\begin{centering}
		\small
		\begin{tabular}{cc|cccccc}
			\hline 
			\hline
			& \multicolumn{1}{c}{} & \multicolumn{3}{c}{Specification 1} & \multicolumn{3}{c}{Specification 2}\tabularnewline
			\cline{3-8} \cline{4-8} \cline{5-8} \cline{6-8} \cline{7-8} \cline{8-8} 
			$\beta_{0}$ & \multicolumn{1}{c|}{} & $m=1$ & $m=2$ & $m=3$ & $\lambda=1$ & $\lambda=2$ & $\lambda=3$\tabularnewline
			\hline 
			$-0.5$ & $n=500$ & 0.1171  & 0.2867  & 0.3663 & 0.0627  & 0.1040  & 0.2124\tabularnewline
			& $n=1000$ & 0.0829 & 0.2271 & 0.3321 & 0.0366 & 0.0614 & 0.1333\tabularnewline
			\hline 
			$-0.3$ & $n=500$ & 0.0924  & 0.1442  & 0.1980  & 0.0658 & 0.0860  & 0.1409 \tabularnewline
			& $n=1000$ & 0.0653 & 0.0965 & 0.1299 & 0.0384 & 0.0513 & 0.0831\tabularnewline
			\hline 
			$0$ & $n=500$ & 0.0810 & 0.1041  & 0.1277  & 0.0662  & 0.0726  & 0.1040 \tabularnewline
			& $n=1000$ & 0.0545 & 0.0653 & 0.0781 & 0.0386 & 0.0420 & 0.0587\tabularnewline
			\hline 
			$0.3$ & $n=500$ & 0.0761  & 0.0972  & 0.1165  & 0.0696  & 0.0767 & 0.1030 \tabularnewline
			& $n=1000$ & 0.0504 & 0.0599 & 0.0704 & 0.0415 & 0.0456 & 0.0603\tabularnewline
			\hline 
			$0.5$ & $n=500$ & 0.0698  & 0.0773 & 0.1008 & 0.0223  & 0.0386 & 0.0728 \tabularnewline
			& $n=1000$ & 0.0198 & 0.0479 & 0.0643 & 0.0068 & 0.0168 & 0.0468\tabularnewline
			\hline 
			& \multicolumn{1}{c|}{} &  &  &  &  &  & \tabularnewline
		\end{tabular}
		\par\end{centering}
	\parbox[c]{6.2in}{ \footnotesize
		Notes: This table shows the average length of confidence
		intervals for $\beta_{0}$ for two models of graph formation $R=5000$.
		The nominal size is $\alpha=0.05$. As expected the average length
		of the confidence interval falls with $n$.}
	\label{table 9}
\end{table}
\bigskip
\bigskip
\bigskip
\bigskip
\bigskip

\bigskip
\bigskip
\bigskip
\bigskip
\bigskip

\begin{center}
	\begin{table}
		\caption{\small Average Length of of Confidence Intervals for $a'\rho_{0}$ for First-Order
			Sophisticated Types}
		
		\begin{centering}
			\small
			\begin{tabular}{cc|cccccc}
				\hline 
				\hline
				  & \multicolumn{1}{c}{} & \multicolumn{3}{c}{Specification 1} & \multicolumn{3}{c}{Specification 2}\tabularnewline
				  \cline{3-8} \cline{4-8} \cline{5-8} \cline{6-8} \cline{7-8} \cline{8-8} 
				 $\beta_{0}$ & \multicolumn{1}{c|}{} & $m=1$ & $m=2$ & $m=3$ & $\lambda=1$ & $\lambda=2$ & $\lambda=3$\tabularnewline
				 \hline 
				 $-0.5$ & $n=500$ & 1.5489  & 1.4912  & 1.6440 & 0.6448  & 1.0615  & 2.1044 \tabularnewline
				 & $n=1000$ & 1.2581 & 1.1732 & 1.2974 & 0.4196 & 0.6760 & 1.2460\tabularnewline
				 \hline 
				 $-0.3$ & $n=500$ & 1.9736  & 2.3093  & 2.6460  & 0.8698  & 1.4064  & 2.9878 \tabularnewline
				 & $n=1000$ & 1.5288 & 1.6772 & 1.8918 & 0.5571 & 0.8749 & 1.6194\tabularnewline
				 \hline 
				 $0$ & $n=500$ & 2.3591  & 2.7335  & 3.1419 & 1.1100  & 1.6186  & 3.2359\tabularnewline
				 & $n=1000$ & 1.6733 & 1.8155 & 2.0145 & 0.6959 & 0.9779 & 1.6898\tabularnewline
				 \hline 
				 $0.3$ & $n=500$ & 2.2385  & 2.6666  & 3.1282  & 1.3922 & 1.8623 & 3.2965 \tabularnewline
				 & $n=1000$ & 1.6314 & 1.8048 & 2.0041 & 0.8720 & 1.1455 & 1.8393\tabularnewline
				 \hline 
				 $0.5$ & $n=500$ & 3.1651  & 2.1409  & 2.5747  & 0.7793  & 1.2397 & 2.3673 \tabularnewline
				 & $n=1000$ & 1.0193 & 1.5059 & 1.7956 & 0.3911 & 0.7031 & 1.5259\tabularnewline
				 \hline 
				 & \multicolumn{1}{c}{} &  &  &  &  &  & \tabularnewline
			\end{tabular}
			\par\end{centering}
		\parbox[c]{6.2in}{ \footnotesize
			Notes: The true $a'\rho_{0}$ is equal to 14. The
			length of confidence intervals tends to be small and substantially
			shortened as the size of the network increases.} 
		\label{table 10}
	\end{table}
	\par\end{center}

\section{Model Selection between Games $\Gamma_0$ and $\Gamma_1$}

It is a matter of econometric model specification to choose between $\Gamma_0$ with simple-type agents or $\Gamma_1$ with the first-order sophisticated type agents as an empirical model. Both models are distinct and nonnested. Here we provide an empirical procedure to select among the two models.\footnote{Note that the two models $\Gamma_0$ and $\Gamma_1$ may not be exclusive of each other because there can be a data generating process such that the payoff graph $G_P$ is a cluster structure where each cluster is a complete graph (so that $N_{P,2}(i) = N_P(i)$), or $\beta_0 = 0$.}

First, we write $v_i(\beta_0)$, $v_i^{\mathsf{FS}}(\beta_0)$, and $\tilde \varphi_i(\beta_0)$ in place of $v_i$, $v_i^{\mathsf{FS}}$, and $\tilde \varphi_i$ to make their dependence on $\beta_0$ explicit. Let $B$ be a set contained in $(-1,1)$ and assumed to contain the true parameter $\beta_0$, and define
\begin{eqnarray*}
	T_{\mathsf{ST}} = \inf_{\beta \in B} T(\beta), \text{ and } T_{\mathsf{FS}} = \inf_{\beta \in B} T^{\mathsf{FS}}(\beta),
\end{eqnarray*}
where $T(\beta_0)$ is as defined in (\ref{Pop Obj}) and $T^{\mathsf{FS}}(\beta_0)$ is similarly defined after (\ref{lambda FS}). Then, we consider the set
\begin{eqnarray*}
	\hat S = \{s \in \{\mathsf{ST},\mathsf{FS}\}: T_s \le c_{1 - \alpha/2}\},
\end{eqnarray*}
where $c_{1 - \alpha/2}$ denotes the $(1-\alpha/2)$-percentile of the distribution of $\chi_{M-d}^2$. Among the two models based on games $\Gamma_0$ and $\Gamma_1$, the set $\hat S$ is the set of the models that are not rejected at $100(1-\alpha)\%$ in the sense to be explained below.

Define
\begin{eqnarray*}
	m_{\mathsf{ST}} &=& \inf_{\beta \in B}\left|\frac{1}{n^*}\sum_{i \in N^*}\mathbf{E}\left[v_i(\beta)|\mathcal{F} \right] \tilde \varphi_i(\beta) \right|, \text{ and }\\
	m_{\mathsf{FS}} &=& \inf_{\beta \in B}\left|\frac{1}{n^*}\sum_{i \in N^*}\mathbf{E}\left[v_i^{\mathsf{FS}}(\beta)|\mathcal{F} \right] \tilde \varphi_i(\beta) \right|.
\end{eqnarray*}
Let $S_0 = \{s \in \{\mathsf{ST},\mathsf{FS}\}: m_s =0\}$. Hence $S_0$ denotes the collection of true models (as distinguished by the moment condition $m_s = 0, s \in \{\mathsf{ST},\mathsf{FS}\}$.) Let $\mathcal{P}_n$ be the collection of the joint distributions of all the observables in the data. For each $\delta>0$, let us define
\begin{eqnarray*}
	\mathcal{P}_{n,1}(\delta) &=& \{P \in \mathcal{P}_n: m_{\mathsf{ST}} = 0, m_{\mathsf{FS}} > \delta \}\\
	\mathcal{P}_{n,2}(\delta) &=& \{P \in \mathcal{P}_n: m_{\mathsf{ST}} > \delta, m_{\mathsf{FS}} =0 \}\\
	\mathcal{P}_{n,3}(\delta) &=& \{P \in \mathcal{P}_n: m_{\mathsf{ST}} = 0, m_{\mathsf{FS}} = 0\}, \text{ and }\\
	\mathcal{P}_{n,4}(\delta) &=& \{P \in \mathcal{P}_n: m_{\mathsf{ST}} > \delta, m_{\mathsf{FS}} > \delta \},
\end{eqnarray*} 
and define
\begin{eqnarray*}
	\mathcal{P}_n(\delta) = \bigcup_{k=1}^4 \mathcal{P}_{n,k}(\delta).
\end{eqnarray*}
Then, the selection rule $\hat S$ can be justified as follows: for each $\delta>0$,
\begin{eqnarray*}
	\liminf_{n \rightarrow \infty} \inf_{P \in \mathcal{P}_n(\delta)} P\{S_0 = \hat S\} \ge 1 - \alpha.
\end{eqnarray*}
We can also make the selection rule a consistent selection rule, by choosing $\alpha = \alpha_n$ to be a sequence so that $c_{1-\alpha_n/2} \rightarrow \infty$ but slowly at a proper rate.

The procedure can be modified to perform model selection with other combinations of the models as long as a testing procedure for moment conditions from each model is available. For example, suppose that $m_{\mathsf{EQ}}=0$ is a moment condition for a complete information game model with equilibrium strategies and a consistent testing procedure (at level $\alpha$) for this moment condition is given by $1\{T_{\mathsf{EQ}} > c_{1 - \alpha}^{\mathsf{EQ}}\}$ for some test statistic $T_{\mathsf{EQ}}$ and critical value $c_{1 - \alpha}^{\mathsf{EQ}}$. Then, one can replace $T_{\mathsf{FS}}$ and $c_{1 - \alpha/2}$ by $T_{\mathsf{EQ}}$ and $c_{1 - \alpha/2}^{\mathsf{EQ}}$ in the previous procedure to select a set of models from $\{\mathsf{ST},\mathsf{EQ}\}$ that are not rejected at $100(1-\alpha)\%$.

Let us provide conditions and a brief proof for the asymptotic justification of the model selection procedure. For brevity, we will provide high level conditions and discussions on how they can be verified using low level conditions. 

Let us first define for each $\delta>0$:
\begin{eqnarray*}
	p_{n,1}(\delta) &=& \inf_{P \in P_{n,1}(\delta)} P \left\{ T_{\mathsf{ST}} \le c_{1 - \alpha/2}, T_{\mathsf{FS}} > c_{1 - \alpha/2} \right\}\\
	p_{n,2}(\delta) &=& \inf_{P \in P_{n,2}(\delta)} P \left\{ T_{\mathsf{ST}} > c_{1 - \alpha/2}, T_{\mathsf{FS}} \le c_{1 - \alpha/2} \right\}\\
	p_{n,3}(\delta) &=& \inf_{P \in P_{n,3}(\delta)} P \left\{ T_{\mathsf{ST}} \le c_{1 - \alpha/2}, T_{\mathsf{FS}} \le c_{1 - \alpha/2} \right\}, \text{ and } \\
	p_{n,4}(\delta) &=& \inf_{P \in P_{n,4}(\delta)} P \left\{ T_{\mathsf{ST}} > c_{1 - \alpha/2}, T_{\mathsf{FS}} > c_{1 - \alpha/2} \right\}.
\end{eqnarray*}
Then, we make the following assumption:
\begin{align}
\label{assump: model sel}
\min\left\{\liminf_{n \rightarrow \infty} p_{n,1}(\delta),\liminf_{n \rightarrow \infty} p_{n,2}(\delta)\right\} &\ge 1 - \alpha/2,\\ \notag
\liminf_{n \rightarrow \infty} p_{n,2}(\delta) &\ge 1 - \alpha,\\ \notag
p_{n,2}(\delta) &\rightarrow 1, \text{ as } n \rightarrow \infty.
\end{align}
The assumptions in (\ref{assump: model sel}) follow if the tests $1\{T^\mathsf{ST}(\beta_0)>c_{1-\alpha/2}\}$ and $1\{T^\mathsf{FS}(\beta_0)>c_{1-\alpha/2}\}$ are asymptotically valid uniformly over the probabilities that satisfy the respective moment conditions $m_s = 0$, and if the tests are consistent under fixed alternatives (i.e., $m_s >\delta$). The uniform validity of the tests can be proved by invoking the uniform boundedness of certain moments and eigenvalues of the variance matrices, and by using Berry-Esseen Lemma. As such arguments are standard, details are omitted here. 

Under the assumptions in (\ref{assump: model sel}), it is not hard to see that
\begin{eqnarray*}
	\liminf_{n \rightarrow \infty} \inf_{P \in \mathcal{P}_n(\delta)} P\{S_0 = \hat S\} \ge 1 - \alpha.
\end{eqnarray*}
Indeed, noting that $\mathcal{P}_n(\delta)$ is partitioned into $\mathcal{P}_{n,k}(\delta), k = 1,...,4$, we can write
\begin{eqnarray*}
	\liminf_{n \rightarrow \infty} \inf_{P \in \mathcal{P}_n(\delta)} P\{S_0 = \hat S\} 
	&=& \liminf_{n \rightarrow \infty} \min_{1 \le k \le 4} \inf_{P \in \mathcal{P}_{n,k}(\delta)} P\{S_0 = \hat S\}\\
	&=& \min_{1 \le k \le 4} \liminf_{n \rightarrow \infty} \inf_{P \in \mathcal{P}_{n,k}(\delta)} P\{S_0 = \hat S\}.
\end{eqnarray*}
The assumptions in (\ref{assump: model sel}) tell us that the last term is bounded from below by $1-\alpha$.

\section{Testing for Information Sharing on Unobservables}
\subsection{The Model with Simple Type Players}\label{testing IU}
One may want to see how much empirical relevance there is for incorporating information sharing on unobservables. Here we explain how one can performa a formal test of information sharing for the case of $\beta_0 \ne 0$. Observe that when $\beta_0 =0$, presence of information sharing on unobservables is not testable. When $\beta_0 = 0$, it follows that
\begin{eqnarray*}
	s_i^{[0]}(\mathcal{I}_{i,0}) = X_i'\rho_0 + v_i,
\end{eqnarray*}
where $v_i = \varepsilon_i + \eta_i$. In this case, it is not possible to distinguish between contributions from $\varepsilon_i$ and $\eta_i$.

Consider the following hypotheses:
\begin{eqnarray*}
	H_0 : \sigma_\varepsilon^2 = 0, \text{ and }
	H_1 : \sigma_\varepsilon^2 > 0,
\end{eqnarray*}
where we recall the definition $\sigma_{\varepsilon}^2 = \text{Var}(\varepsilon_i^2|\mathcal{F})$. The null hypothesis tells us that there is no information sharing on unobservables. Let $\hat v_i (\beta)$, $a_\varepsilon(\beta)$ and $b_\varepsilon(\beta)$ be the same as $\hat v_i$, $a_\varepsilon$ and $b_\varepsilon$ (defined in Appendix D) only with $\beta_0$ replaced by generic $\beta$. From here on we assume that $\beta_0 \ne 0$.

The main idea for testing the hypothesis is that when $\sigma_\varepsilon^2>0$, this implies cross-sectional dependence of residuals $v_i$. For testing, we need to compute the sample version of the covariance between $v_i$ and $v_j$ for $G_P$-neighbors $i$ and $j$. However, Condition C alone does not guarantee that for each $i \in N^*$, we will be able to compute $\hat v_j$ for some $j \in N_P(i)$, because there may not exist such $j$ for some $i \in N^*$ at all. Thus let us introduce an additional data requirement as follows:
\medskip

\noindent \textbf{Condition D:} For each $i \in N^*$, the econometrician observes a nonempty subset $\widetilde{N}(i) \subset N_P(i)$ (possibly a singleton) of agents where for each $j \in \widetilde{N}(i)$, the econometrician observes $Y_j$, $|N_P(j) \cap N_P(k)|$, $n_P(k)$ and $X_k$ for all $k \in N_P(j)$.
\medskip

Condition D is satisfied if there are many agents in the data set where each agent has at least one $G_P$-neighbor $j$ for which the econometrician observes the outcome $Y_j$, the number of their $G_P$-neighbors, the observed characteristics of their $G_P$-neighbors, and the number of the agents who are both their $G_P$-neighbors and the neighbors of their $G_P$-neighbors. The asymptotic validity of inference is not affected if the researcher chooses a nonempty subset $\widetilde{N}(i)$ in Condition D as a singleton subset, say, ${j(i)} \subset N_P(i)$, $j(i) \in N$, such that we observe $Y_{j(i)}$, $|N_P(j(i)) \cap N_P(k)|$, $n_P(k)$ and $X_k$ for all $k \in N_P(j(i))$ are available in the data, so far as the choice is not based on $Y_i$'s but on $X$ only. While this data requirement can still be restrictive in some cases where one obtains a partial observation of $G_P$, it is still weaker than the usual assumption that the econometrician observes $G_P$ fully together with $(Y_i,X_i')_{i \in N}$.

Now let us reformulate the null and the alternative hypotheses as follows:
\begin{eqnarray}
\label{hyp}
H_0 &:& \frac{1}{n^*} \sum_{i \in N^*} \sum_{j \in \widetilde{N}(i)} \mathbf{E}[v_i v_j|\mathcal{F}] = 0, \text{ and }\\
H_1 &:& \frac{1}{n^*} \sum_{i \in N^*} \sum_{j \in \widetilde{N}(i)} \mathbf{E}[v_i v_j|\mathcal{F}] \ne 0. 
\end{eqnarray}

For testing, we propose the following method. Let $C_{1-(\alpha/2)}^\beta$ be the $(1-(\alpha/2))$-level confidence interval for $\beta$. We consider the following test statistics:
\begin{eqnarray*}
	\widehat{IU} = \inf_{\beta \in C_{1-(\alpha/2)}^\beta} \frac{1}{2 \hat S^4(\beta) n^*}\left(\sum_{i \in N^*} \sum_{j \in \widetilde{N}(i)} \hat v_i (\beta) \hat v_j (\beta)\right)^2,
\end{eqnarray*}
where
\begin{eqnarray*}
	\hat S^2(\beta) = \frac{\widetilde{d}_{av}^{1/2}}{n^*}\sum_{i \in N^*} \hat v_i^2(\beta), \text{ and }
	\widetilde{d}_{av} = \frac{1}{n^*}\sum_{i \in N^*} |\widetilde{N}(i)|.
\end{eqnarray*}
When the confidence set includes zero, the power of the test becomes asymptotically trivial, as expected from the previous remark that information sharing on unobservables is not testable when $\beta_0 =0$.

As for the critical value, we take the $(1- (\alpha/2))$-percentile from the $\chi^2$ distribution with degree of freedom 1, which we denote by $c_{1- (\alpha/2)}$. Then the level $\alpha$-test based on the test statistic $\widehat{IU}$ rejects the null hypothesis if and only if $\widehat{IU} > c_{1- (\alpha/2)}$.

\begin{theorem}
	\label{thm: inform sharing}
	Suppose that the conditions of Theorem \ref{thm: best response} and Assumptions \ref{assump: unobserved heterogeneity} - \ref{assump: degree and moment} hold. Then, under the null hypothesis in (\ref{hyp}),
	\begin{eqnarray*}
		\lim_{n^* \rightarrow \infty} P\left\{ \widehat{IU} > c_{1 - \alpha/2} \right\} \le \alpha,
	\end{eqnarray*}
	as $n^* \rightarrow \infty$.
\end{theorem}

\noindent \textbf{Proof:} First, note that
\begin{eqnarray*}
	\frac{1}{\sqrt{n^*}}\sum_{i \in N^*} \sum_{j \in \widetilde{N}(i)} (\hat v_i \hat v_j - v_i v_j) = O_P(1/\sqrt{n^*}),
\end{eqnarray*}
by following precisely the same proof as that of Lemma \ref{lemma: v tildes}(ii). (Recall that $\widetilde N(i)$ is defined in Condition D in Section \ref{testing IU}.) Now, we let
\begin{eqnarray*}
	\sigma^2 = \text{Var}\left( \frac{1}{\sqrt{n^*}} \sum_{i \in N^*} \sum_{j \in \widetilde{N}(i)} \eta_i \eta_j |\mathcal{F}\right)
\end{eqnarray*}
and write
\begin{eqnarray*}
	\frac{1}{\sigma \sqrt{n^*}}\sum_{i \in N^*} \sum_{j \in \widetilde{N}(i)} v_i v_j
	= \frac{1}{\sqrt{n^*}} \sum_{i \in N^*} r_i,
\end{eqnarray*}
where
\begin{eqnarray*}
	r_i = \frac{1}{\sigma} \sum_{j \in \widetilde{N}(i)} \eta_i \eta_j,
\end{eqnarray*}
because $v_i = \eta_i$ under the null hypothesis. Note that $\mathbf{E}[r_i|\mathcal{F}] = 0$. Let $G_P^*$ be a graph on $N^*$ such that $i$ and $j$ are adjacent if and only if $j \in \widetilde{N}(i)$ or $i \in \widetilde{N}(j)$. Then $\{r_i\}_{i \in N^*}$ has $G_P^*$ as a dependency graph conditional on $\mathcal{F}$. Now we show the following:
\begin{eqnarray}
\label{rate cond 2}
(n^{*})^{-1/4}\sqrt{\mu_3^3} + (n^{*})^{-1/2} \mu_4^2 \rightarrow_P 0,
\end{eqnarray}
where for $p \ge 1$,
\begin{eqnarray*}
	\mu_p = \max_{i \in N^*} \left(\mathbf{E}[|r_i|^p|\mathcal{F}] \right)^{1/p}.
\end{eqnarray*}
Then by Theorem 2.3 of \citesupp{Penrose:03:RandomGeometricGraphs}, we obtain that
\begin{eqnarray*}
	\frac{1}{\sigma \sqrt{n^*}}\sum_{i \in N^*} \sum_{j \in \widetilde{N}(i)} v_i v_j \rightarrow_d N(0,1),
\end{eqnarray*}
as $n^* \rightarrow \infty$.
First, note that
\begin{eqnarray*}
	\sigma^2 &=& \mathbf{E}\left( \left(\frac{1}{\sqrt{n^*}} \sum_{i \in N^*} \sum_{j \in \widetilde{N}(i)} \eta_i \eta_j \right)^2 |\mathcal{F}\right)	\\
	&=& \frac{1}{\sqrt{n^*}} \sum_{i_1 \in N^*} \sum_{j_1 \in \widetilde{N}(i_1)} \sum_{i_2 \in N^*} \sum_{j_2 \in \widetilde{N}(i_2)} \mathbf{E}\left[ \eta_{i_1} \eta_{j_1} \eta_{i_2} \eta_{j_2} |\mathcal{F}\right].
\end{eqnarray*}
Note that in the quadruple sum, $i_1 \ne j_1$ and $i_2 \ne j_2$. There are only two ways the last conditional expectation is not zero: either $i_1 = i_2$ and $j_1 = j_2$ or $j_1 = i_2$ and $i_1 = j_2$, because $\eta_i$'s are independent across $i$'s and its conditional expectation given $\mathcal{F}$ is zero. Hence the last term is equal to
\begin{eqnarray}
\label{expr2}
\frac{2 \sigma_{\eta}^4}{n^*} \sum_{i \in N^*} |\widetilde{N}(i)| = 2 \sigma_{\eta}^4 \widetilde{d}_{av}
\end{eqnarray}
Hence for any $p \ge 2$,
\begin{eqnarray*}
	\mu_p^p =	\frac{1}{\sigma^p} \max_{i \in N^*} \mathbf{E}\left[\left|\sum_{j \in \widetilde{N}(i)} \eta_i \eta_j\right|^p|\mathcal{F}\right] &\le &\frac{\max_{i,j \in N^*} \mathbf{E}[|\eta_i \eta_j|^p|\mathcal{F}]}{\sigma^p}\\
	&\le& \frac{\max_{i,j \in N^*} \mathbf{E}[|\eta_i \eta_j|^p|\mathcal{F}]}{ 2^p \sigma_{\eta}^{2p} \widetilde d_{av}^p}.
\end{eqnarray*}
Note that $\widetilde{d}_{av} \ge 1$ because $\widetilde N(i) \ne \varnothing$ for all $i \in N^*$. Thus (\ref{rate cond 2}) follows. Now, by Lemma \ref{lemma: v tildes}, and in the light of the expression (\ref{expr2}), it is not hard to see that
\begin{eqnarray*}
	2 \hat S^4(\beta_0) = \sigma^2 + o_P(1).
\end{eqnarray*}
The desired result follows from this and the Bonferroni procedure. $\blacksquare$

\subsection{The Model with First Order Sophisticated Players}
Let us develop a test for information sharing on unobservables when the game is populated by the first order sophicated players. When $\beta_0 = 0$, it follows that
\begin{eqnarray*}
	s_{i}^{[1]}(\mathcal{I}_{i,1}) = X_i'\rho_0 + v_i^{\textsf{FS}},
\end{eqnarray*}
where $v_i^{\textsf{FS}} = \varepsilon_i + \eta_i$. Therefore, just as in the case of a simple type model, it is not possible to distinguish between contributions from $\varepsilon_i$ and $\eta_i$. Thus let us assume that $\beta_0 \ne 0$. The presence of cross-sectional correlation of residuals $v_i^{\textsf{FS}}$ serves as a testable implications from information sharing on unobservables. As in the case of a model with agents of simple type, we need to strengthen Condition D as follows:\medskip

\noindent \textbf{Condition D1:} For each $i \in N^*$, the econometrician observes a nonempty subset $\widetilde N(i) \subset N_P(i)$ (possibly a singleton) of agents where for each $j \in \tilde N(i)$, the econometrician observes $Y_j$, $|N_P(j) \cap N_P(k)|$, $n_P(k)$ and $X_k$ for all $k \in N_{P,2}(j)$.\medskip

Similarly as before, we consider the following test statistics:
\begin{eqnarray*}
	\widehat{IU}^{\mathsf{FS}} = \inf_{\beta \in C_{1-(\alpha/2)}^\beta} \frac{1}{2 (\hat S^{\mathsf{FS}}(\beta))^4 n^*}\left(\sum_{i \in N^*} \sum_{j \in \widetilde{N}(i)} \hat v_i^{\mathsf{FS}} (\beta) \hat v_j^{\mathsf{FS}} (\beta)\right)^2,
\end{eqnarray*}
where
\begin{eqnarray*}
	(\hat S^{\mathsf{FS}}(\beta))^2 = \frac{\widetilde{d}_{av}^{1/2}}{n^*}\sum_{i \in N^*} \hat v_i^2(\beta).
\end{eqnarray*}
As before, we reject the null hypothesis of no information sharing on unobservables if and only if $\widehat{IU}^{\mathsf{FS}} > c_{1-(\alpha/2)}$, where $c_{1-(\alpha/2)}$ is the $(1-(\alpha/2))$-percentile of $\chi^2_1$. Asymptotic validity of this procedure can be shown in a similar manner as for the case of simple types.

\section{Convergence of Behavioral Strategies to Equilibrium Strategies}

In this section, we prove Theorem \ref{thm: conv}.\medskip

\noindent \textbf{Proof of Theorem \ref{thm: conv}}: Our proof is in two steps. First, we show the convergence of the behavioral strategies to the equilibrium strategies in a game without private information (without $\eta_i$). In the second step, we use this first result and show it also holds when we extend the game to allow for $\eta_i$.

Let us first consider the game without private information (i.e., $\eta_i = 0$ for all $i \in N$). We denote the behavioral strategies in this complete information game by $\tilde{s}_i^{[m]}$, and the equilibrium strategies as $\tilde{s}_i^{\mathsf{BNE}}$. This notation will allow us to differentiate these strategies from the case with incomplete information. From Theorem \ref{thm: best response2}, without $\eta_i$'s, we have that: 

\begin{eqnarray}
\label{der3}
\tilde{s}_i^{[m+1]}(\mathcal{I}_{i,m+1}) &=& \left(\frac{\beta_{0}}{n_{P}(i)}\sum_{k\in N_{P}(i)}w_{ki}^{[m]}+1\right)\tau_{i}\\ \notag
&&+\frac{\beta_{0}}{n_{P}(i)}\sum_{k\in N_{P}(i)}\sum_{j\in N_{P,m+2}(i)}w_{kj}^{[m]}1\{j\in\overline N_{P,m}(k)\}\label{eq:expr3}\\\notag
&=& \left(\frac{\beta_{0}}{n_{P}(i)}\sum_{k\in N_{P}(i)}w_{ki}^{[m]}+1\right)\tau_{i}\\\notag
&& +\frac{\beta_{0}}{n_{P}(i)}\sum_{k\in N_{P}(i)}\left(\sum_{j\in N_{P,m+1}(k)\backslash\{i\}}w_{kj}^{[m]}\tau_{j}+w_{kk}^{[m]}\tau_{k}\right)\nonumber \\
&=& \tau_{i}+\frac{\beta_{0}}{n_{P}(i)}\sum_{k\in N_{P}(i)}s_k^{[m]}(\mathcal{I}_{k,m}).\nonumber 
\end{eqnarray}
Thus we find that for any $m,m'>0$, 
\begin{align}
\label{rel}
&|\tilde{s}_i^{[m+1]}(\mathcal{I}_{i,m+1})-\tilde{s}_i^{[m'+1]}(\mathcal{I}_{i,m'+1})|\\ \notag
&\le\left|\beta_{0}\right|\frac{1}{n_{P}(i)}\sum_{k\in N_{P}(i)}|\tilde{s}_{k}^{[m]}(\mathcal{I}_{k,m})-\tilde{s}_{k}^{[m']}(\mathcal{I}_{k,m'})|.
\end{align}

Let $\mathscr{F}$ be the collection of all the $\mathcal{I}$-measurable
$\mathbf{R}^{n}$-valued maps $f=(f_{i})_{i\in N}$ such that $\mathbf{E}[f_{i}^{2}]<\infty$
for each $i\in N$. We endow $\mathscr{F}$ with a pseudo metric:
for $f=(f_{i})_{i\in N}$ and $g=(g_{i})_{i\in N}$, 
\begin{eqnarray}
\|f-g\|_{2}=\max_{1\le i\le n}\sqrt{\mathbf{E}[(f_{i}-g_{i})^{2}]}.\label{norm}
\end{eqnarray}
As usual, we view $(\mathscr{F},\|\cdot\|_{2})$ as a collection of
equivalence classes on which $d(f,g)\equiv\|f-g\|_{2}$ is a metric.
Since 
\begin{eqnarray}
\sqrt{\frac{1}{n}\sum_{i\in N}\mathbf{E}[(f_{i}-g_{i})^{2}]}\le\|f-g\|_{2}\le\sqrt{\sum_{i\in N}\mathbf{E}[(f_{i}-g_{i})^{2}]},\label{ineq6}
\end{eqnarray}
the metric space $(\mathscr{F},\|\cdot\|_{2})$ is complete, a property
inherited from the completeness of an $L_{2}$ space.

Each strategy profile $\tilde{s}(\mathcal{I}_{m})^{[m]}(\omega)$ from
game $\Gamma_{m}$ belongs to $(\mathscr{F},\|\cdot\|_{2})$. Consider
a sequence of best response strategy profiles $\{\tilde{s}^{[m]}(\mathcal{I}_{m})\}_{m=1}^{\infty}$.
Certainly by (\ref{rel}) and the fact that $|\beta_{0}|<1$, the
sequence $\{\tilde{s}^{[m]}(\mathcal{I}_{m})\}_{m=1}^{\infty}$
is Cauchy in $(\mathcal{F},\|\cdot\|_{2})$, and has a limit, say,
$\tilde{s}_{\infty}(\mathcal{I})$ in $\mathscr{F}$ by its completeness.
Now, for the first step of the proof, it remains to show that $\tilde{s}_{\infty}(\mathcal{I})$ is identical
to $\tilde{s}^{\mathsf{BNE}}(\mathcal{I})$ almost everywhere, where $\tilde{s}^{\mathsf{BNE}}(\mathcal{I})$
is defined as a fixed point to:

\begin{equation}
\tilde{s}_{i}^{\mathsf{BNE}}(\mathcal{I})=\tau_{i}+\beta_{0}\frac{1}{n_{P}(i)}\sum_{k\in N_{P}(i)}\tilde{s}_{k}^{\mathsf{BNE}}(\mathcal{I}).
\end{equation}

To see this, let us view $\tilde{s}^{[m+1]}(\mathcal{I}_{m+1})$
as an $n$-dimensional column vector of $\tilde{s}_i^{[m+1]}(\mathcal{I}_{i,m+1}),i\in N$
and $A$ an $n\times n$ matrix whose $(i,j)$-th entry is given by
$1\{j\in N_{P}(i)\}/n_{P}(i)$. Then we can rewrite (\ref{der3})
as 
\begin{eqnarray*}
	\tilde{s}^{[m]}(\mathcal{I}_{m})=\tau+\beta_{0}A\tilde{s}^{[m-1]}(\mathcal{I}_{m-1}),
\end{eqnarray*}
where $\tau=(\tau_{i})_{i\in N}$. This implies that 
\begin{eqnarray*}
	\tilde{s}_{\infty}(\mathcal{I})-(\tau+\beta_{0}A\tilde{s}_{\infty}(\mathcal{I}))=\tilde{s}_{\infty}(\mathcal{I})-\tilde{s}^{[m]}(\mathcal{I}_{m})+\beta_{0}A(\tilde{s}^{[m-1]}(\mathcal{I}_{m-1})-\tilde{s}_{\infty}(\mathcal{I})).
\end{eqnarray*}
Thus we have 
\begin{eqnarray*}
	&  & \left\Vert \tilde{s}_{\infty}(\mathcal{I})-(\tau+\beta_{0}A\tilde{s}_{\infty}(\mathcal{I}))\right\Vert _{2}\\
	& \le & \left\Vert \tilde{s}_{\infty}(\mathcal{I})-\tilde{s}^{[m]}(\mathcal{I}_{m})\right\Vert _{2}+|\beta_{0}|\|A\|\left\Vert \tilde{s}^{[m-1]}(\mathcal{I}_{m-1})-\tilde{s}_{\infty}(\mathcal{I})\right\Vert _{2},
\end{eqnarray*}
where $\|A\|=\sqrt{\text{tr}(A'A)}$. Note that $\|A\|<\infty$ and does
not depend on $m$. Hence by sending $m\rightarrow\infty$, we have
\begin{eqnarray}
\left\Vert \tilde{s}_{\infty}(\mathcal{I})-(\tau+\beta_{0}A\tilde{s}_{\infty}(\mathcal{I}))\right\Vert _{2}=0.\label{equiv}
\end{eqnarray}
Since $|\beta_{0}|<1$ and $A$ is row normalized, the matrix $I-\beta_{0}A$
is invertible and the row sums of $(I-\beta_{0}A)^{-1}$ are uniformly
bounded (e.g. see \citesupp{Lee:02:ET}, p.257). Therefore, if we
define 
\begin{eqnarray*}
	\tilde{s}^*(\mathcal{I})=(I-\beta_{0}A)^{-1}\tau,
\end{eqnarray*}
we have $\|\tilde{s}^*(\mathcal{I})\|_{2}<\infty$ by (\ref{moment bound}).
On the other hand, it is not hard to see that $\tilde{s}^*(\mathcal{I})$
is almost everywhere identical to the equilibrium strategy profile
$\tilde{s}^{\mathsf{BNE}}(\mathcal{I})$. Also, by (\ref{equiv}), $\tilde{s}^*(\mathcal{I})$
is almost everywhere identical to $\tilde{s}_{\infty}(\mathcal{I})$. The first part of the proof follows by (\ref{ineq6}) and the fact that 
\begin{eqnarray*}
	\mathbf{E}\left[\max_{i\in N}(f_{i}-g_{i})^{2}\right]\le\sum_{i\in N}\mathbf{E}[(f_{i}-g_{i})^{2}].
\end{eqnarray*}

As a result, we have the convergence of behavioral strategies to equilibrium strategies in the complete information analogue to our incomplete information game.

To complete our proof, we note that the actual behavioral strategies with incomplete information (with potentially nonzero $\eta_i$'s) are given by:
\begin{equation}
s_{i}^{[m]}=\tilde{s}_{i}^{[m]} + \eta_i.
\end{equation}
This follows immediately from using equation (\ref{BR linear}), Theorem \ref{thm: best response2} and Assumption 3.1. Analogously, the equilibrium strategies from game $\Gamma_\infty$ are given by:
\begin{equation}
s_{i}^{\mathsf{BNE}}=\tilde{s}_{i}^{\mathsf{BNE}} + \eta_i.
\end{equation}

As a result, convergence of $\tilde{s}_{i}^{[m]}$ to $\tilde{s}_{i}^{\mathsf{BNE}}$ implies convergence of $s_{i}^{[m]}$ to $s_{i}^{\mathsf{BNE}}$, which completes the proof. $\blacksquare$

\section{Empirical Results Based on a Game with the First Order Sophisticated Players}

In this section, we report the empirical results based on the game $\Gamma_1$ populated by the first order sophisticated players. The results are found in Table \ref{statecapacitytable_fos}. 

Compared to the results with simple types (in Table \ref{statecapacitytable}), the confidence sets for $\beta$ in the game with first order sophisticated types are wider. For all specifications, the confidence sets for $\beta$ for the FOS types includes (most or all of) the confidence set for $\beta$ for the simple type. In general, the average marginal effects are similar across both models.\footnote{A caveat is that the numerical implementation for the specifications in Columns (3) and (4) in Table \ref{statecapacitytable_fos} appear more sensitive than the others for $\beta$ close to -1, relying on how the grid is set for those values. Throughout the empirical results in the paper, we present results for a grid of $\beta \in [-0.75,0.75]$. As can be seen in Table \ref{statecapacitytable_fos}, Columns (3)-(4) include a disjoint subset at the smallest values of the grid. This interval does not show up in any of the other specifications (simple type or first order sophisticated), disappears in other specifications similar to Columns (3)-(4) (e.g. when we restrict the set of covariates for land and river quality) and is not present when we consider a smaller grid. We attribute this to (i) lack of variation as $\beta \to -1$ due to more extensive cross-sectional dependence (recall that we need $\left|\beta\right|$ away from 1 by Assumption 2.1) (ii) less variation coming from the instruments $\tilde{Z}_{i}$ when $\beta$ is smaller.  The positive subset in the confidence interval for $\beta$ is stable across specifications, and we focus on it for the discussion of our results.} We note that the instruments used below are the same as those for simple players: polynomials of $X_{i,1}$ and a set of instruments that captures the cross-sectional dependence along the payoff graph ($\tilde{Z}_{i}=n_P(i)^{-1} \sum_{j \in N_{P}(i)} \lambda_{ij}X_{j,1}$). 

Given the results for the empirical model based on the game $\Gamma_1$, we then conduct the model selection procedure developed in Appendix F. This selects among the simple type and first order sophisticated type models. Table \ref{modelselection} presents the results. As the results show, the data did not reject either of the models $\mathsf{ST}$ and $\mathsf{FS}$ at 5\%.

\bigskip
\bigskip

\begin{table}[t]
	\begin{centering}
		\small
		\caption{\small State Presence and Networks Effects across Colombian Municipalities, First Order Sophisticated Types}
		\label{statecapacitytable_fos}
		\resizebox{\columnwidth}{!}{%
			\begin{tabular}{cccccc}
				\hline 
				\hline 
				& &\multicolumn{4}{c}{Outcome: The Number of State Employees}
				\\
				\cline{2-6}
				&  & Baseline &  Distance to Highway & Land Quality & Rivers\\
				&  & (1) & (2) & (3) & (4) \\
				\hline
				\multicolumn{1}{c}{} & &  &  &  & \\
				$\beta_0$ &  &$[0.17,0.35]$ & $[0.17,0.35]$ & $[-0.75,-0.69]\cup[0.17,0.45]$ & $[-0.75,-0.56]\cup[0.08,0.46]$ \\
				\tabularnewline
				%Confidence Set for $\beta_0$
				$dy_i/d(\text{colonial state}$  & & $[-0.055,0.004]$ &$[-0.046,0.001]$&$[-0.045,0.005]$ &$[-0.032,0.001]$\\
				$~\text{officials})$ & & & & & \\
				\tabularnewline
				Average & & & & & \\
				$dy_i/d(\text{colonial state}$  & & $[-1.222,3.667]$ &$[-1.118,2.611]$&$[-0.926,3.047]$ &$[-3.953,2.564]$\\
				$~\text{agencies})$& & & & & \\
				\tabularnewline
				Average & & & & & \\
				$dy_i/d(\text{distance to}$  & & $[-0.010,0.009]$ &$[-0.008,0.010]$&$[-0.015,0.021]$ &$[-0.013,0.022]$ \\
				$~\text{Royal Roads})$ & & & & & \\
				%&  &  &  &  & \tabularnewline
				\hline 
				$n$ & & 1018 & 1018 & 1003 & 1003\\
				\hline 
				\multicolumn{1}{c}{} & &  &  &  &\\
			\end{tabular}
		}
		\par\end{centering}
	\parbox{6.2in}{\footnotesize
		Notes: Confidence sets for $\beta$ are presented in the table, obtained from inverting the test statistic $T(\beta)$ from Section \ref{sec: moment restrictions} for First Order Sophisticated types, with confidence level of 95\%.  The critical values in the first row come from the asymptotic statistic.  Downweighting is used.  The average marginal effects for historical variables upon state capacity are also shown.  The marginal effect of Colonial State Officials is equal to its $\gamma$ coefficient. The marginal effect for Distance to Royal Roads for municipality $i$ equals $\gamma_{Royal~Roads} + 2\gamma_{Royal~Roads^2}(Royal~Roads)_i$, where $\gamma_{Royal~Roads}$ is the $\gamma$ coefficient of its linear term, and $\gamma_{Royal~Roads^2}$ is the coefficient of its quadratic term, as this variable enters $X_1$ as a quadratic form. The analogous expression holds for the variable Colonial State Agencies.  We show the average marginal effect for these two variables. We then present the confidence set for these marginal effects, computed by the inference procedure on $a'\gamma$ developed in Section 3.
		All specifications include controls of latitude, longitude, surface area, elevation, rainfall, as well as Department and Department capital dummies.  Instruments are constructed from payoff neighbors' sum of the $G_P$ neighbors values of the historical variables Total Crown Employees, Colonial State Agencies, Colonial State Agencies squared, population in 1843, distance to Royal Roads, distance to Royal Roads squared, together with the non-linear function $\tilde{Z}_{i}=n_P(i)^{-1} \sum_{j \in N_{P}(i)} \lambda_{ij}X_{j,1}$.  Column (2) includes distance to current highway in $X_1$, Column (3) expands the specification of Column (2) by also including controls for land quality (share in each quality level). Column (4) controls for rivers in the municipality and land quality, in addition to those controls from Column (1).  
		One can see that the results are very stable across specifications.\bigskip\bigskip}
\end{table}

\begin{table}[t]
	\begin{centering}
		\small
		\caption{\small Model Selection, Simple Type or First Order Sophisticated}
		\label{modelselection}
		\resizebox{0.9\columnwidth}{!}{%
			\begin{tabular}{cccccc}
				\hline 
				\hline 
				& &\multicolumn{4}{c}{Specification}
				\\
				\cline{2-6}
				&  & (1) & (2) & (3) & (4) \\
				\hline
				\\
				\multicolumn{1}{c}{\textit{Test Statistics and p-values (in parentheses)}} & &  &  &  & \\
				$T_{\mathsf{ST}}$ &  &$3.361~(0.762)$ & $ 4.705~(0.582)$ & $0.495~(0.998)$ & $4.756~(0.575)$ \\
				
				\multicolumn{1}{c}{} & &  &  &  & \\
				$T_{\mathsf{FS}}$ &  &$ 4.260~(0.642)$ & $4.897~(0.557)$ & $ 1.018~(0.985)$ & $5.010~(0.543)$ \\
				\\
				\hline
				\tabularnewline
				\multicolumn{1}{c}{\textit{Models Not Rejected}} & &  &  &  & \\
				\\
				$\hat{S}$ &  &$\{\mathsf{ST},\mathsf{FS}\}$ & $\{\mathsf{ST},\mathsf{FS}\}$& $\{\mathsf{ST},\mathsf{FS}\}$ & $\{\mathsf{ST},\mathsf{FS}\}$ \\
				\tabularnewline
				\hline
				\tabularnewline
				
			\end{tabular}
		}
		\par\end{centering}
	\parbox{6.2in}{\footnotesize
		Notes: The table shows the results of the Model Selection test, developed in Appendix F. Here $\mathsf{ST}$ refers to the simple type model, $\mathsf{FS}$ to the First Order Sophisticated. The critical value for the test ($c_{1-\alpha/2}$), with 6 degrees of freedom ($M-d$) and level $\alpha = 0.05$, is 14.449. The specifications in each column are the same as those in Table \ref{statecapacitytable_fos}. The first panel shows the values of the statistics, with the $p$-values in parentheses. The bottom panel shows the set $\hat{S}$ of models that are not rejected by the test.\bigskip\bigskip}
\end{table}

\FloatBarrier

\bibliographystylesupp{econometrica}
\bibliographysupp{matching_networks_A3}

\end{document}